\documentclass[a4paper,11pt]{article}
\pdfoutput=1
\usepackage{jheppub}

\usepackage{epsf}
\usepackage{epsfig}
\usepackage{subfigure}
\usepackage{mathtools}
\usepackage{hhline}
\usepackage{float}
\usepackage{multirow}
\usepackage{nicefrac}
\usepackage{epstopdf}
\usepackage{slashed}
\usepackage{xcolor}
\usepackage{url}

\usepackage{braket}
\usepackage{subdepth}

\newcommand{\be}{\begin{eqnarray}}
\newcommand{\ee}{\end{eqnarray}}

\newcommand{\ba}{\begin{array}}
\newcommand{\ea}{\end{array}}
\newcommand{\bee}{\begin{equation}\ba{c}}
\newcommand{\eee}{\ea\end{equation}}

\newcommand{\bi}{\begin{itemize}}
\newcommand{\ei}{\end{itemize}}

\allowdisplaybreaks

\def\nnb{\nonumber}

\def\as{\tilde{\alpha}_s}

\def\gev{{\rm GeV}}
\def\mev{{\rm MeV}}
\def\be{\begin{equation}}
\def\ee{\end{equation}}
\def\bea{\begin{eqnarray}}
\def\eea{\end{eqnarray}}
\def\nnb{\nonumber}
\def\dps{\displaystyle}

\def\bbuildrel#1_#2^#3{\mathrel{\mathop{\kern 0pt#1}\limits_{#2}^{#3}}}
\def\slash#1{\setbox0=\hbox{$#1$}#1\hskip-\wd0\dimen0=5pt\advance
       \dimen0 by-\ht0\advance\dimen0 by\dp0\lower0.5\dimen0\hbox
         to\wd0{\hss\sl/\/\hss}}
\def\gev{{\rm GeV}}
\def\mev{{\rm MeV}}

\newcommand{\f}{\frac}
\newcommand{\fm}[2]{{\textstyle \frac{#1}{#2}}}
\newcommand{\me}[1]{\langle#1\rangle}

\def\be{\begin{equation}}
\def\ee{\end{equation}}
\def\beq{\begin{eqnarray}}
\def\eeq{\end{eqnarray}}

\def\slash#1{#1 \hskip-0.45em /}

\def\DB0{\partial B_0}

\newcommand{\lsim}
{\;\raisebox{-.3em}{$\stackrel{\displaystyle <}{\sim}$}\;}
\newcommand{\gsim}
{\;\raisebox{-.3em}{$\stackrel{\displaystyle >}{\sim}$}\;}

\def\Cl2{\mbox{Cl}_2}

\def\slash#1{#1 \hskip-0.45em /}

\def\s{\hat s}

\newcommand{\spp}{\vphantom{$\Big($}}

\usepackage{color}

\definecolor{Brown}{rgb}{0.5,0.25,0}

\title{Long distance effects in inclusive rare $B$ decays and phenomenology of $\bar{B}\to X_d \ell^+\ell^-$}
\author{Tobias Huber$^1$,}
\author{Tobias Hurth$^2$,}
\author{Jack Jenkins$^3$,}
\author{Enrico Lunghi$^3$,}
\author{Qin Qin$^{1}$,}
\author{K.~Keri Vos$^1$}
\affiliation{
$^1$Theoretische Physik 1, Naturwissenschaftlich-Technische Fakult\"at, Universit\"at Siegen, Walter-Flex-Strasse 3, D-57068 Siegen, Germany\\
$^2$PRISMA+ Cluster of Excellence and Institute for Physics (THEP) Johannes Gutenberg University, D-55099 Mainz, Germany\\
$^3$Physics Department, Indiana University, Bloomington, IN 47405, USA \\
}
\emailAdd{huber@physik.uni-siegen.de}
\emailAdd{tobias.hurth@cern.ch}
\emailAdd{jackjenk@iu.edu}
\emailAdd{elunghi@indiana.edu}
\emailAdd{qin@physik.uni-siegen.de}
\emailAdd{keri.vos@uni-siegen.de}

\abstract{Rare inclusive $B$ decays such as $\bar{B}\to X_{s(d)} \ell^+\ell^-$ are interesting probes for physics beyond the Standard Model. Due to the complementarity to their exclusive counterparts, they might shed light on the anomalies currently seen in exclusive $b \to s$ transitions. Distinguishing new-physics effects from the Standard Model requires precise predictions and necessitates the control of long distance effects. In the present work we revisit and improve the description of various long distance effects in inclusive decays such as charmonium and light-quark resonances, nonfactorizable power corrections, and cascade decays. We then apply these results to a state-of-the-art phenomenological study of $\bar{B}\to X_d \ell^+\ell^-$, including also logarithmically enhanced QED corrections and the recently calculated five-body contributions. To fully exploit the new-physics potential of inclusive flavour-changing neutral current decays, the $\bar{B}\to X_d \ell^+\ell^-$ observables should be measured in a dedicated Belle II analysis.}
\preprint{
\begin{minipage}{3cm}
\small
\flushright
QFET-2019-07\\
SI-HEP-2019-09\\
MITP/19-047
\end{minipage}}

\begin{document}

\maketitle


\section{Introduction}
\label{sec:introduction}

Since the Higgs discovery at the LHC in 2012~\cite{Aad:2012tfa,Chatrchyan:2012xdj} completed the particle content of the Standard Model (SM) of particle physics,
no new fundamental degrees of freedom have been discovered in direct searches for physics beyond the SM (BSM).
The current situation therefore underlines the importance of indirect searches for BSM particles via virtual effects. The latter requires
precision studies of low-energy observables, most prominently in quark and lepton flavour physics.

Inclusive flavour-changing neutral current (FCNC) decays of $B$ mesons provide a perfect environment for this kind of program for several reasons.
First, FCNC decays are especially sensitive to potential BSM effects because they proceed through loop-suppressed electroweak interactions in the SM.
Second, the necessary precision can be achieved on both the theoretical and experimental side. Theoretically, \emph{inclusive} FCNC $B$-meson decays can be reliably predicted using an Operator Product Expansion (OPE), in which non-perturbative effects appear as corrections to the partonic rate at inverse powers of the heavy $b$-quark mass. 

The theory approach to inclusive FCNC decays is in this sense somewhat different compared to exclusive ones; in particular, the underlying hadronic uncertainties in inclusive modes are
largely independent of those in exclusive transitions. Hence, one useful way to shed light on the nature of the anomalies currently outstanding in exclusive $B$ decays at various experiments~\cite{Lees:2012tva,Wehle:2016yoi,Aaij:2019wad,Aaij:2017vbb,Aaij:2016cbx,Aaij:2016kqt,Aaij:2014pli,Aaij:2015oid,Aaij:2015esa,Aaij:2015xza,Sirunyan:2017dhj,Aaboud:2018krd,Abdesselam:2019lab} is a cross-check via the corresponding observables in the inclusive modes.
Indeed, a study on the combined new-physics sensitivity clearly revealed the synergy and complementarity of exclusive versus inclusive FCNC decays~\cite{Kou:2018nap}.

The FCNC decays that have been studied most intensively are $b \to s$ transitions. The amplitude for these decays contains the three combinations of
Cabibbo-Kobayashi-Maskawa (CKM) elements $V^\ast_{ts}V_{tb}$, $V^\ast_{cs}V_{cb}$, and $V^\ast_{us}V_{ub}$. In an expansion in the Wolfenstein parameter
$\lambda \approx \left| V_{us} \right| \simeq 0.22$ they start at orders ${\cal O}(\lambda^2)$, ${\cal O}(\lambda^2)$, and ${\cal O}(\lambda^4)$, respectively.
Neglecting $V^\ast_{us}V_{ub}$ compared to the other two and using CKM unitarity, the $b \to s$ amplitudes are thus proportional to the {\emph{single}}
combination $V^\ast_{ts}V_{tb}$. In $b \to d$ transitions the situation is different since $V^\ast_{td}V_{tb}$, $V^\ast_{cd}V_{cb}$, and $V^\ast_{ud}V_{ub}$ are all ${\cal O}(\lambda^3)$.
This renders the size of the $b \to d$ rate about two orders of magnitude smaller compared to its $b \to s$ counterpart.
On the other hand, the $b \to d$ unitarity triangle is non-degenerate.
Trading $V^\ast_{cd}V_{cb}$ in favour for the other two via CKM unitarity, one obtains a piece proportional to $V^\ast_{td}V_{tb}$ (which is analogous to the $V^\ast_{ts}V_{tb}$ term in $b \to s$ transitions except for a replacement of the overall CKM factor) and a piece proportional to $V^\ast_{ud}V_{ub}$ that contains the effective operators $P_{1,2}^u$
whose matrix elements are not CKM-suppressed in the $b \to d$ case. 

While $b \to d \ell^+\ell^-$ decays have played little role in the program of flavour experiments so far because of their low statistics, 
they will become accessible in the Belle~II era.
A naive rescaling of the corresponding ${\bar B \to X_{s} \, \ell^+\ell^-}$ errors given in~\cite{Kou:2018nap} (without taking detector efficiencies etc.\ into account) shows promising prospects for this decay at Belle~II. Therefore, it would be worthwhile to carry out a dedicated ${\bar B \to X_{d} \, \ell^+\ell^-}$ analysis at Belle~II. 
Besides serving as a cross-check of inclusive ${\bar B \to X_{s} \, \ell^+\ell^-}$ and exclusive $b\to d \ell^+\ell^-$
measurements it has the potential to yield important information on the phenomenon of CP violation.

On the theoretical side the latest phenomenological study of ${\bar B \to X_{d} \, \ell^+\ell^-}$ dates back fifteen years~\cite{Asatrian:2003vq}. Since it is based on short-distance partonic contributions only and includes neither power corrections nor effects from resonances, it is lacking a lot
of features that are inherent to inclusive semileptonic FCNC decays. In view of the prospects on the experimental side, a new theory analysis of ${\bar B \to X_{d} \, \ell^+\ell^-}$ including nonperturbative features is therefore timely.

In the theoretical description of inclusive ${\bar B \to X_{d} \, \ell^+\ell^-}$ decays, many of the results obtained in inclusive ${\bar B \to X_{s} \, \ell^+\ell^-}$ apply after trivial modifications. The short-distance partonic amplitude of the latter is known to NLO~\cite{Misiak:1992bc,Buras:1994dj} and NNLO~\cite{Bobeth:1999mk,Gambino:2003zm,Gorbahn:2004my,Asatryan:2001zw,Asatrian:2001de,Asatryan:2002iy,Ghinculov:2002pe,Asatrian:2002va,Asatrian:2003yk,Ghinculov:2003bx,Ghinculov:2003qd,Greub:2008cy,Bobeth:2003at,deBoer:2017way} in QCD, and to NLO in QED~\cite{Huber:2005ig,Huber:2007vv,Huber:2015sra}. Power-corrections that scale as $1/m_b^2$~\cite{Falk:1993dh,Ali:1996bm,Chen:1997dj,Buchalla:1998mt}, $1/m_b^3$~\cite{Bauer:1999kf,Ligeti:2007sn}, and $1/m_c^2$~\cite{Buchalla:1997ky} have been analysed. The contributions specific to ${\bar B \to X_{d} \, \ell^+\ell^-}$ decays are available from~\cite{Asatrian:2003vq,Seidel:2004jh}, where two-loop virtual and bremsstrahlung corrections involving $P^u_{1,2}$ have been computed. Recently, also contributions from multi-particle final states at leading power have been calculated analytically~\cite{Huber:2018gii}. In the present work we derive the logarithmically enhanced QED corrections to the matrix elements of $P^u_{1,2}$.

Whereas the theoretical prediction of the branching ratio in the low-$q^2$ region is well under control and a precision below $\sim 8\%-10\%$ can be achieved
the same quantity in the high-$q^2$ region suffers from large uncertainties of ${\cal O}(40\%)$ due to the failure  of the heavy-mass expansion near the kinematic endpoint:  The partonic
rate tends to zero while the local $1/m_b^2$ and $1/m_b^3$ power corrections within the heavy mass expansion approach a finite, non-zero value. It was found in~\cite{Neubert:2000ch,Bauer:2001rc} that the expansion is effectively in inverse powers of $m_b(1-\sqrt{s_{\rm min}}/m_b)$ and depends on the lower dilepton mass cut $s_{\rm min}$. Therefore, only {\emph{integrated}} observables are meaningful in the
high-$q^2$ region. In practice the large uncertainty originates from poorly known HQET matrix elements of dimension five and six operators that scale as $1/m_b^2$ and $1/m_b^3$, respectively.
In the present work we obtain their values and uncertainties from analyses of moments of inclusive charged-current semi-leptonic $B$~\cite{Gambino:2016jkc} and $D$ decays~\cite{Gambino:2010jz}.
We emphasize that the precision of theoretical predictions for semileptonic FCNC decays in the high-$q^2$ region would greatly benefit from further studies and lattice calculations of these HQET matrix elements.
In order to reduce the uncertainties from $1/m_b^2$ and $1/m_b^3$ corrections, it was proposed in~\cite{Ligeti:2007sn} to normalise the ${\bar B \to X_{s} \, \ell^+\ell^-}$ rate to the inclusive semi-leptonic $\bar B^0 \to X_u \ell \nu$ rate {\emph{with the same dilepton mass cut}}. Subsequent phenomenological analyses showed indeed a pronounced reduction of the uncertainties for ${\bar B \to X_{s} \, \ell^+\ell^-}$~\cite{Huber:2007vv,Huber:2015sra} and we confirm this behaviour for ${\bar B \to X_{d} \, \ell^+\ell^-}$ in the present work.

Besides, long distance effects that are not captured by the OPE play an essential role in the phenomenology of inclusive ${\bar B \to X_{s(d)} \, \ell^+\ell^-}$ decays, the most prominent coming from
intermediate charmonium resonances $J/\psi$ and $\psi(2S)$ which show up as large peaks in the dilepton invariant mass spectrum
of any angular observable. For ${\bar B \to X_{d} \, \ell^+\ell^-}$, resonances with a $u\bar{u}$ component such as $\rho$ and $\omega$ are also relevant.
The resonance regions can be removed by appropriate kinematic cuts in the dilepton invariant mass squared $q^2$. This leads to the so-called low-$q^2$ region
$1\,{\rm GeV}^2 < q^2 < 6\,{\rm GeV}^2$ and the high-$q^2$ region with $q^2 > 14.4\,{\rm GeV}^2$.
Whereas the low-$q^2$ region is only affected by the tail of the $c\bar c$ peaks, rather broad resonances are present in the high-$q^2$ region itself.
One way of dealing with the resonances was proposed by Kr\"uger and Sehgal (KS)~\cite{Kruger:1996cv,Kruger:1996dt}.
They relied on the assumption that the $c\bar c$ loop and the $b \to s(d)$ transition factorise into two color-singlet currents,
and used a dispersion relation to connect the electromagnetic vacuum polarisation, whose imaginary part is proportional to the hadronic $R$-ratio,
to the $b\to s(d)c\bar c \to s(d)\ell^+\ell^-$ long distance amplitude.
In the present work, we revisit, refine and improve the KS approach in several respects. We use all available data from BESII and BaBar on $e^+e^- \to$~hadrons
as well as from ALEPH on $\tau \to \nu +$hadrons for a precise description of the imaginary part of the vacuum polarisation.
Moreover, we carefully investigate the impact of the choice of the subtraction point of the dispersive integral, and the replacement of the
perturbative loop functions by the KS functions. Finally, we comment on the size of the uncertainties that originate from the KS integral and their impact
on the ${\bar B \to X_{d} \, \ell^+\ell^-}$ observables.

It has been pointed out in the literature that color-octet production of charmonium resonances can be
sizeable~\cite{Ko:1995iv,Beneke:1998ks,Beneke:2009az,Khodjamirian:2010vf,Lyon:2014hpa}, and that the pure color-singlet treatment by the KS approach does not capture
the full size of the $\psi$ resonances. In the present article we further elaborate on the size and treatment of color-octet $c\bar c$ production and the impact on ${\bar B \to X_{s(d)} \, \ell^+\ell^-}$ observables. To cure  the situation a purely phenomenological factor has been introduced~\cite{Kruger:1996cv} to reproduce
the hadronic branching fraction ${\cal B} (\bar B \to \psi X_s)$. However, as was already argued in refs.~\cite{Ghinculov:2003qd,Huber:2007vv}, the introduction of such kind of factor  leads to a double-counting because nonfactorizable corrections due to a $c \bar c$ loop are already taken into account as one of the so-called  resolved contributions in the low-$q^2$ region. These are nonlocal power corrections which occur when other operators than the leading ones are considered in the effective field theory. They indicate a breakdown of the local heavy mass expansion in $\Lambda_{\rm QCD}/ m_b$.  Recently the factorisation of these nonlocal power corrections in ${\bar B \to X_{s} \, \ell^+\ell^-}$ was analysed within the soft-collinear effective theory
(SCET)~\cite{Hurth:2017xzf,Benzke:2017woq} by systematically computing these resolved contributions in the low-$q^2$ region.
Furthermore, it was shown that in the high-$q^2$ region the dominating power contribution (due to operators beyond the leading ones),
the nonfactorizable $c \bar c$ contribution, can be expanded in local operators  again and can be treated along the lines of~\cite{Voloshin:1996gw,Buchalla:1997ky}. 

The systematic SCET analysis of the resolved power corrections in the low-$q^2$ region allows for another crucial observation: the resolved $u \bar u$ contribution vanishes at order $\Lambda_{\rm QCD}/m_b$ in CP-averaged observables in the low-$q^2$ region~\cite{Hurth:2017xzf,Benzke:2017woq}, which significantly reduces the uncertainties in the ${\bar B \to X_{d} \, \ell^+\ell^-}$. For the CP asymmetry, these long distance effects dominate, making our theoretical prediction less clean. Irrespectively, the CP asymmetry remains an interesting observable because it might receive sizeable contributions from BSM effects.

An additional long distance effect at low-$q^2$ comes from cascade decays $\bar B \to X_1 (c\bar{c} \to X_2 \ell^+ \ell^-)$ through the radiative decay of a narrow charmonium resonance such as $\eta_c, J/\psi, \chi_{cJ}$ etc. They form a background that has to be removed by suitable kinematic cuts. Inclusive radiative charmonium decays have been discussed in the context of $\bar B \to X_s \gamma$~\cite{Buchalla:1997ky} and $\bar B \to X_s \ell^+ \ell^-$~\cite{Beneke:2009az}. Here we revisit and systematically investigate the role these decays play as a background, as well as their dependence on a kinematic cut on the hadronic invariant mass $M_X$.

A cut on the hadronic mass $M_X$ might still be required on the experimental side to remove other sources of background at Belle II. The effect of such an $M_X$ cut was previously analysed in ${\bar B \to X_{s} \, \ell^+\ell^-}$ in~\cite{Lee:2005pk,Lee:2005pwa,Lee:2008xc}.  However,  the authors of ref.~\cite{Bell:2010mg} indicated a conceptual problem in those analyses and the authors of refs.~\cite{Hurth:2017xzf,Benzke:2017woq} showed that the assumption made in refs.~\cite{Lee:2005pk,Lee:2005pwa,Lee:2008xc} that the photon virtuality in the low-$q^2$ region scales as a hard mode in SCET is problematic since the kinematics in the presense of an $M_X$ cut implies the scaling of $q^2$ as (anti-)hard-collinear in the low-$q^2$ region. This leads to a different matching and power counting, as well as to the existence of resolved contributions within SCET. It was shown in refs.~\cite{Hurth:2017xzf,Benzke:2017woq} that the resolved contributions represent an irreducible uncertainty even in the absence of an $M_X$-cut. The results of the numerical analysis of these corrections, as given in~\cite{Hurth:2017xzf,Benzke:2017woq}, are used in the phenomenological part of the present paper. Finally, we emphasize that our predictions are given for the case without a hadronic mass cut, leaving such a study for future work.

This article is organised as follows. In sections~\ref{sec:observables} and~\ref{sec:master} we define the ${\bar B \to X_{d} \, \ell^+\ell^-}$ obs\-ervables under consideration and give master formulas
for their phenomenological implementation, respectively. Section~\ref{sec:long} contains our study of long distance contributions such as the $q\bar q$ resonances,
cascade decays, and resolved contributions. Sections~\ref{sec:inputs} and~\ref{sec:results} contain the extraction of our input parameters and the
phenomenological results, respectively. We conclude in section~\ref{sec:conclusion}. Appendix~\ref{sec:apptwoloop} contains the expressions for certain two-loop functions, while appendix~\ref{sec:appomegaem} contains those of the logarithmically enhanced QED corrections.


\section{Definition of the observables}
\label{sec:observables}

In this work, we consider the CP-averaged branching ratio, forward-backward asymmetry and
the CP asymmetry of the inclusive $\bar{B}\to X_d \ell^+\ell^-$ decay. Additional angular-distribution observables~\cite{Lee:2006gs}
are left for possible future studies in case they become accessible experimentally. Alike for the inclusive $\bar{B}\to X_s \ell^+\ell^-$ decays, appropriate kinematic cuts have to be taken in order to remove the large peaks of the $c\bar{c}$ resonances. Here we focus on two regions, the low dilepton mass region $1\ {\rm GeV}^2 <  q^2  < 6\ \rm{GeV}^2$ and the high dilepton mass region $q^2 > 14.4$ GeV$^2$. 

The definitions of the differential decay width ${d\Gamma}/{dq^2}$ and of the differential
forward-backward asymmetry ${dA_{\rm FB}}/{dq^2}$ are given by
\begin{align}
\frac{d\Gamma}{dq^2} &\equiv \int_{-1}^{+1} \frac{d^2\Gamma}{dq^2 dz} dz \; ,\\
\frac{dA_{\rm FB}}{dq^2} &\equiv \int_{-1}^{+1} \frac{d^2\Gamma}{dq^2 dz} {\rm sign}(z) dz \;,
\label{eq:fba}
\end{align}%
where $z = \cos\theta$ and $\theta$ is the angle between the $\ell^+$ and the $B$ meson in the dilepton rest frame.
The differential forward-backward asymmetry ${dA_{\rm FB}}/{dq^2}$ is related to the angular-distribution
observable $H_A(q^2)$ by~\cite{Lee:2006gs}
\begin{align}\label{eq:afb}
\frac{dA_{\rm FB}}{dq^2} = {3\over4}H_A(q^2) \; .
\end{align}
To keep consistent with \cite{Huber:2015sra}, we give the master formula for $H_A(q^2)$, from which ${dA_{\rm FB}}/{dq^2}$ can be derived.
We will calculate the normalized forward-backward asymmetry $\overline{A}_{\rm FB}$ integrated in a region $q_m^2 < q^2 < q_M^2$
\begin{align}\label{eq:Afw}
\overline{A}_{\rm FB}[q_m^2, q_M^2] \equiv
\frac{\int_{q_m^2}^{q_M^2}dq^2 ({dA_{\rm FB}}/{dq^2}) }{\int_{q_m^2}^{q_M^2}dq^2 ({d\Gamma}/{dq^2}) } \; .
\end{align}
The integrations in eq.~\eqref{eq:Afw} are performed numerically.

In the high-$q^2$ region we include hadronic power corrections up to ${\cal O}(1/m_b^3)$. As we will show, similar to
the $b\to s$ case, the uncertainties on these power corrections dominate in that region.
These uncertainties can be significantly reduced by normalising
the $\bar B \rightarrow X_d \ell^+ \ell^-$ decay rate to the
semileptonic $\bar B \rightarrow X_u \ell \bar\nu$ decay rate with the
same $q^2$-cut~\cite{Ligeti:2007sn}:
\begin{equation}
\label{eq:zoltanR}
{\cal R}(s_0) =
\int_{\hat s_0}^1 {\rm d} \hat s \, {{\rm d} {\Gamma} (\bar B \to X_d \ell^+\ell^-) \over {\rm d} \hat s}\,  /\,
\int_{\hat s_0}^1 {\rm d} \hat s \, {{\rm d} {\Gamma} (\bar B \to X_u \ell \nu) \over {\rm d} \hat s}\, ,
\end{equation}
where $\hat{s} = q^2/m_{b,\rm{pole}}^2$.

We note that the above quantities are all CP averaged. In addition, we also calculate the normalized integrated CP
asymmetry, defined by
\begin{align}
\overline{A}_{\rm CP}[q_m^2, q_M^2] \equiv
\frac{\int_{q_m^2}^{q_M^2}dq^2 \left[{d\Gamma(\bar B \to X_d \ell^+\ell^-)}/{dq^2} - {d\Gamma(B \to X_{\bar{d}} \ell^+\ell^-)}/{dq^2}\right] }
{\int_{q_m^2}^{q_M^2}dq^2 \left[{d\Gamma(\bar B \to X_d \ell^+\ell^-)}/{dq^2} + {d\Gamma(B \to X_{\bar{d}} \ell^+\ell^-)}/{dq^2}\right] } \; .
\end{align}


\section{Master formulas for the observables}
\label{sec:master}

As emphasized earlier, the inclusive ${\bar B \to X_{d} \, \ell^+\ell^-}$ decay distinguishes itself from ${\bar B \to X_{s} \, \ell^+\ell^-}$ since the $u$-quark current-current operators are not CKM suppressed and have to be taken into account. The effective Lagrangian is as follows ~\cite{Buchalla:1995vs,Chetyrkin:1997gb},
\begin{align}\label{Leff}
\allowdisplaybreaks
\mathcal{L}_\text{eff} = &\ \mathcal{L}_{\text{QCD} \times \text{QED}}(u,d,s,c,b,e,\mu,\tau)
-\frac{4 G_F}{\sqrt{2}} \sum\limits_{p=u,c}V^*_{pd}V_{pb}\left(C_1(\mu) P_1^p + C_2(\mu) P_2^p\right)\nonumber\\
&+\frac{4 G_F}{\sqrt{2}} V^*_{td} V_{tb}
\left( \sum_{i=3}^{10} C_i(\mu) P_i + \sum_{i=3}^{6} C_{iQ}(\mu) P_{iQ} + C_b(\mu) P_b \right),
\end{align}
where
\begin{align}\label{ope}
P_1^u   &=   (\bar{d}_L \gamma_{\mu} T^a u_L) (\bar{u}_L \gamma^{\mu} T^a b_L),  &P_5   &=  (\bar{d}_L \gamma_{\mu_1}
                     \gamma_{\mu_2}\gamma_{\mu_3}    b_L)\Sigma_q (\bar{q} \gamma^{\mu_1}\gamma^{\mu_2}\gamma^{\mu_3}     q), \nonumber\\
P_2^u   &=  (\bar{d}_L \gamma_{\mu}  u_L) (\bar{u}_L \gamma^{\mu}     b_L),  &P_6 &  =   (\bar{d}_L \gamma_{\mu_1}
                     \gamma_{\mu_2}\gamma_{\mu_3} T^a b_L)\Sigma_q (\bar{q} \gamma^{\mu_1}\gamma^{\mu_2}\gamma^{\mu_3} T^a q), \nonumber\\
P_1^c   &=   (\bar{d}_L \gamma_{\mu} T^a c_L) (\bar{c}_L \gamma^{\mu} T^a b_L),  &P_7   &= {e/(16 \pi^2)} m_b (\bar{d}_L \sigma^{\mu \nu}     b_R) F_{\mu \nu}, \nonumber\\
P_2^c   &=  (\bar{d}_L \gamma_{\mu}  c_L) (\bar{c}_L \gamma^{\mu}     b_L),  &P_8  & =  {g/(16 \pi^2)} m_b (\bar{d}_L \sigma^{\mu \nu} T^a b_R) G_{\mu \nu}^a, \nonumber\\
P_3   &=  (\bar{d}_L \gamma_{\mu}     b_L) \Sigma_q (\bar{q}\gamma^{\mu}     q),   &P_9  & =  (\bar{d}_L \gamma_{\mu} b_L) \Sigma_l (\bar{l}\gamma^{\mu} l), \nonumber\\
P_4   &=  (\bar{d}_L \gamma_{\mu} T^a b_L) \Sigma_q (\bar{q}\gamma^{\mu} T^a q),   &P_{10}   &=  (\bar{d}_L \gamma_{\mu}     b_L) \Sigma_l (\bar{l}\gamma^{\mu} \gamma_5 l), 
\end{align}
and
\begin{align}
P_{3Q} & =  (\bar{d}_L \gamma_{\mu}     b_L) \Sigma_q Q_q(\bar{q}\gamma^{\mu}     q), \nonumber\\
P_{4Q} & =  (\bar{d}_L \gamma_{\mu} T^a    b_L) \Sigma_q Q_q(\bar{q}\gamma^{\mu}  T^a   q),  \nonumber\\
P_{5Q} & =  (\bar{d}_L \gamma_{\mu_1} \gamma_{\mu_2}\gamma_{\mu_3}    b_L)
\Sigma_q Q_q(\bar{q}\gamma^{\mu_1}\gamma^{\mu_2}\gamma^{\mu_3}    q), \nonumber\\
P_{6Q} & =  (\bar{d}_L \gamma_{\mu_1}\gamma_{\mu_2} \gamma_{\mu_3}T^a    b_L)
\Sigma_q Q_q(\bar{q}\gamma^{\mu_1}\gamma^{\mu_2}\gamma^{\mu_3}  T^a   q),  \nonumber\\
 P_b & = {1\over12}[(\bar{d}_L \gamma_{\mu_1} \gamma_{\mu_2}\gamma_{\mu_3}    b_L)
(\bar{b}\gamma^{\mu_1}\gamma^{\mu_2}\gamma^{\mu_3}  b)- 4(\bar{d}_L \gamma_{\mu} b_L) (\bar{b}\gamma^{\mu}  b)]
\end{align}

\vspace{0.1cm}
\noindent and $q=u, d, s, c, b$ and $l$ runs over the three charged lepton flavours.

Similar to the analyses for $\bar{B}\to X_s\ell^+\ell^-$ \cite{Huber:2005ig,Huber:2007vv}, we make a double expansion in
$\tilde{\alpha}_s= \alpha_s(\mu_b)/(4\pi)$
and $\kappa = \alpha_e(\mu_b)/\alpha_s(\mu_b)$, and in the squared amplitude retain terms up to $\mathcal{O}(\alpha_s^3\kappa^3)$. 
In addition, we normalize our observables to the inclusive $\bar B\to X_c e\bar{\nu}$ decay,
\bea
\mathcal{H}_I & = &
{\cal B} (B\to X_c e \bar\nu)_{\rm exp} \;
\left| \frac{V_{td}^* V_{tb}}{V_{cb}} \right|^2 \;
\frac{4}{C} \; \frac{\Phi_{\ell^+\ell^-}^I(\s)}{\Phi_u} \; .
\label{br}
\eea
As mentioned above, among the angular observables $\mathcal{H}_{T,A,L}$~\cite{Lee:2006gs} we consider only the branching ratio ($I=B=T+L$) and the forward-backward asymmetry ($I=A$). In the following, we give the expressions for $\bar{B}\to X_d \ell^+\ell^-$ only. From these, the CP averaged quantities and the CP asymmetry can be trivially obtained.
Here~\cite{Bobeth:2003at,Gambino:2001ew}
\be
C = \left| \frac{V_{ub}}{V_{cb}} \right|^2
       \frac{\Gamma (\bar B\to X_c e\bar\nu)}{\Gamma (\bar B\to X_u e\bar\nu)} \;,
\ee
and  $\Phi_u$ is defined by \cite{Huber:2005ig}
\be \label{bu}
\Gamma (\bar B\to X_u e\bar\nu) =
\frac{G_F^2 m_{b,{\rm pole}}^5}{192 \pi^3} \left| V_{ub}^{}\right|^2 \; \Phi_u.
\ee
Explicitly~\cite{Huber:2015sra},
\begin{align}
\Phi_u &= 1+ \as \varphi^{(1)} + \kappa \left[\frac{12}{23}\left(1-\eta^{-1}\right)\right] + \as^2 \left[\varphi^{(2)} + 2 \beta_0^{(5)} \varphi^{(1)} \ln\left(\frac{\mu_b}{m_b}\right)\right]
+ \frac{\lambda_1}{2 m_b^2}- \frac{9\lambda_2^{\rm eff}}{2 m_b^2} \nnb \\
& + \frac{77}{6}\frac{\rho_1}{m_b^3} -8 \frac{f_u}{m_b^3}+ {\cal O}(\as^3,\kappa^2,\as \kappa,\as\Lambda^2/m_b^2)\; ,\nnb \\[0.5em]
\varphi^{(1)} &= \frac{50}{3}-\frac{8\pi^2}{3} \; , \nnb \\[0.5em]
\varphi^{(2)} &= n_h \left(-\frac{2048 \zeta_3}{9}+\frac{16987}{54}-\frac{340 \pi^2}{81}\right)+n_l
   \left(\frac{256 \zeta_3}{9}-\frac{1009}{27}+\frac{308 \pi^2}{81}\right) \nnb \\
   &-\frac{41848 \zeta_3}{81}+\frac{578 \pi^4}{81}-\frac{104480 \pi^2}{729}+\frac{1571095}{1458}-\frac{848}{27} \pi^2 \ln(2) \; ,
\end{align}
where the $\mathcal{O}(\tilde\alpha^2_s)$ are taken from \cite{vanRitbergen:1999gs}. Here, $n_h=2$ and $n_l=3$ are the numbers of heavy and light quark flavours, respectively, and $\beta_0^{(5)} = 23/3$ is the one-loop QCD $\beta$-function for five active flavours. We include explicitly the power-suppressed $1/m_b^2$ terms $\lambda_{1}, \lambda_{2}$, the $1/m_b^3$ terms $\rho_1$ and the four-quark matrix element $f_u$. These matrix elements are defined and discussed in more detail in section~\ref{sec:inputs}.

The dimensionless function $\Phi_{\ell^+\ell^-}(\s)$ arises from the matrix elements of all the operators
and is written as
\begin{equation}
\Phi_{\ell^+\ell^-}^I(\s) = \text{Re}\;\left(\sum_{i\leq j}\mathcal{R}^{ij}_\text{CKM}C_i^{\text{eff}*}(\mu_b)C_j^{\text{eff}}(\mu_b)\mathcal{H}^I_{ij}(\hat{s})\right) \ ,
\end{equation}
where $i,j = 1u, 2u, 1c, 2c, 3\ldots 10, 3Q\ldots 6Q, b$. The low-scale Wilson coefficients $C_i^{\text{eff}}$
are given explicitly (both analytically and numerically) in \cite{Huber:2005ig} and $C_i^{\text{eff}}$ are unequal to
$C_i$ only for $i=7,8$. To be specific, we use:
\begin{align}\label{eq:C7C8}
C_7^{\rm eff}(\mu_b) {}& \equiv C_7(\mu_b) - \frac{1}{3} C_3(\mu_b) - \frac{4}{9} C_4(\mu_b) -\frac{20}{3}  C_5(\mu_b)- \frac{80}{9} C_6(\mu_b) \ ,
\nonumber \\ 
C_8^{\rm eff}(\mu_b) {}& \equiv C_8(\mu_b) + C_3(\mu_b) - \frac{1}{6} C_4(\mu_b) +20  C_5(\mu_b) - \frac{10}{3} C_6(\mu_b)  \, .
\end{align}

The different CKM prefactors are given by
\begin{align}
\mathcal{R}^{ij}_\text{CKM} =\begin{cases}
               |\xi_u|^2, &\text{for}\ i, j = 1u, 2u;\\
                |1+\xi_u|^2, &\text{for}\ i, j = 1c, 2c;\\
                 -\xi_u^*(1+ \xi_u), &\text{for}\ i=1u, 2u, j=1c, 2c;\\
               -\xi_u^*,&\text{for}\ i = 1u, 2u, j = 3, ..., 10, 3Q\ldots 6Q, b;\\
               1+\xi_u^*,&\text{for}\ i = 1c, 2c, j = 3, ..., 10, 3Q\ldots 6Q, b;\\
               1, &\text{for}\ i, j = 3, ..., 10, 3Q\ldots 6Q, b,\\
            \end{cases}
\end{align}
with $\xi_u \equiv  ({V_{ud}^* V_{ub}})/({V_{td}^*V_{tb}})$.
For the braching ratio,
 \be
\mathcal{H}^B_{ij} = \left\{ \begin{array}{ll}
{}~\sum~~ |M_i^N|^2 \;S^B_{NN} + M_i^{7*} M_i^{9} \;S^B_{79} +\Delta \mathcal{H}^B_{ii}\;,
& \mbox{for~} i=j \, , \\[-2mm]
{\!\!\!\scriptscriptstyle N=7,9,10}\\[2mm]
{}~\sum~~ 2 M_i^{N*} M_j^{N} \; S^B_{NN}
+ \; \left(M_i^{7*} M_j^{9} + M_i^{9*} M_j^{7} \right) \; S^B_{79} +\Delta \mathcal{H}^B_{ij}\;,
& \mbox{for~} i < j \, . \\[-2mm]
{\!\!\!\scriptscriptstyle N=7,9,10}
\end{array}\right.\label{eq:hijI}
\ee
For $I=A$, only the interference terms contribute:
 \be
\mathcal{H}^A_{ij} = \left\{ \begin{array}{ll}
 \;\; 0\; , & \mbox{for~} i=j \; , \\
{}~\sum~~ \; \left(M_i^{N*} M_j^{10} + M_i^{10*} M_j^{N} \right) \; S^A_{N10} +\Delta \mathcal{H}^A_{ij}\;,
& \mbox{for~} i < j \; .\\[-2mm]
{\,\scriptscriptstyle N=7,9}
\end{array}\right.\label{eq:hijA}
\ee
The matrix elements $M_i$ will be discussed in section~\ref{sec:mes}.
The functions $S_{NM}^I$ can be written as
\begin{align}
{\rm S}^I_{NM}  =&\ \sigma_{NM}^I(\s) \left\{
                   1 + 8 \, \as \, \omega_{NM,I}^{(1)} (\s) + 16 \, \as^2 \, \omega_{NM,I}^{(2)} (\s) \right\}  \nnb \\
		   & + \frac{\lambda_1}{m_b^2} \; \chi_{1,NM}^I(\s) + \frac{\lambda_2^{\rm eff}}{m_b^2} \; \chi_{2,NM}^I(\s)
		   + \frac{\rho_1}{m_b^3} \; \chi_{3,NM}^I(\s) + \frac{f_d}{m_b^3} \; \chi_{4,NM}^I(\s)  \, .
\end{align}
For the relevant combinations, we find
\begin{align}
\sigma_{77}^B(\s) &=  (1-\s)^2(4+8/\s) \, , & \sigma_{710}^A(\s) &= -8 (1-\s)^2 \, , \nnb \\
\sigma_{79}^B(\s) &= 12 (1-\s)^2    \, ,  & \sigma_{910}^A(\s) &= -4 \s (1-\s)^2 \, , \nnb \\
\sigma_{99}^B(\s) &= \sigma_{1010}^B(\s) = (1+2 \s) (1-\s)^2    \, . \label{eq:phasespace}
\end{align}
The one-loop QCD functions for the branching ratio, $\omega_{NM,B}^{(1)} (\s)$, are given by
eqs.~(127) and (129)~--~(131) in~\cite{Huber:2005ig} (see also \cite{Jezabek:1988ja,Misiak:1992bc,Buras:1994dj,Asatryan:2001zw})
and the non-vanishing function for $\mathcal{H}_A$ is given in
eq.~(A.2) of \cite{Huber:2015sra} (see also \cite{Asatrian:2002va,Lee:2006gs}). The two-loop QCD
function $\omega_{99,B}^{(2)} (\s) = \omega_{1010,B}^{(2)} (\s)$ is given by
\be\label{eq:omega299}
\omega_{99,B}^{(2)} (\s) = \beta_0^{(5)}\log\left({\mu_b\over m_b}\right)\omega_{99,B}^{(1)} (\s) + \left\{ \begin{array}{ll}
X_2(\omega=\s)/X_0(\omega = \s)\;,
& \text{for low-$q^2$} \, , \\
X_2(\delta=1-\s)/X_0(\delta =1- \s)\;,
& \text{for high-$q^2$} \, ,
\end{array}\right.
\ee
where $\beta_0^{(5)} = 23/3$. The $\omega=\s$ expanded result for the
low-$q^2$ function $X_2(\omega)$ is given in eq.~(60) of \cite{Blokland:2005vq} and $X_0(\omega)$ in
eq.~(2) of \cite{Blokland:2004ye}. The $\delta=1-\s$ expanded results for the
high-$q^2$ functions $X_2(\delta)$ and $X_0(\delta)$ are given in eq.~(2) and (3) of \cite{Czarnecki:2001cz}.
Note that the normalization of the $X_i$ are different by a factor of 2 between \cite{Blokland:2005vq,Blokland:2004ye}
and \cite{Czarnecki:2001cz}. We have checked the consistency between eq.~\eqref{eq:omega299} and the
fit results for two-loop QCD functions $\omega_{99,T}^{(2)} (\s)$ and $\omega_{99,L}^{(2)} (\s)$ in eq.~(A.3) of \cite{Huber:2015sra}.
For the two-loop QCD function $\omega_{910,A}^{(2)} (\s)$ for $H_A$, we use the fit result given in eq.~(A.3) of \cite{Huber:2015sra},
which was extracted from the fully differential calculation of the inclusive $\bar{B}\to X_u\ell\bar{\nu}_\ell$ decay at two loops in QCD \cite{Brucherseifer:2013cu}.
Other two-loop QCD functions such as $\omega_{79,B}^{(2)}$ and $\omega_{710,A}^{(2)}$ are still unknown.

For the ${\cal O}(\Lambda_{\rm QCD}^2/m_b^2)$ corrections, the functions $\chi_{i,NM}^I(\s)$ ($i=1,2$) are given by
\begin{align}
\chi_{1,77}^B(\s) &= \frac{2}{\s}(1-\s)^2(2+\s) \, ,
& \chi_{1,79}^B(\s) &= 6 (1-\s)^2 \, , \nnb \\
\chi_{1,99}^B(\s) &= \chi_{1,1010}^B(\s) = {1\over2}(1-\s)^2(2\s+1) \, , \nnb \\
\chi_{1,710}^A(\s) &= -\frac{4}{3} \left(3 \s^2+2 \s+3\right) \, ,
& \chi_{1,910}^A(\s) &= -\frac{2}{3} \s \left(3 \s^2+2 \s+3\right) \, , \label{eq:chi1}
\end{align}
\begin{align}
\chi_{2,77}^B(\s) &= \frac{6}{\s} \left(5 \s^3-3 \s-6\right) \, ,
& \chi_{2,79}^B(\s) &= 6 \left(7 \s^2-6 \s-5\right) \, , \nnb \\
\chi_{2,99}^B(\s) &=\chi_{2,1010}^B(\s) = {3\over2} \left(10\s^3-15 \s+1\right) \, , \nnb \\
\chi_{2,710}^A(\s) &= -4 \left(9 \s^2-10 \s-7\right) \, ,
& \chi_{2,910}^A(\s) &= -2 \s \left(15 \s^2-14 \s-9\right) \, , \label{eq:chi2}
\end{align}
\begin{align}
\chi_{3,77}^B(\s) &= \frac{2}{3\s} \left(-22+33\s +24\s^2+5 \s^3\right) - {32\over(1-\s)_+} - 16\delta(1-\s) \, ,  \nnb \\
\chi_{3,79}^B(\s) &= 2 \left(13 + 14\s -3 \s^2\right) - {32\over(1-\s)_+} - 16\delta(1-\s)  \, , \nnb \\
\chi_{3,99}^B(\s) &= \chi_{3,1010}^B(\s) = {1\over6} \left(37 + 24\s +33\s^2 + 10\s^3\right) - {8\over(1-\s)_+} - 4\delta(1-\s) \, , \label{eq:chi3}
\end{align}

\begin{align}
\chi_{4,77}^B(\s) &= - 16\delta(1-\s) \, ,
& \chi_{4,79}^B(\s) &= - 16\delta(1-\s) \, , \nnb \\
\chi_{4,99}^B(\s) &=\chi_{4,1010}^B(\s) = - 4\delta(1-\s) \, . \label{eq:chi5}
\end{align}
The plus distribution can be defined via
\be
\int_{x_0}^1{1\over (1-x)_+}f(x) \equiv  \lim_{\epsilon\to 0}\int_{x_0}^1dx \left[ {\theta(1-x-\epsilon)\over 1-x} +\delta(1-x-\epsilon)\ln\epsilon \right]f(x).
\ee
These expressions have been checked to be consistent with \cite{Falk:1993dh,Ali:1996bm,Buchalla:1998mt,Lee:2006gs,Huber:2015sra}.
The ${\cal O}(1/m_b^3)$ corrections to the forward-backward asymmetries are missing, but fortunately
we are only concerned about the forward-backward asymmetries in the low-$q^2$ region, where
the ${\cal O}(1/m_b^3)$ corrections are negligible.

The quantities $\Delta H^I_{ij}$ contain additional corrections that can be parameterized as
\be\label{eq:DeltaH}
\Delta H^I_{ij} = b^I_{ij} + c^I_{ij} + u^I_{ij} + e^I_{ij} + f^I_{ij} \; ,
\ee
where $b^I_{ij}$ represent finite bremsstrahlung corrections that appear at
NNLO and are given in \cite{Asatryan:2002iy, Asatrian:2003vq} for $I=B$ and in \cite{Asatrian:2003yk, Asatrian:2003vq} for $I=A$. In addition, $c^I_{ij} (u^I_{ij})$ are the non-perturbative $c$($u$)-loop power corrections, while $e^I_{ij}$ are the $\ln(m_b^2/m_\ell^2)$-enhanced electromagetic corrections and $f^I_{ij}$ are five-body contributions. We discuss these contributions in the following subsections.


\subsection{Matrix elements}\label{sec:mes}
The matrix elements entering the master formula in eq.~\eqref{eq:hijI} and \eqref{eq:hijA} are obtained from one-loop penguin contractions of the four-fermion operators. They are given by
\bea
\me{P_i}^{\rm peng} & = &  M_i^9  \me{P_9}_{\rm tree} +
                M_i^7  \frac{\me{P_7}_{\rm tree}}{\as(\mu_b) \kappa(\mu_b)} +
                M_i^{10}  \me{P_{10}}_{\rm tree} \;.
\label{matelPi}
\eea
The coefficients
$ M_i^A$ are summarized in table~\ref{tab:hiA}
\begin{table}[!t]
\begin{center}
\begin{tabular}{|l|c|c|c|} \hline
 & $ M_i^9$ &  $ M_i^7$ &  $ M_i^{10}$ \\ \hline
i=1u, 2u, 1c, 2c &
$\as \kappa \; f_{i} (\hat s) - \as^2 \kappa \; F_i^9 (\hat s) $ &
$- \as^2 \kappa \; F_i^7 (\hat s) $ &
0 \\
i=3-6,3Q-6Q,b &
$\as \kappa \; f_{i} (\hat s) $ &
0 & 0 \\
i=7 &
0&
$\as \kappa$ &0\\
i=8 &
$- \as^2 \kappa \; F_8^9 (\hat s) $ &
$- \as^2 \kappa \; F_8^7 (\hat s) $ & 0 \\
i=9 &
$1+\as \kappa \; f_9^{\rm pen} (\hat s) $ &
0 & 0\\
i=10 &
0 & 0 & 1 \\\hline
\end{tabular}
\caption{Matrix elements $ M_i^A$ discussed in eq.~(\ref{matelPi}).
 \label{tab:hiA}}
\end{center}
\end{table}
in terms of the one-loop functions $f_i, f_9^{\text{pen}}$ and the two-loop functions $F_i^A(\hat s)$.
The one-loop perturbative functions are
\bea
f_i (\hat s) &=&
\gamma_{i} \ln \f{m_b}{\mu_b}
+ \sum_{q=u,d,s,c,b}\rho_i^q \, h(y_q)  + \rho_i^\$ \; .
\label{4melem}
\eea
and
 \begin{equation}
 f_9^{\rm pen} (\hat s)  =
8 \ln \f{m_b}{\mu_b} - 3  h(y_\tau)
       + \f{8}{3} \left( \ln \s - i\pi \right) - \f{40}{9} \; .
       \end{equation}
Here $y_a = 4 (m_{a,{\rm pole}}^2-i\eta)/q^2$ with an infinitesimally small, positive quantity $\eta$ that takes care of the correct analytic continuation. Compared to the widely-used one-loop function $g(y_a)$ in the literature,
we introduce $\displaystyle h(y_a) = g(y_a) + (8/9) \ln (m_b/m_a)$ here.
Contrary to previous studies for ${\bar B \to X_{s} \, \ell^+\ell^-}$, we split the coefficients $\rho_i$ into their different quark flavour contributions $\rho_i^q$, collected in table~\ref{tab:rhos}. These numbers reduce to those presented in table~7 of \cite{Huber:2005ig} once the distinction between the light quark flavours is given up and $h(y_a)$ is traded in for $g(y_a)$.
The perturbative one-loop function is
\bea 
h(y_{a}) & = &\f{8}{9} \ln \f{m_b}{m_{a}} + \f{20}{27} + \f{4}{9} y_a + \f{2i}{9}(2+y_a) \sqrt{y_a-1} \;  H_{+}\!\left(\frac{i}{\sqrt{y_a-1}}\right) \label{eq:hmass}\\[3mm]
&=& \f{8}{9} \ln \f{m_b}{m_{a}} + \f{20}{27} + \f{4}{9} y_a - \f{2}{9}(2+y_a) \sqrt{|1-y_a|}
\left\{ \begin{array}{ll}
\ln \left|\f{1+\sqrt{1-y_a}}{1-\sqrt{1-y_a}}\right|, & {\rm for}~ y_a < 0,\\[0.4em]
2 \arctan \f{1}{\sqrt{y_a-1}},                     & {\rm for}~ y_a \ge 1, \\[0.4em]
\ln \left|\f{1+\sqrt{1-y_a}}{1-\sqrt{1-y_a}}\right| - i \pi, & {\rm for}~ 0<y_a < 1,
\end{array} \right.\nnb
\eea
which holds for massive particles, i.e.\ $a=b, c, \tau$. The harmonic polylogarithm of weight ``$+$'' simply reads $H_{+}(z) = \ln \left(\frac{1+z}{1-z}\right)$.
For the light quarks $u, d, s$, this function reduces to
\begin{equation}\label{eq:hlight}
h(y_{u,d,s}) = \frac{20}{27} + \frac{4i}{9}  \pi  - \frac{4}{9} \ln \s \ .
\end{equation}

\begin{table}[!t]
\begin{displaymath}
\hspace*{15mm}
\begin{array}{|c|rrrrrrrrrrrrr|}
\hline
 & 1c & 2c & 1u & 2u & 3 & 4 & 5 & 6 & 3Q & 4Q & 5Q & 6Q & b \\
\hline
&&&&&&&&&&&&& \\[-3mm]
\rho_i^u & 0 & 0 & \f{4}{3} & 1 & 6 & 0 & 60 & 0 & 4 & 0 & 40 & 0 & 0 \\[2mm]
\rho_i^d & 0 & 0 & 0 & 0 & -\f{7}{2} & -\f{2}{3} & -38 & -\f{32}{3} & \f{7}{6}
       & \f{2}{9} & \f{38}{3} & \f{32}{9} & 0\\[2mm]
\rho_i^s & 0 & 0 & 0 & 0 & -3 & 0 & -30 & 0 & 1 & 0 & 10 & 0 & 0 \\[2mm]
\rho_i^c & \f{4}{3} & 1 & 0 & 0 & 6 & 0 & 60 & 0 & 4 & 0 & 40 & 0 & 0 \\[2mm]
\rho_i^b & 0 & 0 & 0 & 0 & -\f{7}{2} & -\f{2}{3} & -38 & -\f{32}{3} & \f{7}{6}
       & \f{2}{9} & \f{38}{3} & \f{32}{9} & -2\\[2mm]
\rho_i^\$ & -\f{16}{27} & -\f{4}{9} & -\f{16}{27} & -\f{4}{9} & \f{4}{9} & \f{16}{27} & \f{8}{9} & \f{320}{27}
        & -\f{4}{27} & -\f{16}{81} & -\f{872}{27} & -\f{320}{81} & \f{26}{27} \\[1mm]
        \gamma_i  & -\frac{32}{27} & -\f{8}{9} & -\frac{32}{27} & -\f{8}{9} &  -\f{16}{9} & \f{32}{27} & -\f{112}{9} & \f{512}{27} & -\f{272}{27}
        & -\f{32}{81} & -\f{2768}{27} & -\f{512}{81} & \f{16}{9}\\[2mm]
\hline
\end{array}
\end{displaymath}
\vspace*{-8mm}
\begin{center}
\caption{Coefficients $\rho^q_i$ and $\gamma_i$ occurring in the four-quark operator
          matrix elements in eq.~(\ref{4melem}). \label{tab:rhos}}
\end{center}
\vspace*{-12mm}
\end{table}

The two-loop contributions $F_i^A(\hat s)$ for the $c$-operators $P_{1,2}^c$ valid in the low-$q^2$ region are given in~\cite{Asatryan:2001zw} 
as an expansion in $\hat s$ up to ${\cal O}(\hat{s}^3)$. In the high-$q^2$ region they were first calculated in~\cite{Ghinculov:2003qd} using a semi-numerical method and analytically given in~\cite{Greub:2008cy}. For the $u$-operators, which is the massless limit of the charm, analytic functions for all $q^2$ are available~\cite{Seidel:2004jh}
\begin{eqnarray}
F_{1u}^{7} &=& -A(\s),\\
F_{2u}^{7} &=& 6 A(\s),\\
F_{1u}^{9} &=& -B(\s) - 4 C(\s),\\
F_{2u}^{9} &=& 6 B(\s) - 3 C(\s),
\end{eqnarray}
where $A(\s)$, $B(\s)$ and $C(\s)$ are listed eqs.~(29)~--~(31) of~\cite{Seidel:2004jh}. 
In a recent study \cite{deBoer:2017way} (see also \cite{Bell:2014zya}), the analytical expressions of the two-loop functions $F_i^A(\hat s)$ for both the up and charm cases have been worked out for arbitrary $q^2$.
As in our previous studies~\cite{Huber:2005ig,Huber:2007vv} we convert the pole masses of the bottom and charm quark perturbatively to a short-distance mass at the level of the squared amplitude in order to eliminate renormalon ambiguities.

When considering also non-perturbative corrections, the factorizable pieces of these perturbative corrections are replaced by their corresponding Kr{\"u}ger-Sehgal (KS) functions, which we discuss in detail in section~\ref{sec:KS}. Contrary to previous works, we do not only replace the one-loop perturbative corrections $h(y_{u,d,s,c})$ by the KS functions, but also the factorizable pieces of the two-loop corrections $F_i^A$. These two-loop factorizable pieces, defined as $h^{(1)}_q$ with $q=u,d,s,c$, can be found in~\cite{deBoer:2017way} and are listed in Appendix~\ref{sec:apptwoloop}. Explicitly, we replace
\begin{equation}
h^{\rm fact}_q \to h_q^{\rm KS} \ ,
\end{equation}
where
\begin{equation}
h^{\rm fact}_q = h(y_{q}) + \as h^{(1)}_q \ .
\end{equation}


\subsection{Resolved photon corrections at  low $q^2$}\label{sec:resolvedlow}
Contrary to previous studies in ${\bar B \to X_{s} \, \ell^+\ell^-}$~\cite{Huber:2015sra}, we do not use the local description of the nonfactorizable $c \bar c$ power correction 
of order $\mathcal{O}(\Lambda_{\text{QCD}}^2/m_c^2)$ in the low-$q^2$ region~\cite{Voloshin:2001xi,Buchalla:1997ky}. Instead, we follow the recent analysis in~\cite{Hurth:2017xzf,Benzke:2017woq} to adopt a more systematic approach to describe these so-called resolved power corrections, including the effect of non-local shape functions. We relegate the conceptual description of these contributions to section~\ref{section:resolved}, and at this point quote the numerical result for the relevant contributions available up to order $\Lambda_{\rm QCD}/m_b$~\cite{Hurth:2017xzf,Benzke:2017woq}. The sum of the resolved contributions, including also an estimate of a numerically relevant term at quadratic order, leads to an additional uncertainty on the branching ratio of $[-4.9,+5.1]\,\%$. We add this uncertainty to our numerical results in section~\ref{sec:brlowq2}. 
For the forward-backward asymmetry, the first nontrivial resolved contribution is of order $\Lambda_\text{QCD}^2/m_b^2$ and yet unknown, which leads us to add an uncertainty of $\pm 5 \%$ to our final result until an explicit estimate is available~\cite{Benzkeworkinprogress}.

We emphasize that in the low-$q^2$ region the nonfactorizable $u \bar u$ power correction to CP-averaged observables vanish due to specific properties of the corresponding shape functions. This contribution previously represented the main uncertainty in $b \to d$ decay (for more details see section~\ref{section:resolved}). 


\subsection{Nonfactorizable power corrections at high $q^2$}\label{sec:resolvedhigh}
For the high-$q^2$ region the power corrections  from nonfactorizable  $c \bar c$ and  
$u \bar u $  loops are available. They can be described as a local power correction as shown in~\cite{Buchalla:1997ky} and discussed in more detail in section~\ref{section:local}. 
For the $c \bar c$ contributions the coefficients are given by \cite{Buchalla:1997ky}
\bea
c^B_{i2c} &=& - \as \kappa \frac{8\lambda_2}{9 m_c^2} \, (1-\s)^2 \, F(r) \left[\frac{1+6\s-\s^2}{\s} \, M_i^{7\ast} + (2+\s) \, M_i^{9\ast}\right] \;, \hskip 0.5cm {\rm for} \; i = 1u,2u \;,\nnb\\
c^B_{i1c} &=& -\fm{1}{6}  \; c^B_{i2c}\;, \hskip 1.5cm {\rm for} \; i = 1u,2u \;,\nnb\\
c^B_{2cj} &=& - \as \kappa \frac{8\lambda_2}{9 m_c^2} \, (1-\s)^2 \, F^\ast(r) \left[\frac{1+6\s-\s^2}{\s} \, M_j^{7} + (2+\s) \, M_j^{9}\right] \;, \hskip 0.5cm {\rm for} \; j \neq 1u,2u,1c \;,\nnb\\
c^B_{1cj} &=& -\fm{1}{6}  \; c^B_{2cj}\;, \hskip 1.5cm {\rm for} \; j \neq 1u,2u,1c,2c \;,\nnb\\
c^B_{1c1c} &=& + \as \kappa \frac{4\lambda_2}{27 m_c^2} \, (1-\s)^2 \, F^\ast(r) \left[\frac{1+6\s-\s^2}{\s} \, M_{1c}^{7} + (2+\s) \, M_{1c}^{9}\right] \;, \nnb\\
c^B_{1c2c} &=& - \as \kappa \frac{8\lambda_2}{9 m_c^2} \, (1-\s)^2 \! \left[F(r) \left(\frac{1+6\s-\s^2}{\s} \, M_{1c}^{7\ast} + (2+\s) \, M_{1c}^{9\ast}\right)\right. \nnb\\
&& \hskip 3.3cm \left.-\frac{1}{6} \, F^\ast(r) \left(\frac{1+6\s-\s^2}{\s} \, M_{2c}^{7} + (2+\s) \, M_{2c}^{9}\right)\right] \;, \nnb\\[1.5em]
c^A_{2c10} &=& +\as \kappa  \frac{4\lambda_2}{9 m_c^2} (1-\s)^2
                  (1+3\s)\,
		  F^*(r)  \;,\nnb\\
c^A_{1c10} & = &  -\fm{1}{6}  \; c^A_{2c10} 
\label{eq:cfunc}\; ,
\eea
where $r= q^2/(4m_c^2)>1$ for the high-$q^2$ region and \cite{Buchalla:1997ky}
\begin{equation}
F(r)   = \frac{3}{2r}\left[ \frac{1}{2\sqrt{r(r-1)}} \left(\ln\frac{1-\sqrt{1-1/r}}{1+\sqrt{1-1/r}}+i\pi\right)-1 \right] \; .
\end{equation}

For the $u \bar u$ contribution, the results are obtained by taking the $m_c \to 0$ limit of the $c_{ij}^I$ in eq.~(\ref{eq:cfunc}). Explicitly, 
\bea
u^B_{2uj} &=& \as \kappa \frac{16\lambda_2}{3 q^2} \, (1-\s)^2 \,  \left[\frac{1+6\s-\s^2}{\s} \, M_j^{7} + (2+\s) \, M_j^{9}\right] \;, \hskip 1.5cm {\rm for} \; j \neq 1u \;,\nnb\\
u^B_{1uj} &=& -\fm{1}{6}  \; u^B_{2uj}\;, \hskip 1.5cm {\rm for} \; j \neq 1u,2u \;,\nnb\\
u^B_{1u1u} &=& - \as \kappa \frac{8\lambda_2}{9 q^2} \, (1-\s)^2 \, \left[\frac{1+6\s-\s^2}{\s} \, M_{1u}^{7} + (2+\s) \, M_{1u}^{9}\right] \;, \nnb\\
u^B_{1u2u} &=& + \as \kappa \frac{16\lambda_2}{3 q^2} \, (1-\s)^2 \! \left[\left(\frac{1+6\s-\s^2}{\s} \, M_{1u}^{7\ast} + (2+\s) \, M_{1u}^{9\ast}\right)\right. \nnb\\
&& \hskip 3.3cm \left.-\frac{1}{6} \, \left(\frac{1+6\s-\s^2}{\s} \, M_{2u}^{7} + (2+\s) \, M_{2u}^{9}\right)\right] \;, \\
u^A_{2u10} &=& -\as \kappa  \frac{8\lambda_2}{3 q^2} (1-\s)^2  (1+3\s)\, \;, \hskip 1.5cm u^A_{1u10}  =   -\fm{1}{6}  \; c^A_{2u10} \; .
\eea


\subsection{Logarithmically enhanced electromagnetic corrections}\label{sec:emcorrections}
The functions $e^I_{ij}$ describe the logarithmically enhanced electromagnetic corrections. For $I=B$, we find
\begin{align}
e^B_{77} & = 8 \, \as^3 \kappa^3 \, \sigma_{77}^B(\s) \, \omega_{77}^{\rm (em)}(\s)^* \; , &
e^B_{79} & = 8 \, \as^2 \kappa^2 \, \sigma_{79}^B(\s) \, \omega_{79}^{\rm (em)}(\s)^* \; , \nnb \\
e^B_{99} & = 8 \, \as \kappa \, \sigma_{99}^B(\s) \, \omega_{99}^{\rm (em)}(\s)^* \; , &
e^B_{1010} & = e^B_{99}\; , \nnb \\
e^B_{2a9} & = 8 \, \as^2 \kappa^2 \, \sigma_{99}^B(\s) \,\omega_{2a9}^{\rm (em)}(\s)^* \, , &
e^B_{2a2a} & = 8\, \as^3 \kappa^3\, \sigma_{99}^B(\s) \, \omega_{2a2a}^{\rm (em)}(\s)^* \, , \nnb\\
e^B_{1a1a} & =   \fm{16}{9}  \; e^B_{2a2a} \, , &
e^B_{1a2a} & =  \fm{8}{3}   \; e^B_{2a2a} \, ,\nnb\\
e^B_{2a7} & = 8 \,\as^3 \kappa^3 \, \sigma_{79}^B(\s) \, \omega_{2a7}^{\rm (em)}(\s)^* \, , &
e^B_{1aj} & =  \fm{4}{3}  e^B_{2aj} , \nnb\\
e^B_{2u2c} & = 8\, \as^3 \kappa^3\, \sigma_{99}^B(\s) \, \omega_{2u2c}^{\rm (em)}(\s)^* \, , & 
e^B_{1u1c} & = \fm{4}{3}e^B_{1u2c} = \fm{4}{3}e^B_{2u1c} = \fm{4}{3}e^B_{1u1c}\, ;
\end{align}
for $I=A$, we find
\begin{align}
e^A_{710} & = 8 \, \as^2 \kappa^2 \, \sigma_{710}^A(\s) \, \omega_{710}^{\rm (em)}(\s)^* \; , &
e^A_{2a10} & = 8 \, \as^2 \kappa^2 \, \sigma_{910}^A(\s) \, \omega_{2a10}^{\rm (em)}(\s)^* \; ,  \nnb\\
e^A_{910} & = 8 \, \as \kappa \, \sigma_{910}^A(\s) \, \omega_{910}^{\rm (em)}(\s)^* \; , &
e^A_{1a10} & =  \fm{4}{3}  e^A_{2a10},
\end{align}
with $a=u,c$ and $j=7,9$.
The $\sigma$ functions were introduced already in eq.~\eqref{eq:phasespace}.
Exact analytical expressions are available for most of the $\omega_{ij}^{\rm (em)}(\s)$ functions~\cite{Huber:2005ig,Huber:2007vv, Huber:2015sra}
and for completeness are listed in Appendix~\ref{sec:appomegaem}. However, the operators $P_1^u$ and $P_2^u$
induce additional functions which were up to now not available. We derived these formulas following the methods discussed
in \cite{Huber:2005ig}, and listed them in Appendix~\ref{sec:appomegaem}.


\subsection{Five-particle contributions}\label{sec:5body}
The five-particle processes $b\to dq\bar{q}\ell^+\ell^-$ at the partonic level also contribute to
the inclusive ${\bar B \to X_{d} \, \ell^+\ell^-}$ decay. While similar contributions are CKM suppressed
for the $b\to s$ transition, such five-particle contributions are at the same order in the Wolfenstein
expansion compared to the partonic three-particle ones. The branching ratios and the forward-backward
asymmetries of $b\to dq\bar{q}\ell^+\ell^-$ have been calculated at tree level in \cite{Huber:2018gii}.
Correspondingly, the $f_{ij}^I$ functions in eq.~\eqref{eq:DeltaH} summarizing such contributions can be written as
\be
f_{ij}^B = \left\{ \begin{array}{ll}
{}~\mathcal{F}_{ii}(\s)\;,
& \mbox{for~} i=j \, , \\
{}~\mathcal{F}_{ij}(\s) + \mathcal{F}_{ji}(\s)\;,
& \mbox{for~} i < j \, ,
\end{array}\right.\label{eq:fijB}
\ee

\be
f_{ij}^A = \left\{ \begin{array}{ll}
-{4\over3}\mathcal{A}_{i}(\s)\;,
& \mbox{for~} i = 1u,2u, 3, \ldots, 6, j = 10 \, , \\
0\;,
& \mbox{for the others} \, ,
\end{array}\right.\label{eq:fijA}
\ee
where the functions $\mathcal{F}_{ij}(\s)$ and $\mathcal{A}_{i}(\s)$ can be found in eqs.~(31) and (37)
of \cite{Huber:2018gii}, respectively.
The matrix elements involving $1c,2c$ and $b$ vanish, while those of the electroweak penguins $3Q,\ldots,6Q$ are neglected in the $f^{A,B}_{ij}$. The indices
$1u,2u$ in $f^{A,B}_{ij}$ correspond to $1,2$ in $\mathcal{F}_{ij}$ and $\mathcal{A}_{i}$.


\section{Long distance contributions and backgrounds}
\label{sec:long}

If only the operators $P_{7,9,10}$ in the effective Hamiltonian were considered, the local heavy mass expansion would be applicable to ${\bar B \to X_{s(d)} \, \ell^+\ell^-}$ observables integrated over the hadronic mass $M_X$. Then, a local OPE would hold and the hadronic decay would be described in terms of the partonic decay plus local power corrections.
Inclusion of operators other than $P_{7,9,10}$ introduces various other long distance effects, and the purpose of the present section is to categorize them for ${\bar B \to X_{s(d)} \, \ell^+\ell^-}$.

The ${\bar B \to X_{s(d)} \, \ell^+\ell^-}$ decay rate is enormously enhanced through the process $\bar{B} \to X_{s(d)}(J/\psi, \psi(2S) \to \ell^+ \ell^-)$  mediated by the $c$-quark operators $P_{1,2}^c$~\cite{Beneke:2009az}. These resonances are a long-distance feature of the partonic decay; they are not power suppressed. While these two resonances can be removed by appropriate $q^2$-cuts, the persistence of higher charmonium resonances in the high-$q^2$ region renders a purely perturbative prediction unreliable there, even if the large $1/m_b$ corrections are taken into account. Perturbation theory is likewise unreliable in the light-quark resonance region $q^2 \lsim 4~\text{GeV}^2$, affecting low-$q^2$ observables of ${\bar B \to X_{d} \, \ell^+\ell^-}$, while for ${\bar B \to X_{s} \, \ell^+\ell^-}$ they are strongly CKM suppressed. To incorporate the resonances into the phenomenological analysis the Kr\"uger-Sehgal (KS) approach~\cite{Kruger:1996cv} is adopted~\cite{Huber:2007vv, Huber:2015sra}. It connects the factorizable part of the resonant amplitude to the hadronic vacuum polarization which can be extracted from $e^+e^- \to \text{hadrons}$ via a dispersion relation. In the present work we significantly improve the KS approach in several aspects. Following ref.~\cite{Lyon:2014hpa} we use accurate interpolations of $e^+e^- \to \text{hadrons}$ data directly as opposed to parameterizations of the resonances, and in contrast to ref.~\cite{Kruger:1996dt} we show that in order to extract the $u\bar u$, $d\bar d$ and $s\bar s$ correlators we need to use $\tau$-decay data which projects out the $u$-quark vacuum polarization from the rest. We also investigate the uncertainties associated with the KS functions and find them to be small. Moreover, for the first time we properly combine resonant amplitudes and ${\cal O}(\alpha_s)$ corrections. Lastly, we emphasize that the subtraction point of the dispersion relation must be chosen large and negative to avoid sensitivity to vacuum condensates. We thoroughly investigate all these points in section~\ref{sec:KS}.

At this point we want to pick up the issue of color-octet production of charmonium resonances. It was pointed out in the literature that this production mechanism leads to sizeable effects~\cite{Ko:1995iv,Beneke:1998ks,Beneke:2009az,Lyon:2014hpa}, and that the pure color-singlet treatment of the KS approach does not capture the full size of the $\psi$ resonances.
In the case of the narrow $J/\psi$ and $\psi(2S)$ resonances, color octet effects are of course very important at the position of the resonances, but due to their sharpness are confined to the close neighborhood of the peaks. One can therefore expect the low-$q^2$ region to remain unaffected by these effects. The high-$q^2$ region is more delicate in this respect since one integrates over broad resonances. However, there the non-factorisable $c\bar c$-resonances are included in the Voloshin term~\cite{Voloshin:2001xi,Buchalla:1997ky}, which corresponds to a local power correction in the high-$q^2$ region (see below and section~\ref{sec:resolvedhigh}), as long as one considers integrals over sufficiently large dilepton invariant mass intervals. In that case one can --~via global quark hadron duality~-- expect that the color-octet induced ``wiggles'' average out and are effectively taken into account by the partonic description of the Voloshin effect. In total, we reason that the color-singlet resonances are under control with the KS method and for the color-octet ones we correctly include their integral via the Voloshin term, thereby also avoiding double counting.

As pointed out above, the local heavy mass expansion breaks down if operators other than $P_{7,9,10}$ in the effective field theory are considered. This breakdown leads to nonlocal power corrections that can be described in the low-$q^2$ region within SCET using subleading shape functions~\cite{Benzke:2017woq,Hurth:2017xzf}. In the high-$q^2$ region the Voloshin term mentioned already above can be expanded locally~\cite{Voloshin:2001xi,Buchalla:1997ky}. In sections~\ref{section:resolved} and~\ref{section:local} we review the essential conceptual steps which lead to this behaviour in the low and high-$q^2$ region, respectively, while the numerical impact of these findings were already given in sections~\ref{sec:resolvedlow} and~\ref{sec:resolvedhigh}.

Finally, in section \ref{section:cascades} we emphasize that the charmonium cascade decays $\bar{B} \to X_1(c\bar{c} \to X_2 \ell^+ \ell^-)$, where the total $X_d = X_1 + X_2$ is measured, are not captured by other elements of our calculation and would form a large background in the low-$q^2$ region if not for the $M_X$ cut, which is very effective in removing them.


\subsection{The Kr{\"u}ger-Sehgal approach}\label{sec:KS}
Under the assumption that the currents associated with the production of a vector hadronic system $V$ in ${B \to X_d V}$ and the subsequent electromagnetic decay ${V \to \ell^+ \ell^-}$ factorize, the hadronization of the $X_d$ system following the electroweak decay is described by an OPE in $\Lambda_\text{QCD}/m_b$, while the lepton pair production is modified by the quark vacuum polarization amplitudes accessible in hadron spectroscopy experiments. Kr{\"u}ger and Sehgal (KS) used $e^+ e^- \to \text{hadrons}$ data and a dispersion relation for ${B \to X_d \ell^+ \ell^-}$ applications \cite{Kruger:1996dt}, following similar work in ${B \to X_s \ell^+ \ell^-}$ \cite{Lim:1988yu, Kruger:1996cv}. Here we supplement the procedure with data from inclusive hadronic $\tau$ decays for the first time in a data-driven analysis.

The correlation functions between each individual quark current and the electromagnetic current which couples to the leptons are needed. We define the following KS function for each flavour, normalized in accordance to their evaluation in perturbation theory, c.f. eqs.~\eqref{eq:hmass} and~\eqref{eq:hlight}.
\begin{equation}
(q^\mu q^\nu - q^2 g^{\mu \nu}) h^\text{KS}_q(q^2) =  \frac{16 \pi^2}{9 Q_q}~i \int d^4 x ~e^{iqx} \braket{0| T J_q^\mu(0) J^\nu_\text{em}(x)^\dagger |0},
\label{eq:hKS}
\end{equation}
where $Q_q$ is the quark charge and the currents are
\begin{equation}
J_q^\mu = \bar{q}\gamma^\mu q, \hspace{1cm} J_\text{em} = \frac{2}{3} J_u - \frac{1}{3} J_d + \frac{2}{3} J_c - \frac{1}{3} J_s.
\end{equation}
The electromagnetic current in \eqref{eq:hKS} guarantees that the correlator has a transverse structure according to the Ward identity. The contributions from the correlators $\braket{J_{q_1} J_{q_2}}$ between different quark flavours are systematically included here, although they are suppressed in perturbation theory at $\mathcal{O}(\alpha_s^3)$.

The imaginary part of the photon vacuum polarization $\Pi_\gamma$ in
\begin{equation}
(q^\mu q^\nu-q^2 g^{\mu \nu}) \Pi_\gamma(q^2) = i \int d^4 x ~e^{iqx} \braket{0|T J_\text{em}^\mu (0) J_\text{em}^\nu(x)^\dagger |0}
\label{eq:pihad}
\end{equation}
is accessible in the inclusive cross section $\sigma_\text{had}$ for $e^+ e^- \to \text{hadrons}$, represented in terms of the hadronic R-ratio
\begin{equation}
R_\text{had}(s) \equiv \frac{3s}{4\pi \alpha^2}~\sigma_\text{had}(s) = 12 \pi ~\text{Im}[\Pi_\gamma(s)].
\end{equation}
Similarly, the imaginary part of the charged vector current correlator $\Pi_{\bar{u}q}$ in
\begin{equation}
(q^\mu q^\nu-q^2 g^{\mu \nu}) \Pi_{\bar{u}q} (q^2) = i \int d^4 x ~e^{iqx} \braket{0|T J_{\bar{u}q}^\mu (0) {J_{\bar{u}q}^\nu} (x)^\dagger |0},
\label{eq:pitau}
\end{equation}
where $J_{\bar{u}q}^\mu = \bar{u} \gamma^\mu q$ ($q=d,s$), is related to the vector spectral function
\begin{equation}
V_{1q} = 2\pi ~\text{Im}[\Pi_{\bar{u}q}],
\label{eq:spectral-function}
\end{equation}
which in turn parameterizes the nonperturbative effects in inclusive hadronic $\tau$ decays into strange ($V_{1s}^-$) or nonstrange ($V_{1d}^-$) vector final states:
\begin{equation}
\frac{d \mathcal{B}(\tau^- \to V_{1q} \nu_\tau)}{ds} = \frac{6 |V_{uq}|^2}{m_\tau^2} \frac{\mathcal{B}(\tau^- \to e^- \bar{\nu}_e \nu_\tau)}{\mathcal{B}(\tau^- \to V_{1q} \nu_\tau )} \left(1-\frac{s}{m_\tau^2}\right)^2\left(1+\frac{2s}{m_\tau^2} \right) V_{1q}(s).
\label{eq:taudecay}
\end{equation}

In sections \ref{section:charm-resonances} and \ref{section:light-resonances}, we explain under what approximations $\text{Im}[h_q^\text{KS}]$ is obtained from $R_\text{had}$ and $V_{1q}$. In particular, the charge-weighted sum of~\eqref{eq:hKS} is fixed to experiment:
\begin{equation}
\sum_q Q_q^2~\text{Im}[h_q^\text{KS}] = \frac{4 \pi}{27} R_\text{had}.
\end{equation}
Both the real and imaginary parts of $h_q^\text{KS}$ appear in $\bar{B} \to X_d \ell^+ \ell^-$ observables through interference effects with the short distance amplitudes. Therefore, the real parts are obtained through the subtracted dispersion relation
\begin{equation}
h^\text{KS}_q(s) = h^\text{KS}_q(s_0) + \frac{s-s_0}{\pi} \int_0^\infty dt ~ \frac{\text{Im} [ h^\text{KS}_q(t)] }{(t-s_0)(t-s-i\epsilon)}
\label{eq:dispersion-relation}
\end{equation}
where $s_0 < 0$. The subtraction point $s_0$ should be chosen sufficiently large and negative such that $h_q^\text{KS}(s_0)$ is dominated by short distance fluctuations of the correlator \eqref{eq:hKS} and can be reliably computed in perturbation theory. This is especially important for light quark loops, for which the perturbative matrix elements in \eqref{eq:hlight} diverge at $s=0$. In the following we choose $s_0 = -(5 \text{ GeV})^2$ to minimize the impact of higher order perturbative corrections which depend on $\log s_0/\mu_b^2$.

\begin{figure}[ht!]
\centering
	\includegraphics[width=\textwidth]{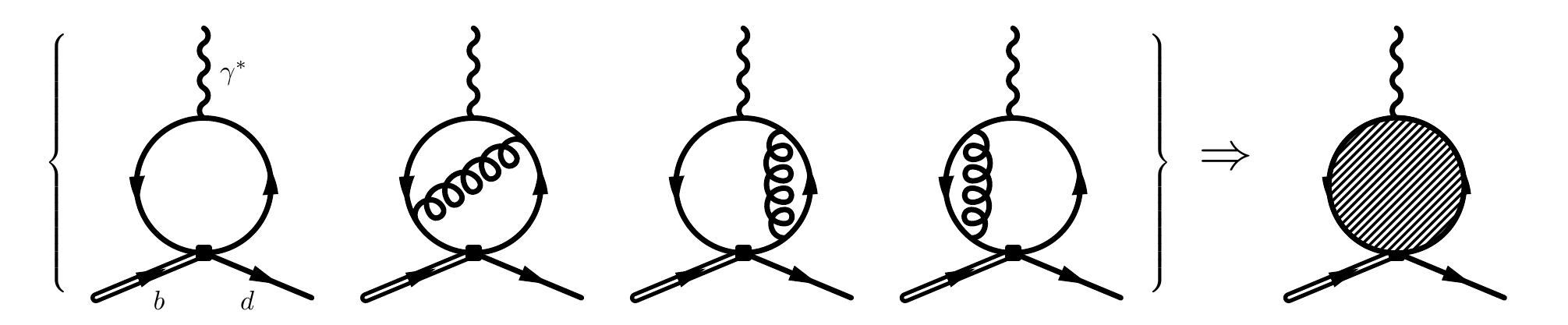} \\
	\includegraphics[width=\textwidth]{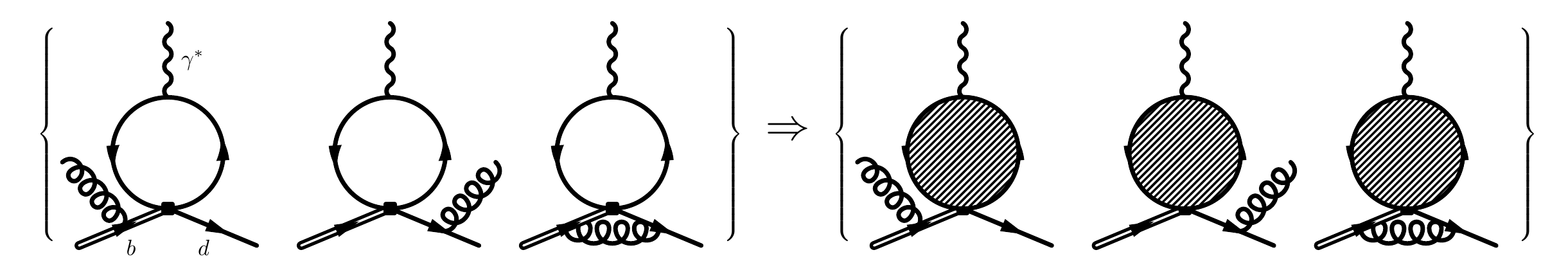}
\caption{The one- and two-loop factorizable quark loop amplitudes and the all-orders factorizable amplitude that replaces them. Nonfactorizable contributions, where the gluon connects the quark-loop and the heavy-light current or radiates off the quark-loop, are kept perturbative and are not replaced by the KS function.}
\label{fig:quarkloop}
\end{figure}

When replacing the perturbative functions by the KS functions there are a number of subtleties at higher orders in the coupling and power expansion. The KS functions encompass factorizable corrections to all orders in $\alpha_s$. Therefore, the KS functions should replace not only the one-loop but also by the $n$-loop factorizable perturbative contributions to avoid double counting. Factorizable QCD corrections are known analytically up to two loops (see~\cite{deBoer:2017way}\footnote{When extracting the factorizable part of the functions in~\cite{deBoer:2017way} one has to keep in mind that $P_1$ and $P_2$ mix under renormalization.} and Appendix \ref{sec:apptwoloop}). The procedure is schematically shown in the first line of figure~\ref{fig:quarkloop}. Via this procedure $\alpha_s$-suppressed corrections shown in the second line of figure~\ref{fig:quarkloop} are replaced as well.

\begin{figure}[ht!]
\centering
\includegraphics[height = 4cm]{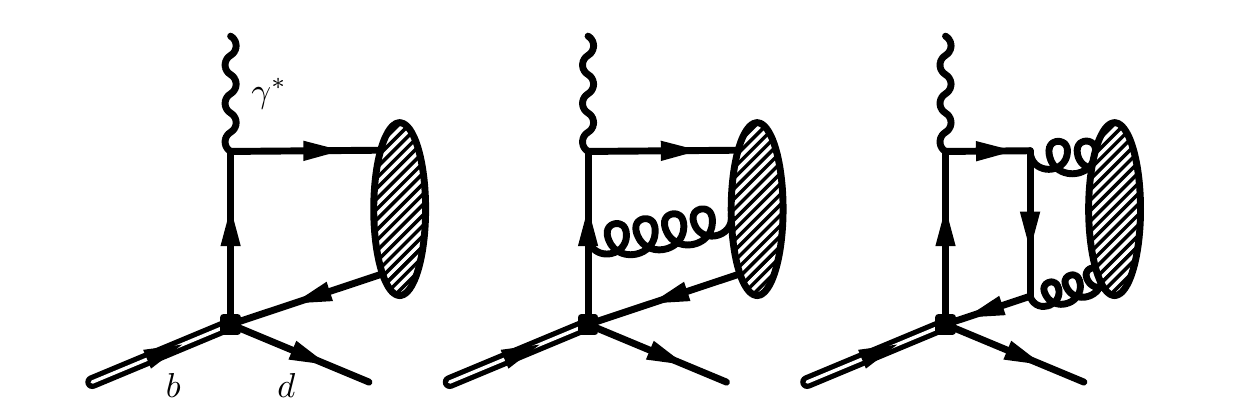}
\caption{Schematic representation of QCD condensates that appear in $\bar B \to X_d \ell^+ \ell^-$ at leading power.}
\label{fig:q2-ope}
\end{figure}

Secondly the function $h(y_q)$ appears at leading power in the $\Lambda_\text{QCD} / m_b$ OPE for $\bar{B} \to X_d \ell^+ \ell^-$, and also at leading power in the $\Lambda_\text{QCD} / Q$ OPE for $e^+ e^- \to \text{hadrons}$. However, there are power-suppressed effects in $e^+ e^- \to \text{hadrons}$ which appear in $\bar B \to X_d \ell^+ \ell^-$ at leading power, depicted in figure \ref{fig:q2-ope}. The light quarks and gluons couple strongly to the QCD vacuum and form $\mathcal{O}(\Lambda_\text{QCD})$ condensates: $\braket{\bar{q} q}, \braket{GG}$, etc. This is captured by the dispersive analysis which evaluates the factorized hadronic ``blob" for both positive and negative $q^2$. This neatly encapsulates why the dispersive analysis improves upon a purely perturbative calculation: it resums not only the coupling expansion but also an implicit power expansion in $\Lambda_\text{QCD} / Q.$

\subsubsection{Experimental inputs}\label{sec:exinputs}
\begin{figure}[t!]
\centering
\includegraphics[width=130mm]{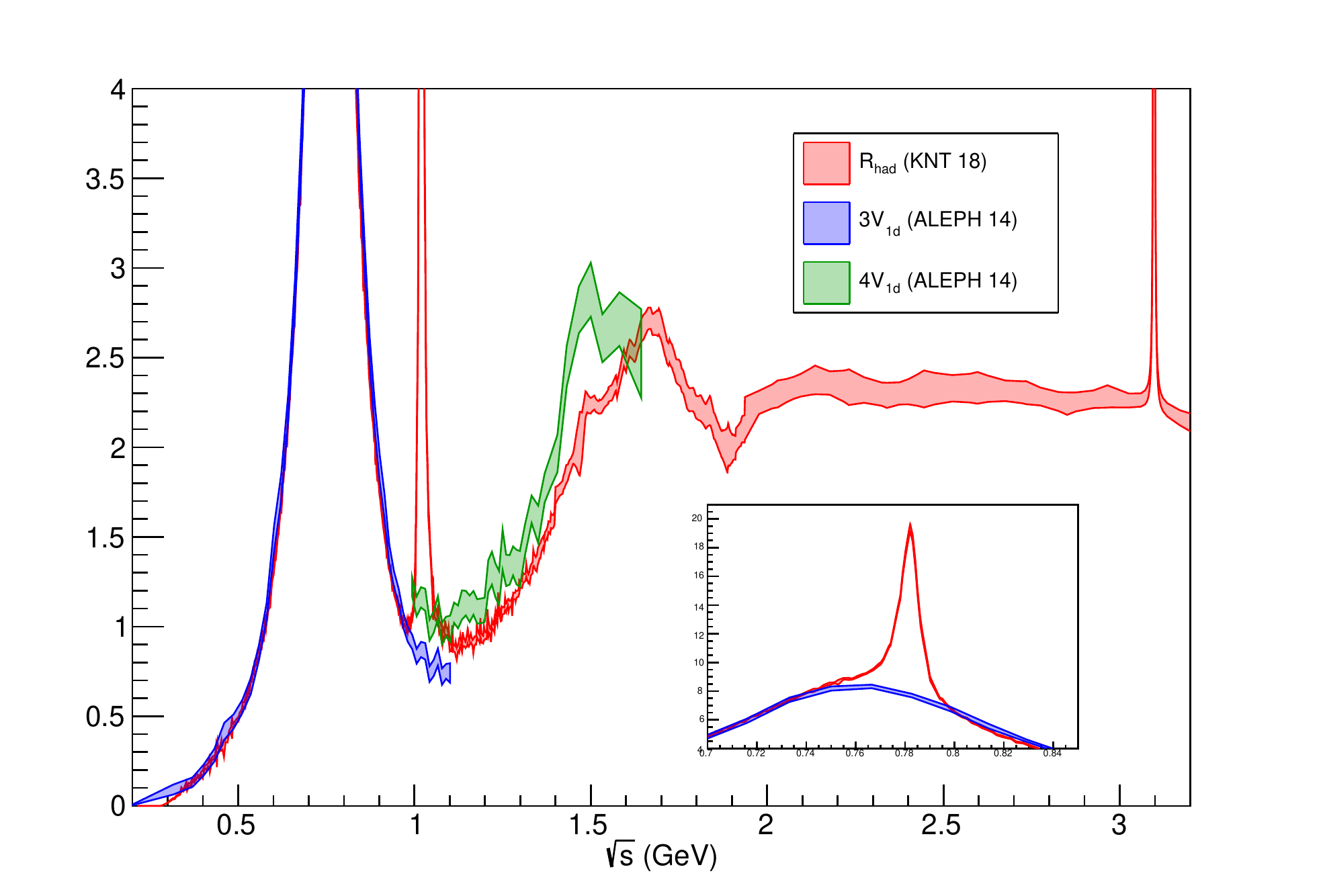}
\caption{The hadronic R-ratio~\cite{Keshavarzi:2018mgv} and spectral function from hadronic $\tau$ decay~\cite{Davier:2013sfa}. Under isospin considerations, $3V_{1d}$ is to be compared with the isovector contribution to $R_\text{had}$ (from the $\rho$ resonance) with the discrepancy due to the $\omega$ shown in the subfigure. Flavour $SU(3)$ symmetry predicts $R_\text{had} = 4 V_{1d}$. For further details see section~\ref{section:light-resonances}.}
\label{fig:data-light}
\end{figure}

For the flavour threshold regions we use a compilation of all available data on the hadronic R-ratio~\cite{Keshavarzi:2018mgv}, in which the data is provided in center of mass energy points with a total point-to-point covariance matrix. The BESII data~\cite{Ablikim:2007gd} dominates the statistics in the charm threshold region. The broad oscillation at $\sqrt{s} \sim 1.6 \text{ GeV}$ due to the phase space enhancements of isobar processes including $e^+ e^- \to \rho^+ \rho^- \to \pi^+ \pi^- \pi^0 \pi^0$ has been resolved by precise measurements of multi-body final states at BABAR~\cite{TheBaBar:2017vzo, Aubert:2005eg,  Aubert:2007ef, TheBABAR:2017aph, Aubert:2006jq, Lees:2014xsh, Aubert:2007ur, Aubert:2007ym, TheBABAR:2018vvb, TheBABAR:2017vgl}. The total nonstrange vector spectral function from $\tau$ decays is taken from ALEPH~\cite{Davier:2013sfa}. A compilation of the data is shown in figure \ref{fig:data-light}. We do not use the strange spectral function because the vector (V) and axial vector (A) contributions are more difficult to distinguish experimentally in this case, and are currently only available in the form V+A.

We supplement this data outside of the resonance region with the results of the program Rhad \cite{Harlander:2002ur} for computing the hadronic R-ratio up to ${\cal O}(\alpha_s^4)$ in perturbation theory. The only inputs into the program are the $\overline{\text{MS}}$ mass $\overline{m}_c(\overline{m}_c) = 1.275(25)$~GeV and $\alpha_s(M_Z) = 0.1181(11)$ from table~\ref{tab:inputs}, $\overline{m}_b(\overline{m}_b) =  4.18(4) \text{ GeV}$ and $\overline{m}_t(\overline{m}_t) = 160(5) \text{ GeV}$~\cite{Tanabashi:2018oca}. The default decoupling scales $\mu_c = 2 m_c$, $\mu_b = m_b$ and $\mu_t = m_t$ are used, and the scale is varied between $\sqrt{s} / 2 < \mu < 2 \sqrt{s}$ to estimate the effect of higher order corrections.
In our data-driven approach, we integrate the data directly rather than fit it to a certain model for the resonances.

\subsubsection{Charm resonances at high $q^2$}\label{section:charm-resonances}
Below open charm threshold, to a good approximation, $\text{Im}[h^\text{KS}_c]$ only has support near the masses of the $\psi$ and $\psi'$ resonances, which may eventually decay into light hadrons but resonate through a $c\bar{c}$ current. Above open charm threshold, and below the matching point to perturbation theory with $n_f = 4$ flavours at $\sqrt{s} = 6 \text{ GeV}$, the contribution to the R-ratio from the light quarks is perturbative and can be subtracted from the measured spectrum.
\begin{equation}
\text{Im}[h^\text{KS}_c] = \left\{
\begin{array}{ll}
0\, , & ~\sqrt{s} < 3 \text{ GeV}, \\ [8pt]
\dfrac{\pi}{3} \left( R_\text{had} - R_{uds}^\text{pert} \right) \, ,& ~3 \text{ GeV}< \sqrt{s} < 6 \text{ GeV}, \\ [10pt]
\dfrac{\pi}{3} R_c^\text{pert}\, , & ~\sqrt{s} >6 \text{ GeV}
\end{array}
\right.
\label{eq:KScharmrule}
\end{equation}

We note that a charmonium resonance can form from a vector current of light quarks in $e^+ e^- \to \text{hadrons}$ through single photon or three gluon exchange. This mixing has a substantial effect for the CP asymmetry in $\bar B \to X_d \psi$ decays \cite{Dunietz:1993cg, Soares:1994vi}. Since the QED correction contributes to the present calculation without logarithmic enhancement, we neglect it and the comparable QCD correction\footnote{In \cite{Dunietz:1993cg} it was stated that three gluon exchange is comparable to single photon exchange because of the similarities of the branching fractions $B(\psi \to \ell^+ \ell^-)$ and $B(\psi \to ggg)$.} and we expect in this case no major nonperturbative enhancement away from the $\psi$ and $\psi'$ resonances.


\subsubsection{Light quark resonances at low $q^2$}\label{section:light-resonances}
The most important new feature for the light quark resonances which we introduce in this paper is to include matrix elements $\langle J_q J_{q^\prime} \rangle_{q\neq q^\prime}$ involving different light-quark currents at very low $q^2$. As a consequence, $R_{\rm had}$ alone is not sufficient to extract $\text{Im}[h^\text{KS}_{u,d,s}(q^2)]$.

\begin{figure}[t!]
\centering
\includegraphics[height = 4cm]{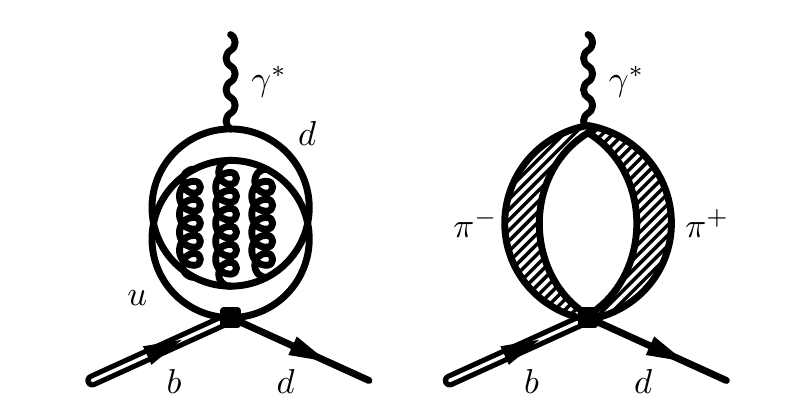}
\caption{``Hairpin-turn'' diagram in QCD which is ${\cal O}(\alpha_s^3)$-suppressed in perturbation theory, and the origin of its nonperturbative enhancement for low momentum transfer.}
\label{fig:hairpin-turn}
\end{figure}

The dominant contributions to the $e^+e^- \to \text{hadrons}$ OPE are from $\langle J_q J_q \rangle$ for $q=u,d,s$ as they enter at $\mathcal{O}(\alpha_s^0)$. The leading power contributions to $\langle J_q J_{q^\prime} \rangle_{q\neq q^\prime}$ (see left panel of figure~\ref{fig:hairpin-turn}) are $\mathcal{O}(\alpha_s^3)$ and therefore very small\footnote{Moreover, in the $SU(3)$ limit, the sum of all these contributions vanishes due to an exact cancellation among light quark charges.}. An expression for these contributions can be found in~\cite{Harlander:2002ur}. This is what lead the authors of ref.~\cite{Kruger:1996dt} to systematically neglect all terms with $q\neq q^\prime$ in \eqref{eq:hKS}.
Unfortunately, at low $q^2$ the OPE for $e^+ e^- \to \text{hadrons}$ breaks down and we have to adopt a sum-over-hadrons picture. For instance, at very low $q^2$ $(\lsim 1 \text{ GeV}^2)$ the dominant hadronic final states are two and three pions~--~corresponding essentially to the $\rho$ and $\omega$ resonances~--~and $\langle J_u J_d \rangle \sim \langle J_u | \pi\pi (\pi) \rangle \langle \pi\pi (\pi)| J_d \rangle \sim \langle J_u | \pi\pi (\pi) \rangle \langle \pi\pi (\pi)| J_u \rangle \sim \langle J_u J_u \rangle$.
At larger $q^2$ $(\gsim 4 \text{ GeV}^2)$ there is a proliferation of multiparticle intermediate states and the OPE result that $\braket{J_u J_d} \sim 0$ is recovered via dramatic cancellations between various exclusive final states~\cite{Lipkin:1984sw, Lipkin:1986av} as confirmed by lattice-QCD calculations~\cite{Isgur:2000ts}.

To quantify these effects, it is convenient to work in terms of a basis of neutral isospin currents
\begin{equation}
J_0 = \frac{J_u + J_d}{\sqrt{2}}, \hspace{0.5cm} J_1 = \frac{J_u - J_d}{\sqrt{2}},  \hspace{0.5cm} J_s,
\label{eq:isocurrents2}
\end{equation}
where $J_0, J_s$ are singlets under isospin and $J_1$ transforms as a vector. The correlation functions of these currents describe the propagation of the relevant degrees of freedom in the low energy resonance region:
\begin{equation}
(q^2 g^{\mu \nu} - q^\mu q^ \nu)\Pi_{ab}(q^2) = i \int d^4 x~ e^{iqx} \braket{0| T {J_a}^\mu(0) {J_b}^\nu(x) |0}.
\label{eq:isocorrelators}
\end{equation}
Note that if the six correlators $\Pi_{ab}$ were known exactly, the six correlators between quark currents $\braket{J_{q_1} J_{q_2}}$ and finally the KS functions~\eqref{eq:hKS} would be determined exactly through simple relations at the operator level. We note in particular that the electromagnetic current and the $u-$quark current are given exactly by
\begin{equation}
J_u = \frac{J_1+J_0}{\sqrt{2}}, \hspace{5mm} J_\text{em} = \frac{3J_1 + J_0 -\sqrt{2}J_s}{3\sqrt{2}}.
\label{eq:jem}
\end{equation}
In the isospin limit, the correlators $\Pi_{10}$ and $\Pi_{1s}$ vanish, \eqref{eq:pihad} and \eqref{eq:pitau} simplify to
\begin{equation}
\Pi_\gamma = \frac{9 \Pi_{11} + \Pi_{00}+2 \Pi_{ss}-2\sqrt{2} \Pi_{0s}}{18}, \hspace{5mm} \Pi_{\bar{u}d} = \Pi_{11},
\label{eq:pigamma2}
\end{equation}
and the KS functions simplify to
\begin{align}
h^\text{KS}_u &= \frac{4\pi^2}{9}(3\Pi_{11}+\Pi_{00}-\sqrt{2} \Pi_{0s}), \nnb \\
h^\text{KS}_d &= \frac{4\pi^2}{9}(6\Pi_{11} -  2\Pi_{00} +2 \sqrt{2} \Pi_{0s}), \nnb \\
h^\text{KS}_s &= \frac{4\pi^2}{9} (4\Pi_{ss}-2\sqrt{2} \Pi_{0s}).
\label{eq:hKSsu2}
\end{align}
Since the KS functions in the isospin limit depend on the four correlators $\Pi_{11}, \Pi_{00}, \Pi_{ss}, \Pi_{0s}$ and only two observables $R_\text{had}$ and $V_{1d}$ are available, additional assumptions are required whose range of applicability depends on the energy. 
\begin{table}[t]
	\begin{center}
	\begin{displaymath}
	\begin{tabular}{|ccccc|}
		\hline\spp
		 $\sqrt{s}$ (GeV)& $[0,0.99]$ & $[0.99, 1.13] $ & $[1.13, 1.65]$ & $[1.65, 3]$\\
		 \hline
		\rule{0pt}{19pt}$\text{Im}[h^\text{KS}_u]$ & $\dfrac{3}{2} \hat R_\text{had} -3 \hat V_{1d}$ & $\dfrac{3}{2} \hat V_{1d}$ & $\dfrac{1}{2} \hat R_\text{had} + \left(\dfrac{1}{2} \hat R_\text{had} - 2 \hat V_{1d}\right)\delta_u$& $\dfrac{1}{2} \hat R_\text{had}$ \\[8pt]
		$\text{Im}[h^\text{KS}_d]$ & $12 \hat V_{1d} - 3 \hat R_\text{had}$ & $3 \hat  V_{1d}$ & $\dfrac{1}{2} \hat R_\text{had} + \left(\dfrac{1}{2} \hat R_\text{had} - 2 \hat V_{1d}\right)\delta_d$ & $\dfrac{1}{2}  \hat R_\text{had}$ \\[8pt]
		$\text{Im}[h^\text{KS}_s]$ & 0 & $3 \hat R_\text{had} - 9 \hat V_{1d}$ & $\dfrac{1}{2} \hat R_\text{had} + \left(\dfrac{1}{2} \hat R_\text{had} - 2 \hat V_{1d}\right)\delta_s$& $\dfrac{1}{2} \hat R_\text{had}$ \\[8pt]
		\hline
	\end{tabular}
	\end{displaymath}
	\end{center}
	\caption{Imaginary parts of the KS functions in various regions as determined from experimental data. One has $\hat R_\text{had} = 4\pi/9 R_\text{had}$ and $\hat V_{1d} = 4\pi/9 V_{1d}$. \label{tab:KSbreakdown}}
\end{table}

Below $K\bar{K}$ threshold, in addition to isospin symmetry, we assume that the hidden strange contributions to the final states $\pi^+ \pi^-$ and $\pi^+ \pi^- \pi^0$ are small ($\text{Im}[\Pi_{ss}] = \text{Im}[\Pi_{0s}] = 0$). Then the Kr{\"u}ger-Sehgal functions as well as $R_\text{had}$ and $V_{1d}$ depend only on $\Pi_{00}$ and $\Pi_{11}$, and inverting these equations yields the second column of table \ref{tab:KSbreakdown}.

The $\phi$ resonance, which we identify as the region between $K\bar{K}$ and $K\bar{K}\pi$ thresholds, decays predominantly into $\pi^+ \pi^- \pi^0$ and $K\bar{K}$ final states, and these contributions are understood as contributions to $\Pi_{ss}$ up to rescattering effects suppressed in $\Pi_{0s}$ and further suppressed in $\Pi_{00}$, which we both neglect. The isovector background dominated by the tail of the $\rho$ is understood from the $\tau$ data, using the isospin correspondence $R^{I=1}_\text{had} \to 3 V_{1d}$ (see figure~\ref{fig:data-light}). This yields the third column of table \ref{tab:KSbreakdown}.

Above $K\bar{K}\pi$ threshold, all four correlators $\Pi_{ab}$ appearing in \eqref{eq:hKSsu2} are important. To proceed, we consider the consequences of enlarging the symmetry group to flavour $SU(3)$, introducing the currents:
\begin{equation}
J_0^{(3)} = \frac{J_u + J_d + J_s}{\sqrt{3}}, \hspace{0.3cm}J_3^{(3)} = \frac{J_u - J_d}{\sqrt{2}},  \hspace{0.3cm} J_8^{(3)} = \frac{J_u + J_d - 2 J_s}{\sqrt{6}}.
\label{eq:isocurrents3}
\end{equation}
where $J_0^{(3)}$ is an $SU(3)$ singlet and $J_{3,8}^{(3)}$ transform as vectors (the subscripts refer to Gell-Mann matrix indices). In the flavour symmetry limit, the correlators $\Pi_{03}^{(3)}$ and $\Pi_{08}^{(3)}$ vanish, $\Pi_{38}^{(3)}$ vanishes by isospin symmetry, and $\Pi_{33}^{(3)} = \Pi_{88}^{(3)}$. Only two independent correlators remain: $\Pi_{33}^{(3)}$ and $\Pi_{00}^{(3)}$. The vacuum polarization and spectral function, however, are independent of $\Pi_{00}^{(3)}$. Therefore flavour symmetry predicts $R_\text{had} = 4 V_{1d}$, to be compared with experiment (figure~\ref{fig:data-light}). The difference between $R_\text{had}$ and $4V_{1d}$ corresponds to the breaking of flavour symmetry, which is apparent but moderate. In the flavour symmetry limit the Kr{\"u}ger-Sehgal functions are independent of flavour, but there is a systematic error associated to the difference between $R_\text{had}$ and $4 V_{1d}$. We account for this by introducing standard normal variables $\delta_{u,d,s}$ which are varied in the error analysis (see fourth column of table~\ref{tab:KSbreakdown}). For $\sqrt{s} > 1.65 \text{ GeV}$, where there is no reliable $\tau$ data, we assume flavour symmetry, see last column of table \ref{tab:KSbreakdown}. The perturbative result from Rhad \cite{Harlander:2002ur} is used for $\sqrt{s}> 3 \text{ GeV}$.

\vspace*{5pt}

\begin{figure}[t!]
\centering
\includegraphics[width=75mm]{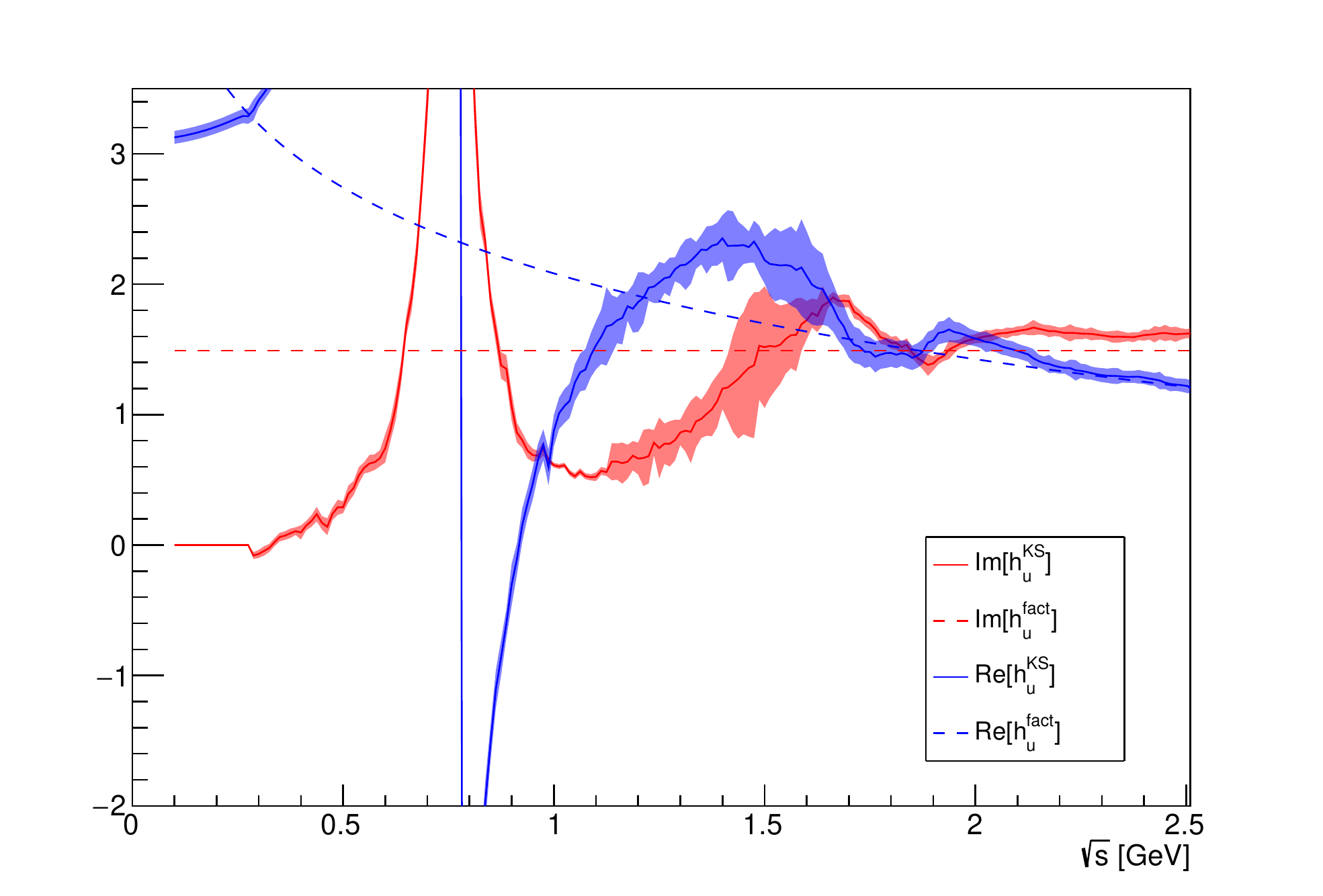} \includegraphics[width=75mm]{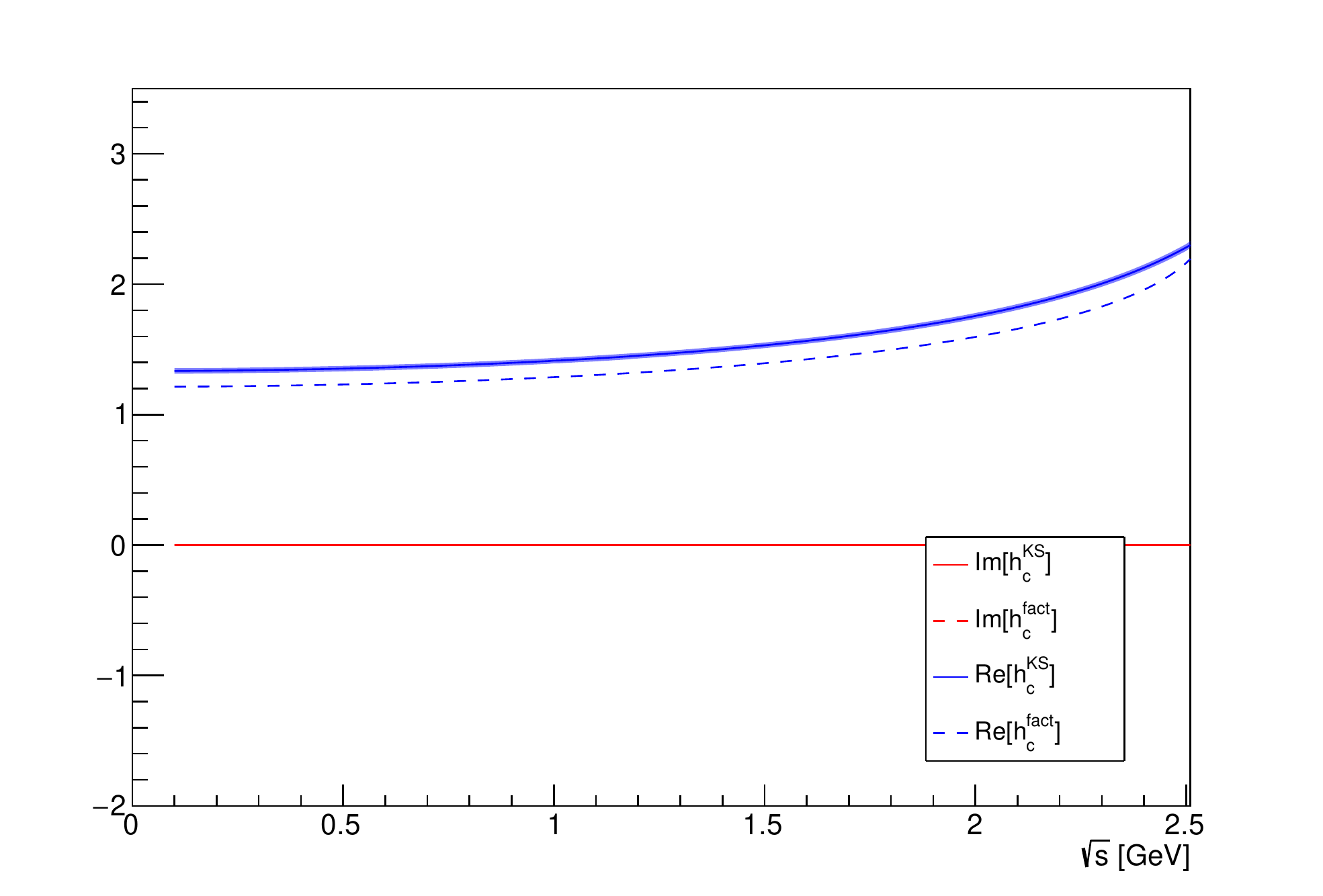} \\
\includegraphics[width=75mm]{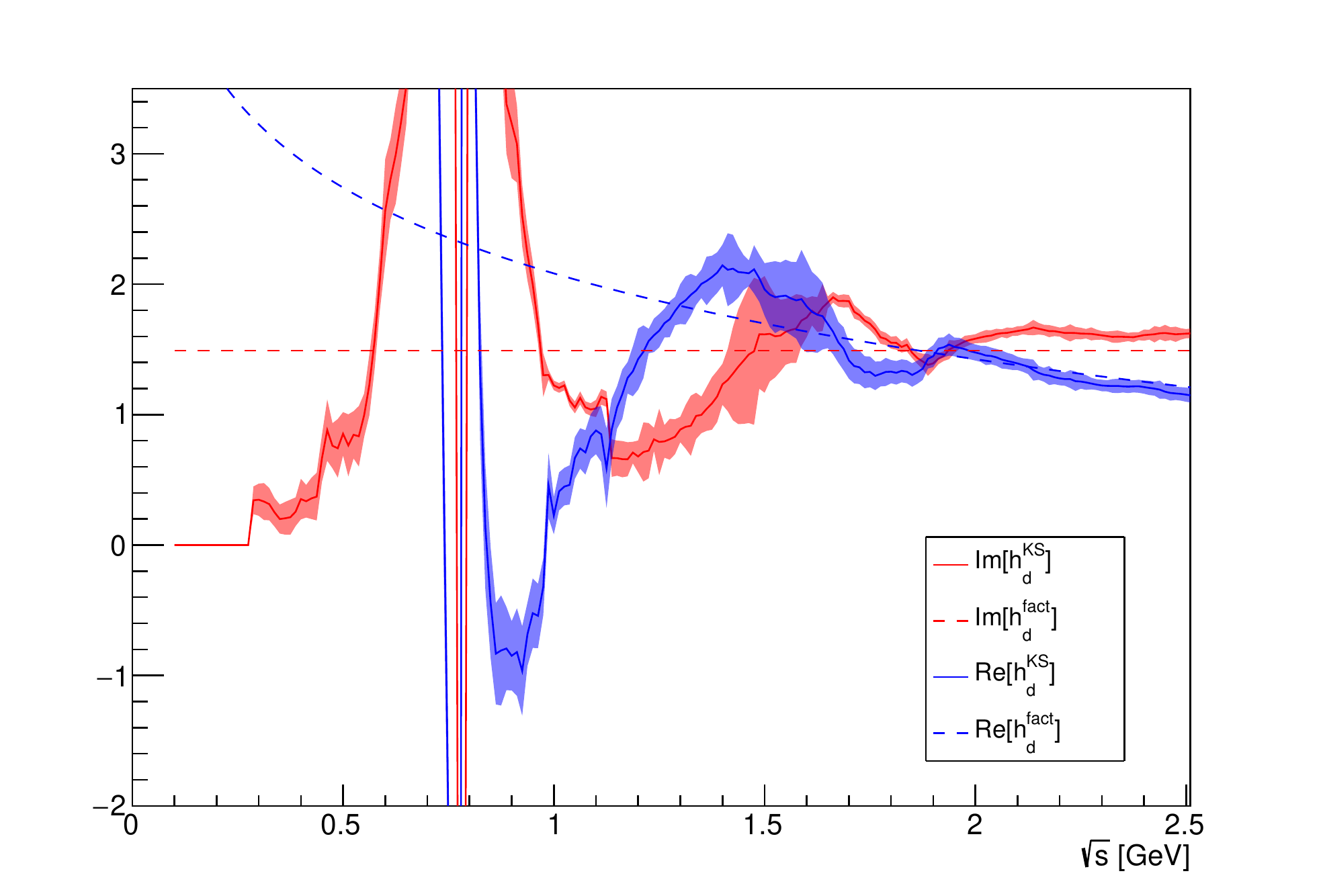} \includegraphics[width=75mm]{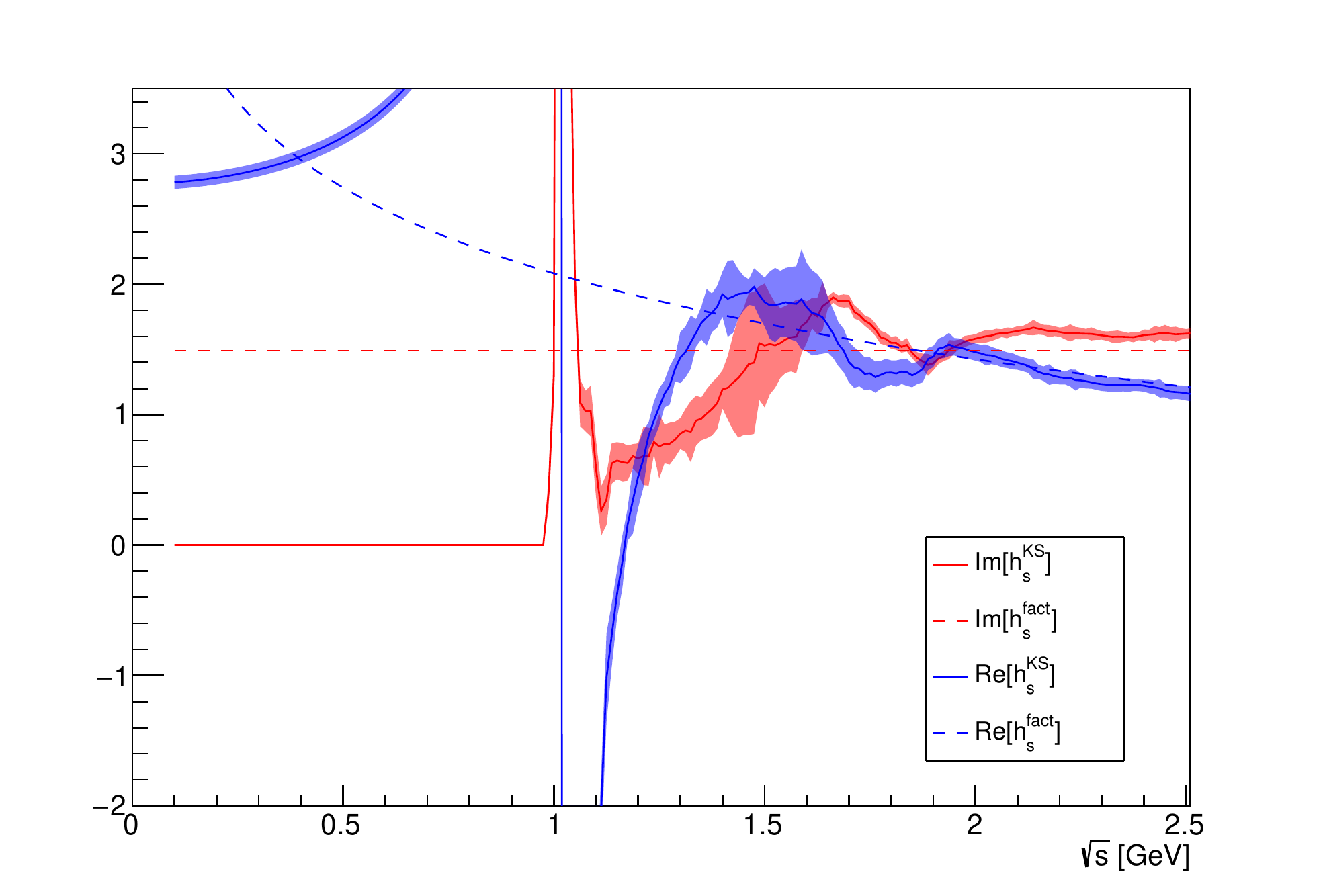} \\
\includegraphics[width=75mm]{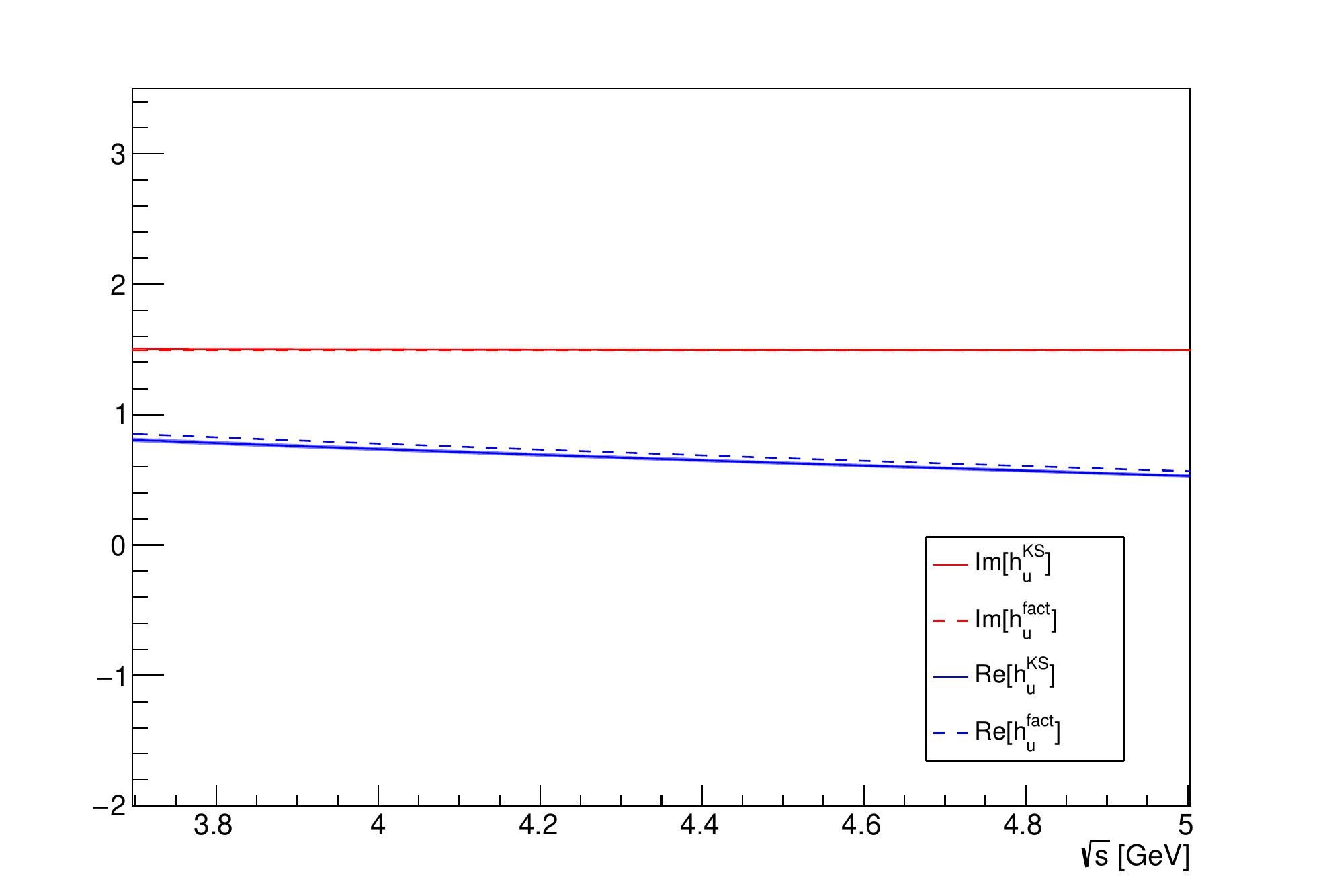} \includegraphics[width=75mm]{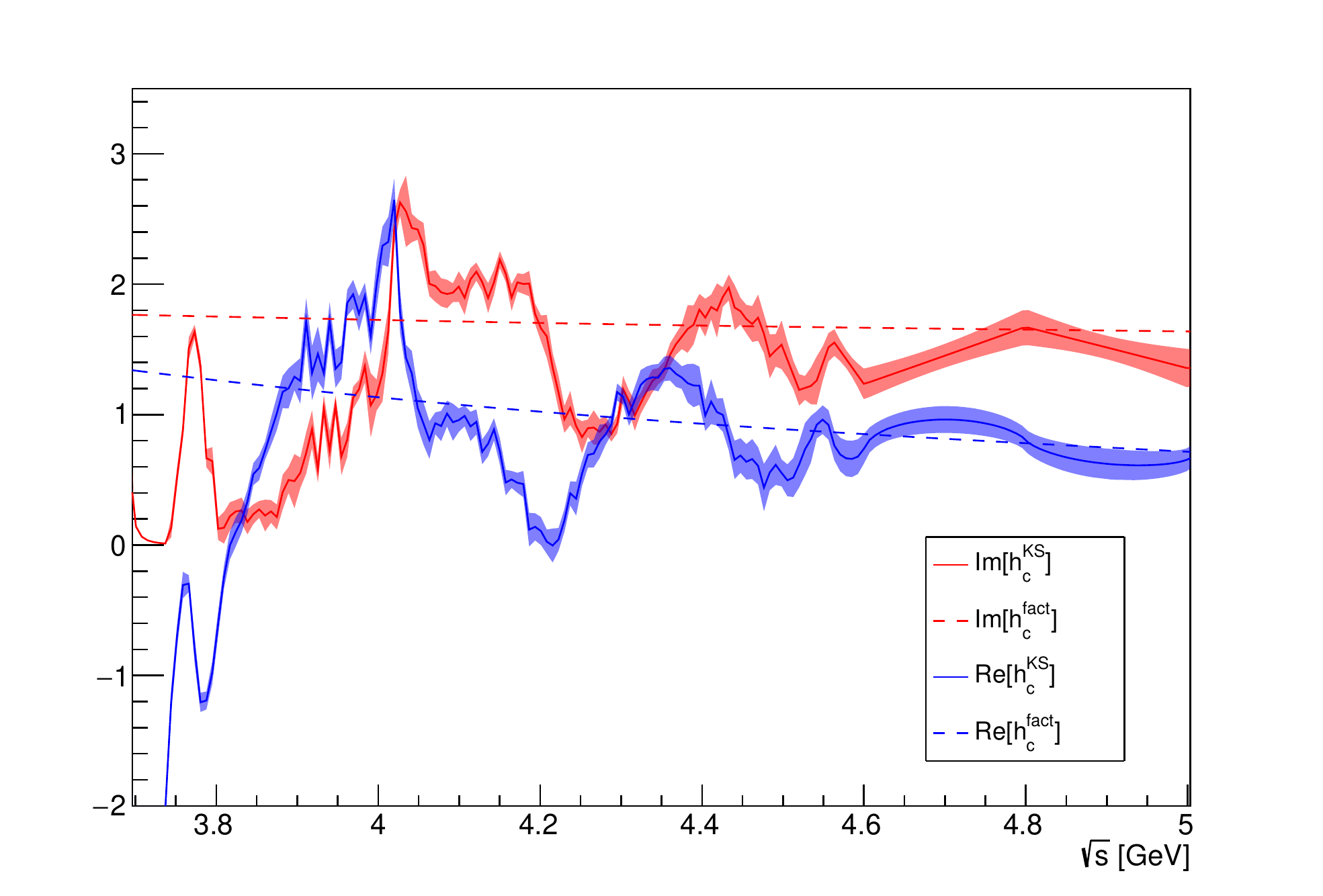} \\
\caption{KS functions and the corresponding perturbative functions to two loops in the low- and high-$q^2$ region. See text for further details.}
\label{fig:light-KS}
\end{figure}

By means of this procedure, we obtain the KS functions which are displayed in figure~\ref{fig:light-KS} together with the perturbative functions up to two loops.
Uncertainties are propagated by generating samples of the data, and for each sample calculating $\text{Im}[h_q^\text{KS}]$ using table \ref{tab:KSbreakdown} and eq.~\eqref{eq:KScharmrule}, and subsequently calculating the real part via the integral~\eqref{eq:dispersion-relation}. Note that in this way, estimates of $SU(3)$ breaking and uncertainties in charmonium resonance parameters are included in the error analysis. We investigated the dependence on the subtraction point $s_0$ and found it to be small.

\subsection{Resolved contributions at low $q^2$}\label{section:resolved}
As discussed at the beginning of this section, the local heavy mass expansion  breaks down
if one includes operators beyond the leading ones in the effective field theory. One then finds  nonlocal power corrections in the low-$q^2$ region which can be systematically analysed within
soft-collinear effective theory (SCET).  The resolved photon contributions to the inclusive decay
$\bar B \rightarrow X_{s(d)} \ell^+ \ell^-$ contain subprocesses in which the virtual photon couples to light partons instead of connecting directly to the effective weak-interaction vertex. These resolved contributions of the $\bar B \rightarrow X_{s} \ell^+ \ell^-$ decay were calculated in SCET in the presence of an $M_X$ cut to order $1/m_b$~\cite{Hurth:2017xzf, Benzke:2017woq}.  
They  can be represented as the convolution integrals of a jet-function, characterizing the hadronic final state $X_{s(d)}$, and of a soft (shape) function which is defined by a non-local heavy-quark effective theory matrix element.  The hard contribution is factorized into the Wilson coefficients.
It was explicitly shown~\cite{Hurth:2017xzf, Benzke:2017woq}, that the resolved contributions stay nonlocal when the hadronic cut is released and thus, represent an irreducible uncertainty. The support properties of the shape function imply that the resolved contributions (besides the ${\cal O}_{8g} - {\cal O}_{8g}$ one\footnote{In this subsection we follow the notation of refs.~\cite{Hurth:2017xzf, Benzke:2017woq} which uses the BBL basis with operators ${\cal O}_i$.}) are almost cut-independent.

Within the inclusive decay {\bf $\bar B \to X_{d} \ell^+ \ell^-$},  there are four resolved contributions at leading order in $1/m_b$ for the decay rate,  namely from the interference terms ${\cal O}_{7\gamma} - {\cal O}_{8g}$,\, ${\cal O}_{8g} - {\cal O}_{8g}$, and ${\cal O}^{c}_{1} - {\cal O}_{7\gamma}$, but also
${\cal O}^{u}_1 - {\cal O}_{7\gamma}$ . For the $b \to d$ case the resolved contributions need some obvious modifications compared to the $b \to s$ case which was calculated in Refs.~\cite{Hurth:2017xzf,Benzke:2017woq}: The CKM parameter combinations $\lambda_i^s = V_{is}^\ast V_{ib}$ have to be replaced by $\lambda_i^d = V_{id}^\ast V_{ib}$ and $s$-quark fields have to be replaced by $d$-quark fields in the shape functions. These modifications only  change the numerical results.

It is well-known that the ${\cal O}^{u}_{1} - {\cal O}_{7\gamma}$ contribution is CKM-suppressed in the $b \to s$ case, but not in the $b \to d$ case. 
However, both in $b \to d$ and in $b \to s$,  this  contribution from the $u$-quark loop vanishes within the CP averaged quantities at the order $1/m_b$ as one can derive from the results given in Ref.~\cite{Benzke:2017woq}: If we start with the ${\cal O}^c_{1} - {\cal O}_{7\gamma}$  contribution in eq.~(6.3) of ref.~\cite{Benzke:2017woq} and consider the penguin functions given in eqs.~(4.4) and~(4.5) of that reference, which enter the jet function, we find  in the limit $m_c \to m_u = 0$  that the $\omega_1$ integral reduces to 
\begin{equation}
\int d\omega_1 \frac{1}{\omega_1 + i \epsilon} \times \frac{1}{\omega_1}\left[\,\, \omega_1\,\, \right]
\,g_{17}(\omega,\omega_1,\mu)\,.
\end{equation}
The trace formalism of HQET (see ref.~\cite{Benzke:2010js}) implies that
\begin{equation}\label{g17int}
   \int_{-\infty}^{\bar\Lambda}\!d\omega\,
    g_{17}(\omega,\omega_1,\mu)
   = \int_{-\infty}^{\bar\Lambda}\!d\omega\,
    ( g_{17}(\omega,-\omega_1,\mu) )^* .
    \end{equation}
Moreover, it is a consequence of PT invariance that $g_{17}$ is real. Thus, the integration of $\omega_1$ leads to the result that the interference term ${\cal O}^u_{1} - {\cal O}_{7\gamma}$ vanishes within the integrated CP averaged rate. This is a crucial result for all  CP-averaged inclusive $b \to d \ell^+ \ell^-$ quantities because previously no estimate for this up-quark loop of order $\Lambda_{\rm QCD}/m_b$  was available (see ref.~\cite{Buchalla:1997ky}) and thus represented the main uncertainty in the inclusive $b \to d \ell^+ \ell^-$ observables.
Further insight into the moments of $g_{17}$ was recently given in~\cite{Gunawardana:2019gep}.

The calculation of the other (nontrivial) resolved contributions given in refs.~\cite{Hurth:2017xzf, Benzke:2017woq} starts with the explicit form of the shape functions as HQET matrix elements and  derives general properties of those.
One can then use various model functions which have all these properties to get conservative estimates of the resolved contributions by maximizing the value of the convolution integral of the subleading shape function with the perturbatively calculable jet function (for more details see ref.~\cite{Benzke:2017woq}).  We are  interested in  the relative magnitude of the resolved contributions compared to the total decay rate. We finally get for the various contributions at order $1/m_b$ in the $b \to s$ and $b \to d$ decays:
\begin{eqnarray}
{\mathcal F}^s_{17}\in [-0.5,+3.4]\,\%,&{\mathcal F}^d_{17}\in [-0.6,+4.1]\,\%,
{\mathcal F}^{d,s}_{78}  \in [-0.2,-0.1]\,\%,&{\mathcal F}^{d,s}_{88}  \in [0,0.5]\,\%\,. \nnb \\ \label{eq:Fijresolved}
\end{eqnarray}
Summing them up in a conservative way we  arrive at
 \begin {eqnarray}
 {\mathcal F}^d_{{1/m_b}}\in [-0.8,+4.5]\,\%,\,\, {\mathcal F}^s_{{1/m_b}}\in [-0.7,+3.8]\,\%.
\end{eqnarray}

It was found~\cite{Hurth:2017xzf,Benzke:2017woq}
that at leading order in $1/m_b$ there is no resolved contribution to the forward-backward asymmetry. This starts at order $1/m_b^2$ only with an interference term of ${\cal O}^c_{1} - {\cal O}_{10}$ for example.   Also the resolved ${\cal O}^c_{1} - {\cal O}_{9}$ term, contributing to the rate, only occurs at the subleading $1/m_b^2$ order. This is a consequence of the fact that the virtual photon is hard-collinear and not hard in the low-$q^2$ region as explicitly shown in refs.~\cite{Hurth:2017xzf,Benzke:2017woq}. On the other hand, these $1/m_b^2$ terms might be numerically relevant due to the large ratio $ |C_{9/10}| \sim 13  |C_{7\gamma}|$ of Wilson coefficients which necessitates their calculation~\cite{Benzkeworkinprogress}.

Because of the opposite sign of $C_9$ compared to $C_7$ one can also expect the same behaviour of
the resolved ${\cal O}^c_{1} - {\cal O}_{9}$ term with respect to ${\cal O}^c_{1} - {\cal O}_{7\gamma}$. We therefore estimate the interval of the missing ${\cal O}^c_{1} - {\cal O}_{9}$ piece to be reversed with respect to eq.~(\ref{eq:Fijresolved}) and add it linearly to
the interval of the corresponding ${\cal O}^c_{1} - {\cal O}_{7\gamma}$ term,
\begin{eqnarray}
{\mathcal F}^s_{1(7+9)}\in [-3.9,+3.9]\,\%,&{\mathcal F}^d_{1(7+9)}\in [-4.7,+4.7]\,\%.
\end{eqnarray}

In our final calculation we combine this result with the ${\cal O}_{7\gamma} - {\cal O}_{8g}$ and ${\cal O}_{8g} - {\cal O}_{8g}$ interferences from~\eqref{eq:Fijresolved} to obtain
\begin{eqnarray}
{\mathcal F}^s\in [-4.1,+4.3]\,\%,&{\mathcal F}^d\in [-4.9,+5.1]\,\%.
\end{eqnarray}
For the first nontrivial resolved contribution to the forward backward asymmetry from the  ${\cal O}^c_{1} - {\cal O}_{10}$ term at order $1/m_b^2$  we add an error of $\pm 5 \%$  in our final result  before an explicit estimate is available~\cite{Benzkeworkinprogress}.


\subsection{Nonfactorizable power contributions at high $q^2$}\label{section:local}
Power corrections due to operators beyond the leading ones also exist  in the high-$q^2$ region. 
The only available pieces are  the nonfactorizable charm- and up-loop diagrams of the four-quark operator ${P}_{1,2}$ with a soft gluon which interacts with the spectator cloud. However, in the high-$q^2$ region  the dilepton mass $q^2$ is a hard momentum and any  cut on the hadronic mass has no  influence in the high-$q^2$ region. Thus, the kinematic situation is a different one compared to the low-$q^2$ region, in particular there is no nonlocal shape function involved. In this case the original
treatment  of Voloshin~\cite{Voloshin:1996gw} is applicable which leads to a local expansion again~\cite{Buchalla:1997ky}.

Here we  briefly recall the crucial issues of the  calculational details presented in ref.~\cite{ Buchalla:1997ky}.
The nonperturbative effect due to the $\bar  s b \gamma g$ vertex is represented by a form factor $\overline F$ which depends on the two variables $ r= q^2 / (4 m_c^2)$ and $t= k\cdot q / (2 m_c^2)$ where $k$ denotes the soft gluon
momentum ($k^2 = 0$) and $q$ the virtual photon momentum.
The form factor ${\overline F}(r+t,t)$ is given in eq.~(4.28) of ref.~\cite{Buchalla:1997ky}.
One may  expand ${\overline F}$ in powers of $t$ . In the high-$q^2$ region $q$ is a hard momentum (of order $m_b$) and $k$ is a soft momentum. Thus,  if  $m_b \Lambda_{QCD} / (2 m_c^2)$ is small the first term in the expansion about $t=0$ can be regarded as dominating.  Moreover,  one  may  additionally expand the form factor also in $1/r$ which  is of order  $4 m_c^2/m_b^2$.  The authors of ref.~\cite{Buchalla:1997ky} then keep only the leading term in $1/r$ in each of the coefficients of $t^n$ and find
\begin{equation}\label{fstr}
{\overline F}(r+t,t)\approx\sum^\infty_{n=0}(-1)^{n+1}
\frac{3}{(n+2) r^{n+1}}t^n=-\frac{3}{2r}+\frac{t}{r^2}+\ldots
\end{equation}
This means that the leading corrections to the $t=0$ result are suppressed by $t/r = 2 k\cdot q / q^2$. 
An additional numerical test~\cite{Buchalla:1997ky} suggests that  the  $t=0$ term is the dominating one in the high-$q^2$ region. The concrete results for the leading $1/m_c^2$ term are given in section~\ref{sec:resolvedhigh}.  If we consider the corresponding nonfactorizable contribution with an up-quark loop in the high-$q^2$ region, one finds that the leading term is of order
$\Lambda^2_\text{QCD}/q^2$ and  corrections are suppressed by powers of $t/r\sim \Lambda_\text{QCD}/ (2 \sqrt{q^2})$~\cite{Buchalla:1997ky}.  The leading order results for the up-quark are also given in section~\ref{sec:resolvedhigh}.

\begin{figure}[t!]
\centering
\includegraphics[height=6cm]{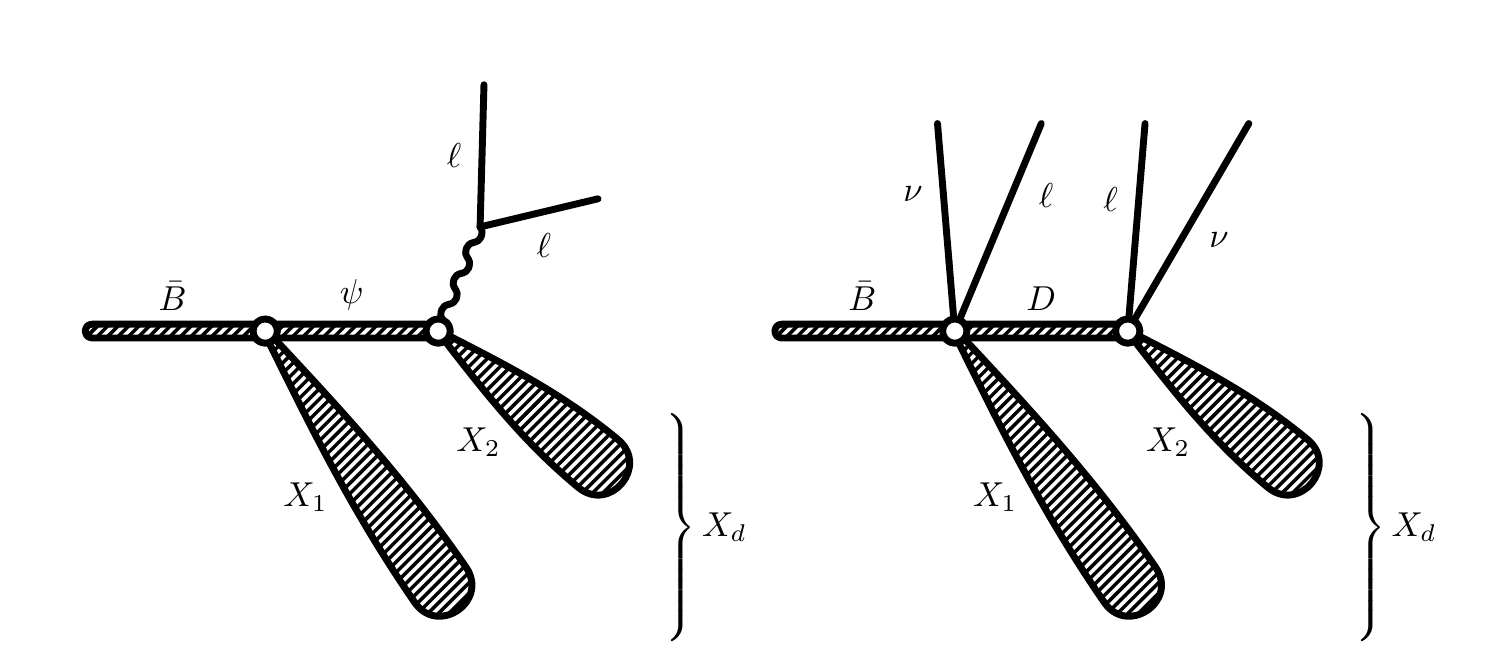}
\caption{Long distance backgrounds not removed by $q^2$ cuts. Left: Charmonium cascade lowering $q^2$ to the perturbative window.
Right: Double semileptonic background through sequential weak decays.}
\label{fig:cascade}
\end{figure}


\subsection{Charmonium cascade backgrounds}\label{section:cascades}
Another long distance effect at low $q^2$ are the cascade decays $\bar B \to X_1 (c\bar{c} \to X_2 \ell^+ \ell^-)$ through the radiative decay of a narrow charmonium resonance $\eta_c, \eta_c', \psi, \psi', \chi_{cJ}, h_c$ or exotic XYZ state, collectively referred to as $c\bar{c}$, 
as depicted in the left panel of figure \ref{fig:cascade}. In contrast to the infamous $\psi \to \ell^+ \ell^-$, for example the decay $\psi \to \eta' \ell^+ \ell^-$ completely escapes the upper cut $q^2=\text{6 GeV}^2$ in $\bar B \to X_q \ell^+ \ell^-$:
\begin{equation}
4m_\ell^2 < q^2 < (M_{c\bar{c}} - M_{X_2})^2.
\label{eq:q2-from-mx}
\end{equation}
In the following we focus on $\bar B \to X_s (c\bar{c} \to X_2 \ell^+ \ell^-)$ with the understanding that the relative effect of the cascades on $\bar B \to X_d \ell^+ \ell^-$ and $\bar B \to X_s \ell^+ \ell^-$ are roughly the same due to the CKM scaling of charmonium production:
\begin{equation}
\frac{\Gamma(B \to X_s c\bar{c})} {\Gamma(B \to X_d c\bar{c})} \sim \frac{\Gamma(B \to X_s \ell^+ \ell^-)} {\Gamma(B \to X_d \ell^+ \ell^-)} \sim \left| \frac{V_{cb} V_{cs}}{V_{cb} V_{cd}} \right|^2.
\label{eq:CKM-scaling}
\end{equation}
Charmonium production from $B$-decays is reasonably well described by an expansion in the heavy quark velocity (NRQCD)~\cite{Beneke:1998ks, Beneke:1999gq} and has been investigated by several experiments, summarized in table~\ref{tab:cascade-table}. The inclusive spectra from $c\bar{c} \to X \ell^+ \ell^-$ are not yet available, although the decays $\psi \to (\pi^0, \eta, \eta') \ell^+ \ell^-$ have been measured at BESIII \cite{Ablikim:2014nro, Ablikim:2018eoy, Ablikim:2018bhf}, and happen to be the most important. We note that the dilepton decays between charmonium states \cite{Ablikim:2017kia, Ablikim:2019jqp, Aaij:2017vck} are not pertinent to $\bar B \to X_s \ell^+ \ell^-$ because the leptons in this case come with invariant mass below the difference in charmonium masses, which is less than $1 \text{ GeV}$.

Radiative and dilepton charmonium decays have been mentioned in the context of $\bar B \to X_s \gamma$~\cite{Misiak:2000jh, Buchalla:1997ky,Beneke:2009az} and $\bar B \to X_s \ell^+ \ell^-$~\cite{Buchalla:1997ky}. The background in $\bar B \to X_s \ell^+ \ell^-$ could simply be subtracted by vetoing events where the two leptons and any permutation of light hadrons reconstruct any of the charmonium masses. However, only the direct leptonic decay of $\psi$ and $\psi'$ were interpreted as background to $\bar B \to X_s \ell^+ \ell^-$ at Belle~\cite{Sato:2014pjr, Iwasaki:2005sy} and BaBar~\cite{Lees:2013nxa}. This is problematic because there are also cascade decays of the type $\bar B\to X_1 \psi \to X_1 X_2 \ell^+\ell^-$ that form a reducible background in the limit in which interference between the cascade and the genuine short distance $\bar B\to X_s \ell^+\ell^-$ amplitudes is negligible. On general grounds this interference is expected to be much smaller than the square of the cascade amplitude. If estimates of the cascade contributions are low enough, we can argue that interference effects can be neglected, implying that these cascades are a reducible background that can be either separately calculated and subtracted or experimentally removed. As we show below, cascades in $\bar B\to X_s \ell^+\ell^-$ satisfy this requirement, but only after taking into account the cut on the invariant mass of the $X_s$ system which is required experimentally to remove the double semileptonic background, as shown in the right panel of figure \ref{fig:cascade}.
In the rest of this subsection we show how to estimate the impact of the $M_X$ cuts on a generic cascade.
\begin{table}[t!]
	\centering
	\begin{tabular}{|l|l||l|l|}
		\hline
 \rule{0pt}{14pt}& $\mathcal{B} \times 10^3$ & & $\mathcal{B} \times 10^5$ \\
		\hline
		$\bar B\to X_s \psi $ 		   & $7.8 \pm 0.4$ & $\psi \to \eta\ell^+\ell^-$ & $1.43 \pm 0.07$ \\
		$\bar B\to X_s \psi' $ 		& $3.07 \pm 0.21$  & $\psi \to \eta'\ell^+\ell^-$ & $6.59 \pm 0.18$ \\
		$\bar B\to X_s \chi_{c1}$    & $3.09\pm 0.22$  	             & $\psi \to \pi^0\ell^+\ell^-$ & $0.076 \pm 0.014$ \\
		$\bar B\to X_s\chi_{c2}$ 	   & $0.75 \pm 0.11$ & $\psi' \to \eta'\ell^+\ell^-$ & $0.196\pm 0.026$ \\
      $\bar B\to X_s \eta_c $ 		& $4.88\pm 0.97$ \cite{Aaij:2014bga} & & \\
      $\bar B\to X_s \chi_{c0} $ 	& $3.0 \pm 1.0$ \cite{Aaij:2017tzn} 		& & \\
		$\bar B\to X_s h_c $ 		   & $2.4 \pm 1.0^\dagger$ \cite{Beneke:1998ks} & & \\
      $\bar B\to X_s \eta_c'$ 		& $0.12 \pm 0.22^\dagger$ \cite{Fan:2011aa} & & \\
		\hline
	\end{tabular}
	\caption{Branching ratios of (direct) inclusive $B$-decay into charmonium, and of vector charmonium dilepton decay to light pseudoscalars. Numbers marked with $\dagger$ are NRQCD estimates, and unless otherwise stated are taken from the PDG~\cite{Tanabashi:2018oca}.}
	\label{tab:cascade-table}
\end{table}

The relative momentum between $X_1$ and $X_2$ implies that the cascade events come with somewhat large total $M_X$ when the two systems are combined. Since it is already necessary to measure $\bar B \to X_s \ell^+ \ell^-$ with an $M_X$ cut to remove the double semileptonic background, if cascade events were efficiently removed by this cut, then there would be no need for their further consideration. The invariant mass of the $X$ system is given  by
\begin{align}
M_X^2 &= M_{X_1}^2 + M_{X_2}^2 + 2 E_{X_1} E_{X_2}\left(1- \frac{|\vec{p}_{X_1}||\vec{p}_{X_2}|}{E_{X_1} E_{X_2}} \cos \theta_X \right) , \label{eq:mx_tot} \\
E_{X_1} &= \frac{M_B^2 - M_{c\bar{c}}^2 - M_{X_1}^2}{2 M_{c\bar{c}}},\label{eq:EX1}\\
E_{X_2} &= \frac{M_{c\bar{c}}^2 + M_{X_2}^2 - q^2}{2 M_{c\bar{c}}}, \label{eq:EX2}
\end{align}
where $\theta_X$ is the angle between the $X_1$ and $X_2$ systems in the charmonium rest frame.
It is interesting to consider the minimum value of $M_{X_2}$ such that $M_{X} > M_{X}^\text{cut}$ for all $q^2$ and $M_{X_1}$. This is given in closed form by
\begin{equation}
M_{X_2}^\text{cut} = \frac{M_{c\bar{c}} M_{X}^\text{cut}}{M_B}
\label{eq:hard-cut}
\end{equation}
and corresponds to the extreme values $\theta_X = 0$, $M_{X_1}=0$ and $q^2 =0$.
States with $M_{X_2}$ heavier than \eqref{eq:hard-cut} are completely removed by the $M_X$ cut. In the case of $\psi$-decays with $M_X^\text{cut} = 2 \text{ GeV}$, $M_{X_2}^\text{cut} \sim 1.2 \text{ GeV}$ (the minimum mass $M_{X_1} = M_K$ and cut at $q^2=1\text{ GeV}^2$ causes this to be slightly smaller). The decays $\psi \to (\pi, \eta, \eta')\ell^+ \ell^-$ and nonresonant S- or D-wave $\psi \to 2\pi \ell^+ \ell^-$ are therefore of interest, while the resonances $\psi \to (f_0,f_2) \ell^+ \ell^-$ are cut away. Inferring from the photon energy spectrum in $\psi \to 2\pi \gamma$, nonresonant $\psi \to 2\pi \ell^+ \ell^-$ is probably very small. Similarly $h_c \to (\eta, \eta',2\pi)\ell^+ \ell^-$ and $\chi_{c1} \to (\rho, \omega) \ell^+\ell^-$ are of interest but they are suppressed by an order of magnitude compared with the $\psi$ decays in the real photon case:
\begin{equation}
	\frac{\mathcal{B}(B \to X \chi_{c1}) \mathcal{B} (\chi_{c1} \to \rho \gamma)}{\mathcal{B}(B \to X \psi) \mathcal{B}(\psi \to \eta \gamma)} \sim 0.08 \, , \hspace{1cm} \frac{\mathcal{B}(B \to X h_c) \mathcal{B} (h_c \to \eta \gamma)}{\mathcal{B}(B \to X \psi) \mathcal{B}(\psi \to \eta \gamma)} \sim 0.14 \ . \label{eq:tensor-resonances}
\end{equation}
The sequence $\bar B \to X_1\psi'$, $\psi' \to 2\pi \psi$, $\psi \to \eta(\eta') \ell^+ \ell^-$ is also of interest but the total $M_X$ from $X_1$, $2\pi$ and $\eta(\eta')$ is largish. Finally, we observe that the branching ratio of $\psi \to \pi \ell^+ \ell^-$ is about two orders of magnitude smaller compared to $\psi \to \eta(\eta') \ell^+ \ell^-$ (see table~\ref{tab:cascade-table}). Hence the conclusion is that the direct decay $\psi \to \eta \ell^+ \ell^-$ and $\psi \to \eta' \ell^+ \ell^-$ dominate the background to $\bar B \to X_s \ell^+ \ell^-$ from all charmonium radiative decays in the presence of a cut $M_X < 2 \text{ GeV}$.

\begin{figure}[t!]
\centering
\includegraphics[width=\linewidth]{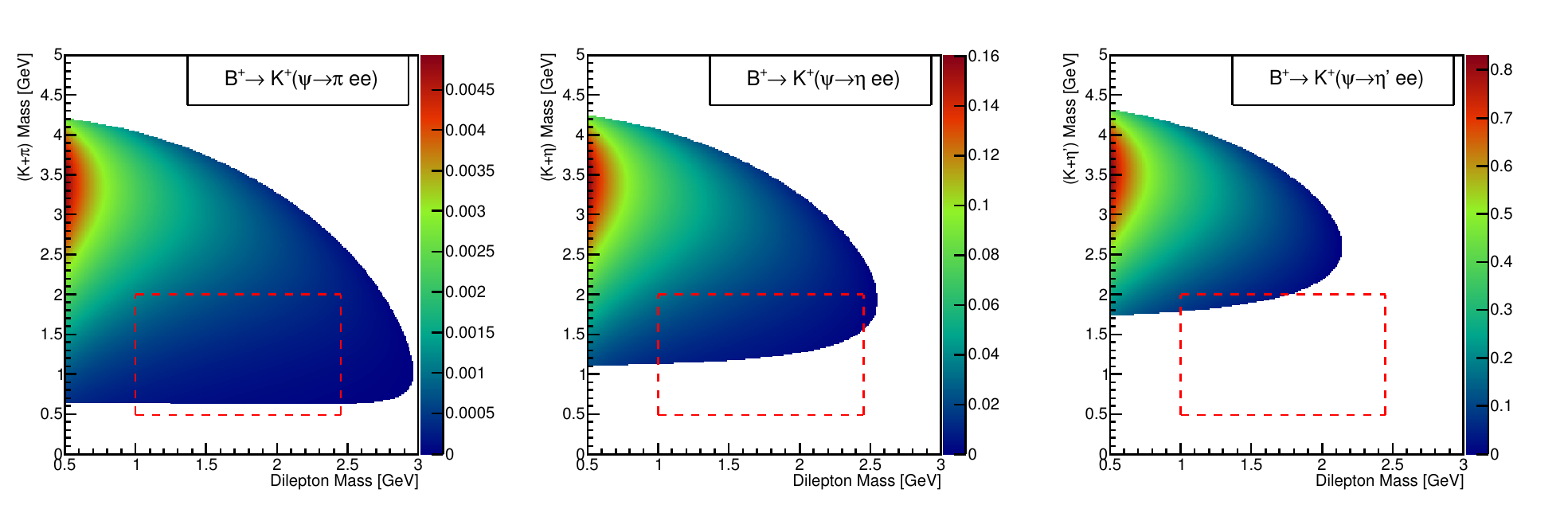}
\caption{The background from $B^+\to K^+(\psi \to h \ell^+ \ell^-)$ with $h = \pi, \eta, \eta'$ on the ${\bar B \to X_s \ell^+ \ell^-}$ phase space. The color bar corresponds to the branching ratio differential in dilepton mass ($\sqrt{s}$) and hadronic mass ($M_X$), in units of $10^{-8}~\text{GeV}^{-2}$. The outlined box indicates the $1~\text{GeV}^2<q^2<6~\text{GeV}^2$ and $M_K < M_X< 2~\text{GeV}$ cuts. A single-pole parameterization of the transition form factor~\cite{Landsberg:1986fd} with the same $\Lambda = 3.686 \text{ GeV}$ for $\pi, \eta, \eta'$ was used to generate the plots. The fact that the cuts are less efficient for $\psi \to \pi$ is not problematic due to the small rate of this channel.}
\label{fig:cascade-dalitz}
\end{figure}

The helicity-projected rates for $\psi \to \eta(\eta') \ell^+ \ell^-$ normalized to the rate $\psi \to \eta(\eta') \gamma$ can be calculated from first principles in terms of a single $q^2$-dependent form factor~\cite{Landsberg:1986fd}. The angular distribution is simply given in terms of the polarization $\alpha$ as
\begin{align}
  \frac{d\Gamma(\psi\to\eta\ell^+\ell^-)}{d\cos\theta_X} &\propto 1 + \alpha \cos^2\theta_X  \; .
  \label{eq:psi-pol}
\end{align}
Due to the V$-$A coupling of the underlying transition $b \to s c\bar{c}$, $\bar B \to \psi X_1$ prefers the longitudinal polarization~\cite{Anderson:2002md, Aubert:2002hc} with corrections quantified in NRQCD \cite{Fleming:1996pt}; this fact reduces the background from the cascades because $\eta(\eta')$ and $X_1$ cannot be collinear through the dominant longitudinal polarization. The resulting distribution in the $[q^2,M_X]$ plane for the decays into $\eta$ and $\eta'$ are presented in figure~\ref{fig:cascade-dalitz}, where we take for simplicity a constant value $\alpha=-0.59$ corresponding to the low $M_{X_1}$ bin of the BaBar measurement: this causes the $M_X$ cut to be more efficient by about 20\%.

The contributions of the cascade into $\eta$ and $\eta'$ to the total $\bar B \to X_s \ell^+\ell^-$ branching ratio before any $M_X$ cut are $1\%$ and $5\%$, respectively. After imposing $M_X < 2 \; {\rm GeV}$ these effects are diluted to $0.05\%$ and $0.0005\%$. Keeping in mind that the impact of the $M_X$ cut on the non-resonant decay is about $60\%$, we conclude that the experimental cut on $M_X$ completely removes any pollution from cascade charmonium decays. This conclusion persists as long as the $M_X$ cut is at most $3 \; {\rm GeV}$.


\section{Inputs}
\label{sec:inputs}
The numerical inputs used in the phenomenological analysis are presented in table~\ref{tab:inputs}. Most of the quantities listed in the table have been determined with great accuracy and will not be discussed further. The required HQET matrix elements, on the other hand,  necessitate a more in depth discussion.

\begin{table}[t]
	\begin{center}
		\begin{displaymath}
		\begin{tabular}{|l|l|}
		\hline\spp
		$\alpha_s (M_z) = 0.1181 (11)$ &
		$m_e = 0.51099895 \;\mev $ \\ \spp
		$\alpha_e (M_z) =  1/127.955 $ &
		$m_\mu = 105.65837 \;\mev$ \\ \spp
		$s_W^2 \equiv \sin^2\theta_W = 0.2312$ &
		$m_\tau = 1.77686 \;\gev$ \\ \spp
		$|V_{ts}^* V_{tb}/V_{cb}|^2 = 0.96403 (87)$~\cite{Charles:2004jd} &
		$\overline{m}_c(\overline{m}_c) = 1.275 (25) \;\gev$ \\\spp
		$|V_{ts}^* V_{tb}/V_{ub}|^2 = 123.5 (5.3) $~\cite{Charles:2004jd} &
		$m_b^{1S} = 4.691(37) \;\gev$~\cite{Amhis:2012bh,Schwanda:2013bg} \\ \spp
		$|V_{td}^* V_{tb}/V_{cb}|^2 = 0.04195 (78) $~\cite{Charles:2004jd} &
		$|V_{us}^* V_{ub}/(V_{ts}^*V_{tb})| = 0.02022(44)$~\cite{Charles:2004jd} \\ \spp
		$|V_{td}^* V_{tb}/V_{ub}|^2 = 5.38(26)$~\cite{Charles:2004jd} &
		$\arg \left[V_{us}^* V_{ub}/(V_{ts}^*V_{tb})\right] = 115.3(1.3)^{\circ} $~\cite{Charles:2004jd}\\ \spp
		$\mathcal{B}(B\to X_c e \bar\nu)_{\rm exp}=0.1065 (16) $~\cite{Amhis:2016xyh} &
		$|V_{ud}^* V_{ub}/(V_{td}^*V_{tb})| = 0.420(10)$ \\ \spp
	   $m_B = 5.2794\;\gev$ &
      $\arg \left[V_{ud}^* V_{ub}/(V_{td}^*V_{tb})\right] = -88.3(1.4)^{\circ}$ \\ \spp
      $M_Z = 91.1876\;\gev$ & $m_{t,{\rm pole}}= 173.1 (0.9) \;\gev$\\ \spp
		$M_W = 80.379\;\gev$ &
		$C = 0.568 (7)(10)$~\cite{Alberti:2014yda} \\ \spp
		$\mu_b = 5^{+5}_{-2.5}\;\gev$ & $\mu_0 = 120^{+120}_{-60}\;\gev$ \\ \spp
		$f_\text{NV} = (0.02 \pm 0.16)\;\gev^3$ &
      $\lambda_2^{\rm eff} = 0.130 (21)\;\gev^2$ \cite{Gambino:2016jkc} \\ \spp
		$f_\text{V}-f_\text{NV} = (0.041 \pm 0.052)\;\gev^3$ &
      $\lambda_1 = - 0.267 (90)\;\gev^2$~\cite{Gambino:2016jkc} \\ \spp
		$[\delta f]_{SU(3)} = (0 \pm 0.04)\;\gev^3$ &
      $\rho_1 = 0.038 (70)\;\gev^3$~\cite{Gambino:2016jkc} \\ \spp
      $[\delta f]_{SU(2)} = (0 \pm 0.004)\;\gev^3$ &
       \\ \hline
		\end{tabular}
		\end{displaymath}
		\caption{Numerical inputs used in the phenomenological analysis. Unless specified otherwise, they are taken from PDG~\cite{Tanabashi:2018oca}. In order to avoid somewhat uncontrolled non-perturbative effects, we use the pole mass of the top quark obtained from cross section measurements. The CKM matrix elements have been obtained by propagating the uncertainties on the four CKM Wolfenstein parameters ($\lambda$, $A$, $\bar\rho$ and $\bar \eta$) taken from the global fit as of Summer 2018 presented by the CKMfitter Group~\cite{Charles:2004jd}. All HQET matrix elements are calculated between physical $B$ mesons. Only the combination $\lambda_2^{\rm eff} \equiv \lambda_2 -\rho_2/m_b$ enters in $\bar B\to X_{d,s} \ell^+\ell^-$. The annihilation matrix elements required for $\bar B\to X_d \ell^+\ell^-$ and $\bar B\to X_s \ell^+\ell^-$ are $(f_u, f_d)$ and $(f_u^0, f_u^\pm, f_s)$, respectively, where we use the notation $f_q\equiv (f_q^0+f_q^\pm)/2$. As explained in the text we express them in terms of the valence and non-valence matrix elements, $f_\text{V}$ and $f_\text{NV}$, and the flavour $SU(3)$ and the isospin breaking differences $[\delta f]_{SU(3)}$ and $[\delta f]_{SU(2)}$. }
		\label{tab:inputs}
	\end{center}
\end{table}

For our phenomenological study, we need the matrix elements of the following dimension five and six operators:
\begin{align}
\lambda_1 &\equiv \frac{1}{2m_B} \langle B | \bar h_v (i  D)^2 h_v | B \rangle \; , \\
\lambda_2 &\equiv \frac{1}{12m_B} \langle B | \bar h_v  (-i\sigma_{\mu\nu}) G^{\mu\nu} h_v | B \rangle  \; , \\
 \rho_1 &\equiv \frac{1}{2m_B}
\langle B | \bar h_v i D_\mu (i v\cdot D) i D^\mu h_v | B \rangle\; , \\
 \rho_2  &\equiv \frac{1}{6m_B}
\langle B | \bar h_v i D^\mu (i v\cdot D) i D^\nu h_v  (-i\sigma_{\mu\nu})| B \rangle\; , \\
f^a_q &\equiv \frac{1}{2 m_B} \langle B^a | Q_1^q - Q_2^q | B^a \rangle \; ,
\end{align}
where $a=0,\pm$ denotes the charge of the meson, $q=u,d,s$ is the flavour of the spectator quark and \cite{Voloshin:2001xi}
\begin{align}
Q_1^q &= \bar h_v \gamma_\mu (1-\gamma_5) q \; \bar q \gamma^\mu (1-\gamma_5) h_v \; , \\
Q_2^q &= \bar h_v  (1-\gamma_5) q \; \bar q (1+\gamma_5) h_v \; .
\end{align}
The leading matrix elements $\lambda_{1,2}$ and $(\rho_{1,2}, f_q^a)$ scale as $m_b^{-2}$ and $m_b^{-3}$ respectively. The matrix elements $\lambda_2$ and $\rho_2$ appear only in the combination
\begin{align}
\lambda_2^{\rm eff} &\equiv \lambda_2 - \frac{\rho_2}{m_b} \; .
\end{align}

Note that we consider exclusively HQET matrix elements between physical $B$ mesons. Matrix elements in the infinite mass limit are independent of the heavy quark mass and are sometimes used when combining fits involving both $b$- and $c$-hadrons. For the two leading dimension five operators the relation between these matrix elements is:
\begin{alignat}{3}
\lambda_1 &\equiv \frac{1}{2m_B} \langle B | \bar h_v (i  D)^2 h_v | B \rangle
  = \frac{1}{2m_B} \langle H_\infty | \bar h_v (i D)^2 h_v |  H_\infty \rangle + \frac{\tau_1 + 3 \tau_2}{m_b} \; , \\
\lambda_2 &\equiv\frac{1}{12m_B}  \langle B | \bar h_v (-i\sigma_{\mu\nu}) G^{\mu\nu} h_v | B \rangle
 = \frac{1}{12m_B}  \langle H_\infty | \bar h_v (-i\sigma_{\mu\nu}) G^{\mu\nu} h_v |  H_\infty \rangle + \frac{\tau_3 +3 \tau_4}{m_b} \; ,
\end{alignat}
where the non-local matrix elements $\tau_i$ can be found, for instance, in ref.~\cite{Gremm:1996df}.

The $\lambda_i$ and $\rho_i$ matrix elements can be extracted from measurements of several leptonic and hadronic moments of the inclusive $\bar B\to X_c \ell\nu$ spectrum, under the assumption that its shape is unaffected by new physics effects. The most recent analysis has been presented in ref.~\cite{Gambino:2016jkc}, where the results are expressed in the kinetic scheme~\cite{Bigi:1994ga, Bigi:1997fj, Bigi:1996si, Gambino:2004qm}. In this scheme the renormalized matrix elements at a scale $\mu=1$ GeV are connected to the usual pole-scheme ones by calculating several leptonic $\bar B\to X_c \ell\nu$ moments in the small velocity (SV) limit (see ref.~\cite{Bigi:1994ga} for a pedagogical review) and using $\mu$ as a Wilsonian cut-off; this implies that the difference between pole and kinetic scheme matrix elements is proportional to powers of $\mu$ and not just logarithms (as in the $\overline{\text{MS}}$ scheme). When using these matrix elements in the calculation of any other observable (e.g. in $\bar B\to X_q \ell^+\ell^-$) one has to modify the perturbative part of the calculation accordingly by introducing the same Wilsonian cut-off in both virtual and real corrections. The alternative, which we adopt, is to convert the matrix elements to the pole scheme (which corresponds to setting $\mu=0$) and keep the rest of the calculation unchanged.

The explicit expressions that we use are (see eqs.~(9) of ref.~\cite{Gambino:2004qm} and eqs.~(11)~-~(13) of ref.\cite{Gambino:2007rp}):
\begin{alignat}{3}
-\lambda_1 &\equiv \mu_\pi^2(0) && = \mu_\pi^2(\mu) - [ \mu_\pi^2(\mu)]_{\rm pert} \, \label{kin1}\\
3 \lambda_2 &\equiv \mu_G^2(0) && = \mu_G^2(\mu) - [ \mu_G^2(\mu)]_{\rm pert} \, \label{kin2}\\
\rho_1 &\equiv \rho_D^3 (0) && = \rho_D^3(\mu) - [ \rho_D^3(\mu)]_{\rm pert} \, \label{kin3}\\
3 \rho_2 &\equiv \rho_{LS}^3 (0) && = \rho_{LS}^3(\mu) - [ \rho_{LS}^3(\mu)]_{\rm pert} \, \label{kin4}
\end{alignat}
where
\begin{align}
[ \mu_\pi^2(\mu)]_{\rm pert} =&\ C_F \frac{\alpha_s(m_b)}{\pi} \mu^2 \left[ 1 + \frac{\alpha_s(m_b)\beta_0}{2\pi} \left(\log \frac{m_b}{2\mu} + \frac{13}{6}\right) - \frac{\alpha_s(m_b)}{\pi} C_A \left(\frac{\pi^2}{6} -\frac{13}{12} \right) \right] \nnb \\
&+ \mathcal{O}\left(\frac{\mu^3}{m_b}\right) \;, \label{kin11}\\
[ \mu_G^2(\mu)]_{\rm pert} =&\  \mathcal{O}\left(\frac{\mu^3}{m_b}\right)\;, \label{kin12}\\
[ \rho_D^3(\mu)]_{\rm pert} =&\ \frac{2}{3} C_F \frac{\alpha_s(m_b)}{\pi} \mu^3 \left[ 1 + \frac{\alpha_s(m_b)\beta_0}{2\pi} \left(\log \frac{m_b}{2\mu} + 2\right) - \frac{\alpha_s(m_b)}{\pi} C_A \left(\frac{\pi^2}{6} -\frac{13}{12} \right) \right] \nnb \\
&+\mathcal{O}\left(\frac{\mu^4}{m_b}\right)\; ,\label{kin13} \\
[ \rho_{LS}^3(\mu)]_{\rm pert} =&\ \mathcal{O}\left(\frac{\mu^4}{m_b}\right)\; ,\label{kin14}
\end{align}
with $\beta_0=9$ (three active flavours), $C_F=4/3$ and $C_A=3$.
Note that following ref.~\cite{Gambino:2007rp} (see footnote above eq.~(13)) we omit terms of order $\mu^3$ in eqs.~(\ref{kin11}) and (\ref{kin12}); this is necessary in order to convert the matrix elements extracted from the fit presented in ref.~\cite{Gambino:2016jkc} (which differ from the kinetic scheme matrix elements by terms of order $\mu^3$) into the pole scheme. A discussion of the absence of $\mu^2$ terms in $[\mu_G^2(\mu)]_{\rm pert}$ can be found in ref.~\cite{Uraltsev:2001ih}. The inputs summarized in table~\ref{tab:inputs} are obtained from the results presented in table~II of ref.~\cite{Gambino:2016jkc} with the help of eqs.~(\ref{kin1})~-~(\ref{kin14}); we estimate the uncertainty due to missing ${\cal O}(\alpha_s^3)$ corrections in eqs.~(\ref{kin11}) and (\ref{kin13}) by assuming that the relative magnitudes of NNLO and NNNLO terms are identical.

The discussion of the weak annihilation matrix elements $f_q^a$ is greatly simplified by isospin and flavour $SU(3)$ considerations:
\begin{align}
f_{\rm V}  &\equiv f_u^\pm \stackrel{SU(2)}{=} f_d^0 \\
f_{\rm NV} &\equiv f_u^0 \stackrel{SU(2)}{=} f_d^\pm \stackrel{SU(3)}{=} f_s^0 \stackrel{SU(2)}{=} f_s^\pm
\end{align}
where we indicate the valence and non-valence terms with respect to external $B^{0,\pm}$ states. Therefore, up to isospin and flavour $SU(3)$ breaking effects, the six matrix elements needed for $\bar B\to X_{s,d}\ell^+\ell^-$ reduce to two. In the vacuum saturation approximation these matrix elements vanish: they can be written as $f_\text{A} = 2\pi^2 f_B^2 m_B( B_1^\text{A} - B_2^\text{A})$ with $B_1^\text{V} = B_2^\text{V} = 1$ and  $B_1^\text{NV} = B_2^\text{NV} = 0$~\cite{Voloshin:2001xi, Ligeti:2007sn}. Assuming violations of this approximation at the $\delta B^\text{A} \sim O(0.1)$ we find $f_\text{V} \sim f_\text{NV} \lesssim 0.4$. As numerical inputs we adopt upper limits extracted from branching ratios and the first two moments of semileptonic $D^{0,\pm}$ and $D_s$ decays rescaled by a factor $m_B f_B^2/(m_D f_D^2)$. The result of a  re-analysis of semileptonic $D$ decay data from the CLEO-c Collaboration \cite{Asner:2009pu} following closely ref.~\cite{Gambino:2010jz}\footnote{The only difference in the analysis is that we included correlations induced by the charm quark mass, and added explicit nuisance parameters to describe $SU(3)$ breaking and end-point smearing of the annihilation contributions to the first two moments.} are summarized in table~\ref{tab:inputs}, where we present the two largely uncorrelated quantities $f_{\rm NV}$ and $f_{\rm V} - f_{\rm NV}$. Additionally, following ref.~\cite{Ligeti:2007sn}, we assume $SU(3)$ and $SU(2)$ breaking effects at the level of $[\delta f]_{SU(3)} = 0.04$ and $[\delta f]_{SU(2)} = 0.004$, respectively.

Note that we calculate $\Gamma(\bar B\to X_{s(d)} \ell^+\ell^-)/\Gamma(\bar B\to X_u \ell\nu)$; therefore, for the $X_s$ case we need $f_s^{0,\pm}$ and $f_u^{0,\pm}$ and for the $X_d$ one we need $f_d^{0,\pm}$ and $f_u^{0,\pm}$. The required inputs for the various observables are ($f_q = (f_q^0+f_q^\pm)/2$):
\begin{align}
{\mathcal B}(B\to X_s\ell^+\ell^-) &\Longrightarrow
\begin{cases}
f_s = f_\text{NV}\cr
f_u = (f_\text{V} + f_\text{NV})/2 \cr
\end{cases} ,\\
{\cal R}(s_0,B\to X_s\ell^+\ell^-) &\Longrightarrow
\begin{cases}
(f_s + f_u^0)/2 = f_\text{NV} \cr
f_s - f_u^0  = [\delta f]_{SU(3)} \cr
\end{cases} ,\\
{\mathcal B}(B\to X_d\ell^+\ell^-) \text{ and } {\cal R}(s_0,B\to X_d\ell^+\ell^-)
&\Longrightarrow
\begin{cases}
(f_d+ f_u)/2 = (f_\text{V} + f_\text{NV})/2\cr
f_d - f_u = [\delta f]_{SU(2)} \cr
\end{cases} .
\end{align}
In conclusions, we need the four quantities $f_\text{NV}$, $(f_\text{V}+ f_\text{NV})/2$, $[\delta f]_{SU(3)}$ and $[\delta f]_{SU(2)}$.


\section{Phenomenological results}
\label{sec:results}

In this section, we present the final numerical results, for which we use the inputs defined in table~\ref{tab:inputs}.
We give the results for the branching ratios integrated over the low-$q^2$ region $(1~\text{GeV}^2 < q^2 < 6~\text{GeV}^2$)
and over the high-$q^2$ region $q^2 < 14.4~\text{GeV}^2$. The corresponding CP asymmetries are given as well.
In order to reduce large uncertainties from power corrections in the high-$q^2$ region we compute the ratios
$\mathcal{R}(s_0)$. The forward-backward asymmetries $A_\text{FB}$ and the related angular observable $H_A$ are computed
for the low-$q^2$ region. Due to a zero-crossing in the differential $A_\text{FB}$ and $H_A$, we subdivide the low-$q^2$ region into two bins,
bin 1 $(1~\text{GeV}^2 < q^2 < 3.5~\text{GeV}^2$) and bin 2 $(3.5~\text{GeV}^2 < q^2 < 6~\text{GeV}^2$) when presenting
the results for these two observables. In addition, we give the position of the zero crossing.
As is customary, we present our results for both electron and muon final states separately.

To obtain our phenomenological results, we expand our observables up to $\as^3$ and $\kappa^3$,
and neglect all higher terms. In addition, we expand up to linear terms in the power-correction parameters
$\lambda_{1,2}, \rho_1, f_u^{0,\pm}, f_s$ and drop all higher powers and product of these parameters. For the low-$q^2$ region, we neglect $1/m_b^3$ corrections.

Below, we give the central values of all the observables with uncertainties from different sources. These uncertainties are
obtained by varying the inputs within their ranges indicated in table~\ref{tab:inputs}, where we assume that $m_c$ and $C$ are
fully anti-correlated. The total uncertainties are obtained by adding the individual ones in quadrature. The uncertainties from the Kr\"uger-Sehgal functions are always below the
percent level of the central values and are therefore not included.
We present our results up to 2 decimal digits, however in some cases where this would lead to 0.00,
we give the first significant number. We emphasize that the contribution of $\lambda_1$ to the error budget is tiny in both low and high-$q^2$ and therefore it is not displayed.


\subsection{Branching ratio, low-$q^2$ region}\label{sec:brlowq2}
The branching ratios for the low-$q^2$ region are found to be nearly $10^{-7}$, smaller than the
$\bar{B}\to X_s\ell^+\ell^-$ number by about 2 orders of magnitude mainly due to the CKM suppression.
The total uncertainties are about 8\%.

\begin{align}\label{eq:BRee}
\dps {\cal B}[1,6]_{ee} =&\  (7.81\pm 0.37_{\text{scale}} \pm 0.08_{m_t}\pm 0.17_{C,m_c}\pm 0.08_{m_b} \pm0.04_{\alpha _s} \pm 0.15_{\text{CKM}}\nonumber \\
& \hspace*{40pt} \pm 0.12_{\text{BR}_{\text{sl}}}\pm 0.05_{\lambda _2} \pm 0.39_{\text{resolved}} )\cdot 10^{-8} = (7.81\pm 0.61) \cdot 10^{-8} \; .
\end{align}
\begin{align}\label{eq:BRmumu}
\dps {\cal B}[1,6]_{\mu\mu} =&\ (7.59 \pm 0.35_{\text{scale}}\pm 0.08_{m_t}\pm 0.17_{C,m_c}\pm 0.09_{m_b} \pm 0.04_{\alpha _s} \pm 0.14_{\text{CKM}} \nonumber \\
& \hspace*{40pt} \pm 0.11_{\text{BR}_{\text{sl}}}\pm 0.05_{\lambda _2} \pm 0.38_{\text{resolved}} )\cdot 10^{-8} = (7.59\pm 0.59) \cdot 10^{-8}  \; .
\end{align}

These results include  Kr\"uger-Sehgal corrections as described in section~\ref{sec:KS}. Comparing with the pure perturbative results,
the central value being $\mathcal{B}[1,6]=7.46 (7.23)\cdot 10^{-8}$ for electrons (muons), we find that the inclusion of the KS functions
shifts the branching ratio by about $+5\%$. The other sizable corrections include the log-enhanced electromagnetic corrections (about
4\% for electrons and 2\% for muons) and the five-particle contributions (about 1\%).
In comparison, the $1/m_b^2$ and bremsstrahlung corrections are only of $\mathcal{O}(0.5\%)$.
We note that dominant uncertainties arise from the scale variation.
As discussed in section~\ref{section:resolved}, we add an additional $5\%$ uncertainty due to the resolved-photon contributions.


\subsection{Branching ratio, high-$q^2$ region}\label{sec:brhighq2}
In the high-$q^2$ region, the $1/m_b^{2,3}$ power-corrections are very pronounced. The large uncertainties on their hadronic input parameters, as discussed in section~\ref{sec:inputs}, dominate the uncertainty on the branching ratio which is $\mathcal{O}(40\%)$.

\begin{align}
\dps {\cal B}[>14.4]_{ee} =&\ ( 0.86\pm 0.12_{\text{scale}}\pm 0.01_{m_t}\pm 0.01_{C,m_c}\pm 0.08_{m_b}\pm 0.02_{\text{CKM}} \pm 0.02_{\text{BR}_{\text{sl}}}\nonumber \\[0.5em]
   & \hspace*{25pt} \pm 0.06_{\lambda _2}\pm 0.25_{\rho _1}\pm 0.25_{f_{u,d}}) \cdot 10^{-8} = ( 0.86 \pm 0.39 ) \cdot 10^{-8} \; , \\
\dps {\cal B}[>14.4]_{\mu\mu} =&\ ( 1.00 \pm 0.12_{\text{scale}}\pm 0.01_{m_t}\pm 0.02_{C,m_c}\pm 0.09_{m_b}\pm 0.02_{\text{CKM}} \pm 0.02_{\text{BR}_{\text{sl}}}\nonumber \\[0.5em]
   & \hspace{25pt} \pm 0.05_{\lambda _2}\pm 0.25_{\rho _1}\pm 0.25_{f_{u,d}}) \cdot 10^{-8}
   = (1.00\pm 0.39) \cdot 10^{-8} \; .
\end{align}

Here we do not quote the uncertainty coming from the variation of $\alpha_s$ as this is negligible.
We quote the uncertainty coming from $f_u$ and $f_d$ together, by summing quadratically individual uncertainties from
variation of $f_\text{NV}$, $f_\text{V}-f_\text{NV}$ and $[\delta f]_{SU(2)}$, where $f_{\text{NV}}$ gives the dominant uncertainty.


\subsection{The ratio ${\cal R}(s_0)$}\label{sec:ratioRhighq2}
Comparing the ratio ${\cal R}(s_0)$ with the branching ratio in the high-$q^2$ region, we see a large reduction
of the total uncertainties from ${\cal O}(40\%)$ to $9\%$ and $6\%$ in the electron and muon channel, respectively:

\begin{align}
  {\cal R}(14.4)_{ee} =&\ ( 0.93 \pm 0.02_{\text{scale}} \pm 0.01_{m_t} \pm 0.01_{C,m_c} \pm 0.002_{m_b} \pm 0.01_{\alpha
   _s} \pm 0.05_{\text{CKM}} \nonumber \\[0.5em]
   & \hspace*{25pt} \pm 0.004_{\lambda _2} \pm 0.06_{\rho_1} \pm 0.05_{f_{u,d}} )  \times 10^{-4}
                       =( 0.93 \pm 0.09 )  \times 10^{-4} \; ,\\[1.0em]
  {\cal R}(14.4)_{\mu\mu} =&\ ( 1.10  \pm 0.01_{\text{scale}} \pm 0.01_{m_t} \pm 0.01_{C,m_c} \pm 0.002_{m_b} \pm 0.01_{\alpha
   _s} \pm 0.05_{\text{CKM}} \nonumber \\[0.5em]
   & \hspace*{25pt} \pm 0.002_{\lambda _2} \pm 0.04_{\rho_1} \pm 0.02_{f_{u,d}} )  \times 10^{-4}
                       = ( 1.10 \pm 0.07 )  \times 10^{-4} \; .
\end{align}
Besides the uncertainties arising from power corrections, also the scale uncertainty and the one due to $m_b$
get significantly reduced. The largest source of uncertainty arises from the CKM elements, especially from $V_{ub}$.


\subsection{Forward-backward asymmetry, low-$q^2$ region}\label{sec:fbalowq2}
The integrated $\mathcal{H}_{\text{A}}$ rate and the forward-backward asymmetry are tiny when integrated over the full low-$q^2$ bin, as was already observed in the  $\bar B\to X_s \ell^+\ell^-$ case~\cite{Huber:2015sra}. This is because of a zero-crossing in $\mathcal{H}_{\text{A}}$ which occurs close to the middle of the low-$q^2$ region. Therefore, we separate our results in two additional bins:
\begin{align}
\dps {\cal H}_\text{A}[1,3.5]_{ee} =& \ (-0.41\pm 0.02_{\text{scale}}\pm 0.004_{m_t}\pm 0.002_{C,m_c}\pm 0.01_{m_b} \pm 0.01_{\alpha _s} \pm 0.01_{\text{CKM}}
\nonumber \\
& \hspace*{14pt}\pm 0.01_{\text{BR}_{\text{sl}}}\pm 0.0001_{\lambda _2}\pm 0.02_{\text{resolved}})\cdot 10^{-8}=( -0.41\pm 0.04)\cdot 10^{-8}  \; , \nonumber\\[0.5em]
\dps {\cal H}_\text{A}[3.5,6]_{ee} =& \ (0.40\pm 0.06_{\text{scale}}\pm 0.004_{m_t}\pm 0.02_{C,m_c}\pm 0.02_{m_b}  \pm 0.01_{\alpha _s} \pm 0.01_{\text{CKM}}
\nonumber \\
& \hspace*{14pt}\pm 0.01_{\text{BR}_{\text{sl}}}\pm 0.002_{\lambda _2}\pm 0.02_{\text{resolved}})\cdot 10^{-8}=( 0.40\pm 0.07)\cdot 10^{-8}  \; , \nonumber\\[0.5em]
\dps {\cal H}_\text{A}[1,6]_{ee} =& \ (-0.01\pm 0.08_{\text{scale}}\pm 0.0002_{m_t}\pm 0.01_{C,m_c}\pm 0.03_{m_b} \pm 0.02_{\alpha _s} \pm 0.002_{\text{CKM}}
\nonumber \\
& \hspace*{14pt}\pm 0.0002_{\text{BR}_{\text{sl}}}\pm 0.002_{\lambda _2}\pm 0.001_{\text{resolved}})\cdot 10^{-8}=( -0.01\pm 0.09)\cdot 10^{-8}  \; .
\end{align}
\begin{align}
\dps {\cal H}_\text{A}[1,3.5]_{\mu\mu} =&  (-0.44\pm 0.02_{\text{scale}}\pm 0.004_{m_t}\pm 0.003_{C,m_c}\pm 0.01_{m_b}  \pm 0.01_{\alpha _s}
\pm 0.01_{\text{CKM}}\nonumber \\
& \hspace*{14pt}\pm 0.01_{\text{BR}_{\text{sl}}}\pm 0.0002_{\lambda _2}\pm 0.02_{\text{resolved}})\cdot 10^{-8}=( -0.44\pm 0.04)\cdot 10^{-8}  \; , \nonumber\\[0.5em]
\dps {\cal H}_\text{A}[3.5,6]_{\mu\mu} =&  (0.37\pm 0.06_{\text{scale}}\pm 0.004_{m_t}\pm 0.02_{C,m_c}\pm 0.02_{m_b} \pm 0.01_{\alpha _s}\pm 0.01_{\text{CKM}}\nonumber \\
& \hspace*{14pt}\pm 0.01_{\text{BR}_{\text{sl}}}\pm 0.002_{\lambda _2}\pm 0.02_{\text{resolved}})\cdot 10^{-8}=( 0.37\pm 0.07)\cdot 10^{-8}  \; , \nonumber\\[0.5em]
\dps {\cal H}_\text{A}[1,6]_{\mu\mu} =&  (-0.07\pm 0.08_{\text{scale}}\pm 0.0003_{m_t}\pm 0.01_{C,m_c}\pm 0.03_{m_b}  \pm 0.02_{\alpha _s}\pm 0.003_{\text{CKM}}\nonumber \\
& \hspace*{14pt}\pm 0.001_{\text{BR}_{\text{sl}}}\pm 0.003_{\lambda _2}\pm 0.004_{\text{resolved}})\cdot 10^{-8}=( -0.07\pm 0.09)\cdot 10^{-8}  \; .
\end{align}
As discussed in sec.~\ref{section:resolved}, at order $1/m_b^2$, ${\cal H}_\text{A}$ receives
resolved-photon contributions from the interference between $P^c_{1,2} - P_{10}$. Since an explicit estimate
of such contributions is not yet available~\cite{Benzkeworkinprogress}, we added an additional uncertainty of 5\% to our results.

For completeness, we also quote the value of the normalized forward-backward asymmetry,
which can be obtained from $\mathcal{H}_{\text{A}}$ by using eqs.~\eqref{eq:afb} and~\eqref{eq:Afw},
\begin{align}
\dps \overline{A}_{\rm FB}[1,3.5]_{ee} =& \ (-7.10\pm 0.67_{\text{scale}}\pm 0.01_{m_t}\pm 0.11_{C,m_c}\pm 0.22_{m_b} \pm 0.19_{\alpha _s}
\nonumber \\
& \hspace*{5pt}\pm 0.02_{\text{CKM}}\pm  0.04_{\lambda _2} \pm 0.50_\text{resolved})\%=(-7.11\pm 0.89)\% \; , \nonumber\\[0.5em]
\dps \overline{A}_{\rm FB}[3.5,6]_{ee} =& \ (8.60\pm 0.74_{\text{scale}}\pm 0.01_{m_t}\pm 0.13_{C,m_c}\pm 0.37_{m_b}  \pm 0.18_{\alpha _s}
\nonumber \\
& \hspace*{5pt}\pm 0.02_{\text{CKM}}\pm 0.11_{\lambda _2} \pm 0.61_\text{resolved})\%= (8.60\pm 1.06)\% \; , \nonumber\\[0.5em]
\dps \overline{A}_{\rm FB}[1,6]_{ee} =&\  (-0.12\pm 0.77_{\text{scale}}\pm 0.004_{m_t}\pm 0.13_{C,m_c}\pm 0.29_{m_b} \pm 0.20_{\alpha _s}
\nonumber \\
& \hspace*{5pt}\pm 0.02_{\text{CKM}}\pm 0.02_{\lambda _2} \pm 0.01_\text{resolved})\%= (-0.12\pm 0.86)\% \; ,
\end{align}
\begin{align}
\dps \overline{A}_{\rm FB}[1,3.5]_{\mu\mu} =&\  (-7.97\pm 0.69_{\text{scale}}\pm 0.01_{m_t}\pm 0.11_{C,m_c}\pm 0.22_{m_b} \pm 0.20_{\alpha _s}
\nonumber \\
& \hspace*{2pt}\pm 0.02_{\text{CKM}}\pm 0.05_{\lambda _2} \pm 0.56_\text{resolved})\%= (-7.97\pm 0.95)\% \; , \nonumber\\[0.5em]
\dps \overline{A}_{\rm FB}[3.5,6]_{\mu\mu} =& \ (8.16\pm 0.82_{\text{scale}}\pm 0.01_{m_t}\pm 0.13_{C,m_c}\pm 0.39_{m_b}\pm 0.19_{\alpha _s} \nonumber \\
& \hspace*{2pt}\pm 0.02_{\text{CKM}}\pm 0.11_{\lambda _2} \pm 0.58_\text{resolved}) \%= (8.16\pm 1.10)\%  \; , \nonumber\\[0.5em]
\dps \overline{A}_{\rm FB}[1,6]_{\mu\mu} =& \ (-0.70\pm 0.82_{\text{scale}}\pm 0.004_{m_t}\pm 0.13_{C,m_c}\pm 0.30_{m_b}\pm 0.21_{\alpha _s} \nonumber \\
& \hspace*{2pt}\pm 0.02_{\text{CKM}} \pm 0.02_{\lambda _2} \pm 0.05_\text{resolved})\%= (-0.70\pm 0.91)\%  \; .
\end{align}
The forward-backward asymmetries are obtained by taking ${\cal H}_\text{A}$ normalized by the
corresponding branching ratios, both of which receive resolved-photon contributions. Since the
the contributions to ${\cal H}_\text{A}$ and the corresponding branching ratios are induced by different
operators, {\it i.e.} $P^c_{1,2} - P_{10}$ and $P^c_{1,2} - P_{7,9}$, respectively,
we have assumed that the resolved-photon uncertainties of the branching ratios and ${\cal H}_\text{A}$
are independent. We emphasize that the uncertainties stemming from the scale and $\alpha_s$ are very pronounced. This is caused by the opposite effect of the scale and $\alpha_s$ variation in $\mathcal{H}_{\text{A}}$ and the branching ratios.

Finally, we also give the zero-crossing $q_0^2$ (in units of GeV$^2$) for the forward-backward asymmetry $\mathcal{A}_\text{FB}$ and
equivalently $\mathcal{H}_{\text{A}}$:
\begin{align}\label{eq:zero}
\dps (q_0^2)_{ee} =& \ 3.28\pm 0.11_{\text{scale}}\pm 0.001_{m_t}\pm 0.02_{C,m_c}\pm 0.05_{m_b} \nonumber \\
& \hspace*{25pt} \pm 0.03_{\alpha _s}\pm 0.004_{\text{CKM}}\pm 0.001_{\lambda _2}\pm 0.06_\text{resolved} =3.28\pm 0.14  \; , \\[0.5em]
\dps (q_0^2)_{\mu\mu}=& \ 3.39\pm 0.12_{\text{scale}}\pm 0.001_{m_t}\pm 0.02_{C,m_c}\pm 0.05_{m_b} \nonumber \\
& \hspace*{25pt} \pm 0.03_{\alpha _s}\pm 0.004_{\text{CKM}}\pm 0.002_{\lambda _2}\pm 0.06_\text{resolved}=3.39\pm 0.14  \; .
\end{align}


\subsection{CP asymmetry}\label{sec:cpa}
The CP asymmetries in the low- and high-$q^2$ regions are of the order of $1\%$, with perturbative 
and parametric uncertainties of about $50\%$ and $40\%$, respectively, and dominated by the scale and $\alpha_s$.

In the low-$q^2$ region:
\begin{align}
\dps \overline{A}_\text{CP}[1,6]_{ee} =& \ (-1.45\pm 0.75_{\text{scale}}\pm 0.02_{m_t}\pm 0.02_{C,m_c}\pm 0.05_{m_b} \nonumber \\
& \hspace*{25pt} \pm 0.15_{\alpha _s}\pm 0.03_{\text{CKM}}\pm 0.002_{\lambda _2})\cdot 10^{-2}=( -1.45\pm 0.77)\cdot 10^{-2}  \; , \\
\dps \overline{A}_{\text{CP}}[1,6]_{\mu\mu} =& \ (-1.32\pm 0.71_{\text{scale}}\pm 0.01_{m_t}\pm 0.02_{C, m_c}\pm 0.05_{m_b}\nonumber \\
& \hspace*{25pt}\pm 0.15_{\alpha _s}\pm 0.03_{\text{CKM}}\pm 0.002_{\lambda _2})\cdot 10^{-2}= (-1.32\pm 0.72)\cdot 10^{-2} \; .
\end{align}
Not included here are resolved photon 
contributions where, contrary to CP-averaged quantities, the up-loop contribution does not vanish. An estimate of the size of these effects is not yet available.
A corresponding study for $\bar{B} \to X_{s(d)}\gamma$~\cite{Benzke:2010tq} has revealed that these contributions induce an uncertainty that can exceed the central value by
a large factor. An analogously large uncertainty is also possible here.

In the high-$q^2$ region, the perturbative uncertainty is drastically reduced. However, as for the branching ratio,
large uncertainties arise from the non-perturbative $1/m_b^{2,3}$ corrections.
\begin{align}
\dps \overline{A}_\text{CP}[>14.4]_{ee} =& \ (-1.91\pm 0.14_{\text{scale}}\pm 0.01_{m_t}\pm 0.12_{C,m_c}\pm 0.12_{m_b} \pm 0.05_{\alpha _s} \pm 0.05_{\text{CKM}}\nonumber \\
& \hspace*{22pt} \pm 0.28_{\lambda _2} \pm 0.34_{\rho_1}\pm 0.55_{f_{u,d}})\cdot 10^{-2}=( -1.91\pm 0.74)\cdot 10^{-2}  \; , \\
\dps \overline{A}_{\text{CP}}[>14.4]_{\mu\mu} =& \ (-1.84\pm 0.21_{\text{scale}}\pm 0.01_{m_t}\pm 0.09_{C, m_c}\pm 0.11_{m_b}\pm 0.04_{\alpha _s}\pm 0.05_{\text{CKM}}\nonumber \\
& \hspace*{22pt}\pm 0.24_{\lambda _2}\pm 0.30_{\rho_1}\pm 0.45_{f_{u,d}})\cdot 10^{-2}= (-1.84\pm 0.64)\cdot 10^{-2} \; .
\end{align}
We emphasize that for high-$q^2$, the nonfactorizable contributions from both charm and up-loops are taken into account (see section~\ref{sec:resolvedhigh}). We found that the contributions of the latter are negligible.


\section{Conclusion}
\label{sec:conclusion}

As a FCNC process the inclusive $\bar{B}\to X_d\ell^+\ell^-$ decay provides many observables sensitive 
to BSM physics. Contrary to the more frequently studied $\bar{B}\to X_s\ell^+\ell^-$ channel, $\bar{B}\to X_d\ell^+\ell^-$
receives contributions from the operators $P_{1,2}^u$ without CKM suppression and can thus 
yield more, complementary, information. In particular, the CP violation in $\bar{B}\to X_d\ell^+\ell^-$ is 
expected to be much larger than that of $\bar{B}\to X_s\ell^+\ell^-$. In the present work, we perform a state-of-the-art 
phenomenological analysis of $\bar{B}\to X_d\ell^+\ell^-$, providing the SM predictions for observables 
including the branching ratio, the forward-backward asymmetry and the CP asymmetry, which are quite promising 
to be studied at Belle~II.

Disentangling potentially small new-physics effects from SM uncertainties requires both precise theoretical predictions
and accurate experimental measurements. For inclusive FCNC decays, this not only necessitates the inclusion of perturbative
and local power corrections associated to the partonic rate, but also requires attention to additional long distance contributions on which we put
particular emphasis in the present work.

The most prominent among the long distance contributions arises from intermediate charmonium and light-quark resonances such
as $J/\psi$, $\rho$ and $\omega$, which are not captured by the local OPE. Even with kinematic cuts, the resonances may still 
have sizable effects on the observables, especially the high-$q^2$ ones. To handle the color-singlet resonance 
contributions, we adopt the Kr\"uger-Sehgal (KS) approach with improvement in several aspects compared to previous studies. The 
most up-to-date $e^+e^-\to\text{hadrons}$ and $\tau\to\nu+\text{hadrons}$ data are used to interpolate the imaginary 
parts of the KS functions. In the dispersion integral to obtain the real parts, we choose the subtraction point 
$s_0 = -(5 \mathrm{GeV})^2$ large and negative, which is not only far away from the charm and light-quark resonances but also 
avoids large logarithms $\log(s_0/\mu_b^2)$ from higher-order perturbative corrections.
Besides the flavoured $u$- and $c$-quark KS functions we also obtain the $d$- and $s$-quark KS functions which are also featured in $c \to u$ transitions.
At last, the one- and two-loop factorizable perturbative functions together get replaced by the corresponding KS functions. We find that 
the asymptotic behaviour of the perturbative and the KS functions match very well. We also study the 
uncertainties associated with the KS functions and find that they turn out to be negligible in the numerical results of the observables.

Beyond the KS treatment, the nonfactorizable resonance contributions to $\bar{B}\to X_d\ell^+\ell^-$ can also be considerable.
For the high-$q^2$ observables we adopt the description of the nonfactorizable charm-loop diagrams 
with a soft gluon connecting the quark loop and the spectator cloud from~\cite{Buchalla:1997ky}. It was pointed out
in~\cite{Buchalla:1997ky,Benzke:2017woq} that for large $q^2$ the effects due to nonfactorizable charm-loop diagrams are
local and hence easy to handle. The corresponding nonfactorizable up-loop contribution is obtained by taking the $m_c\to 0$ limit 
of the charm result. It turns out that both the nonfactorizable charm- and up-loop contributions are very 
small in the high-$q^2$ region.
In the low-$q^2$ region, the procedure in~\cite{Buchalla:1997ky} does not apply 
because the local heavy mass expansion fails. Such nonfactorizable contributions, including the one from charm loops, have been 
systematically studied within SCET by calculating resolved-photon contributions up to order $1/m_b$~\cite{Benzke:2017woq}.
We also discuss that the effects from up loops vanish in all CP-averaged quantities,
but might give rise to a large uncertainty in the CP asymmetry. 
Combing the results of~\cite{Benzke:2017woq} with a conservative estimate of the potentially large $1/m_b^2$ resolved 
contributions we find that they lead to an additional 5\% uncertainty in the branching ratio and the forward-backward asymmetry.

Finally, we identify another long distance effect contributing to the inclusive $\bar{B}\to X_{s,d}\ell^+\ell^-$
decays in the low-$q^2$ region, the cascade decays $\bar{B}\to X_1(c\bar{c}\to X_2\ell^+\ell^-)$.
In the context of inclusive FCNC decays they are considered for the first time in the present work.
We thoroughly investigate their kinematics and phase space distributions. It is found that under a typical cut $M_X < 2$~GeV 
taken in the experiments to suppress the double semileptonic background, the potentially most important cascade 
channels, $\bar B\to X_1(\psi\to \eta\ell^+\ell^-)$ and $\bar B\to X_1(\psi\to \eta'\ell^+\ell^-)$, contribute only 0.05\% respectively 0.0005\% to the 
total $\bar{B}\to X\ell^+\ell^-$ branching ratio. Therefore a hadronic mass cut effectively removes the pollution from the cascade decays as long as
$M_X < 3$~GeV. As mentioned earlier, the theoretical predictions in the present work are given for the case without a hadronic mass cut.
We leave the thorough theoretical and phenomenological study of the $\bar B\to X_{s(d)}\ell^+\ell^-$ decays with an $M_X$ cut to a future project.

In our calculation, we take into account all available perturbative and power corrections.
While many of the expressions for $\bar{B}\to X_d\ell^+\ell^-$ can be taken over from $\bar{B}\to X_s\ell^+\ell^-$,
effects from the current-current operators $P_{1,2}^u$ are new. In particular we derive the 
logarithmically-enhanced QED corrections associated with these operators and also include the 
partonic multi-particle contribution recently calculated in~\cite{Huber:2018gii}.
As for the power corrections, we extract the relevant HQET matrix elements that scale as $1/m_b^2$ and 
$1/m_b^3$ from inclusive semileptonic $B$- and $D$-decay data. We find that the poorly determined 
HQET matrix elements dominate the uncertainties in all high-$q^2$ observables. We therefore stress again that
lattice calulations of these HQET matrix elements should be performed, as they would significantly improve the theoretical precision
of semileptonic FCNC decays in the high-$q^2$ region.

We update the SM predictions for the branching ratio, the unnormalized and normalized forward-backward 
asymmetry, the zero crossing of the forward-backward asymmetry and the CP asymmetry of $\bar{B}\to X_d\ell^+\ell^-$
in the low-$q^2$ region. The uncertainties of the CP-averaged quantities are in general from 5\% to 20\%, except for the 
forward-backward asymmetries in the entire low-$q^2$ region for which the central values are small
due to the zero crossing in the middle of the low-$q^2$ region. The low-$q^2$ CP asymmetry may receive large and unknown
uncertainties from the resolved contributions in addition to the parametric and perturbative ones.
For the high-$q^2$ region, we give the predictions for the branching ratio, the observable
$\mathcal{R}(14.4)$ and the CP asymmetry. Owing to the poorly determined hadronic parameters characterising the $1/m_b^{2,3}$
power corrections, the branching ratio and the CP asymmetry have large relative uncertainties of ${\cal O}(40\%)$.
On the other hand, the uncertainties arising from the hadronic parameters get largely cancelled in the ratio
$\mathcal{R}(14.4)$ such that its uncertainty is smaller than 10\%.

In light of the anomalies persistent in exclusive $b \to s$ transitions, a cross-check via the corresponding inclusive
$b \to s$ and $b \to d$ modes is very much desired. The complementarity between inclusive and exclusive $b \to s$ decays
in the search for new physics was already pointed out in~\cite{Kou:2018nap}, and $b \to d$ transitions will yield additional useful
insights. The $\bar{B}\to X_d \ell^+\ell^-$ observables should therefore be measured in a dedicated Belle II analysis.


\subsubsection*{Acknowledgements}

We would like to thank Stefan de Boer, Paolo Gambino, Matteo Fael, Emilie Passemar and Alex Keshavarzi for useful discussions and correspondence. The work of JJ and EL was supported in part by the U.S. Department of Energy under grant number DE-SC0010120. T.~Huber, QQ, and KKV were supported by the Deutsche Forschungsgemeinschaft (DFG) within research unit FOR 1873 (QFET). The work of T.~Hurth was supported by the Cluster of Excellence `Precision Physics, Fundamental Interactions, and Structure of Matter' (PRISMA+ EXC 2118/1) funded by the German Research Foundation (DFG) within the German Excellence Strategy (Project ID 39083149).  He also thanks the CERN theory group for its hospitality during his regular visits to CERN where part of this work was written.   


\begin{appendix}


\section{Factorizable perturbative two-loop functions}
\label{sec:apptwoloop}

In order to take non-perturbative effects into account we use the KS-functions defined in sec.~\ref{sec:KS}. Explicitly, we replace
\begin{equation}
h^{\rm fact}_q \to h^{\rm KS}_q \ ,
\end{equation}
where the factorizable perturbative function is given by
\begin{equation}
h^{\rm fact}_q = h(y_{q}) + \as h^{(1)}_q \ .
\end{equation}
Here $h^{(1)}_q$
are the factorizable perturbative two-loop functions.
For the $q=u,d,s$ cases, the analytical expression is available:
\begin{equation}
h^{(1)}_{u,d,s} = \frac{-16 \log{(s/\mu_b^2)}}{9} - \frac{64 \zeta_3}{9}+\frac{196}{27} + i \frac{16\pi}{9} \ .
\end{equation}
For the charm case the functions can be found in~\cite{deBoer:2017way}. The fits in $s=q^2$ for our default value of $m_c$ read
\begin{equation}
h^{(1)}_c= \left\{ \begin{array}{ll} 
4.04 + 0.939 s - 0.0421s^2 + 0.0178 s^3  + \frac{32}{9} \log{\mu_b} \, , & \qquad \text{low-$q^2$}, \\
78.5 - 11.1 s + 0.472s^2 - 0.00683 s^3 + \frac{32}{9} \log{\mu_b} & \\
+i(105 - 9.42 s + 0.354 s^2 - 0.00475 s^3) \, ,& \qquad \text{high-$q^2$} ,
\end{array} \right.
 \end{equation}
where $s$ and $\mu_b$ have to be inserted in units of GeV$^2$ and GeV, respectively. We have checked that these functions are consistent with the corresponding two-loop 
photon self-energy functions calculated in \cite{Broadhurst:1993mw}.


\section{Log-enhanced electromagnetic corrections}
\label{sec:appomegaem}

Here we list the exact analytical expressions for the log-enhanced electromagnetic corrections 
as calculated in Ref.~\cite{Huber:2005ig,Huber:2007vv,Huber:2015sra}.
\bea \label{om1010em}
\omega_{1010}^{\rm (em)}(\s) & = &
\ln \left(\frac{m_b^2}{m_\ell^2}\right)\,\left[ 
\ln (1 - \s)
- \frac{1 + 4\,\s - 8\,\s^2}{
 6\,\left( 1 - \s \right)\,\left( 1 + 2\,\s \right) }
- \frac{\left( 1 - 6\,\s^2 + 4\,\s^3 \right) \,\ln \s}{
   {2\,\left( 1 - \s \right) }^2\,\left( 1 + 2\,\s \right) } \right] ,\\
\omega_{77}^{\rm (em)}(\s) & = &
\ln \left(\frac{m_b^2}{m_\ell^2}\right)\,\left[\frac{\s}{2\,{\left( 1 - \s \right) }\,\left( 2 + \s \right) } +
  \ln (1 - \s) -
  \frac{\s\,\left( -3 + 2\,\s^2 \right) }{2\,{\left( 1 - \s \right) }^2\,\left( 2 + \s \right) }\,\ln (\s)\right] ,\\
\omega_{79}^{\rm (em)}(\s) & = & \ln \left(\frac{m_b^2}{m_\ell^2}\right)\,\left[-\frac{1}{2\,( 1- \s)} + \ln (1 - \s) + \frac{\left(
-1 + 2\,\s - 2\,\s^2 \right) }{2\,{\left( 1 - \s \right) }^2}\,\ln (\s)\right], \\
\dps \omega^{\rm (em)}_{710}(\s)  &= & \dps
\ln \left(\frac{m_b^2}{m_\ell^2}\right)\,\left[ \frac{7 - 16\,\sqrt{\s} + 9\,\s}{
 4\,\left( 1 - \s \right)} + \ln (1 - \sqrt{\s}) + \frac{1+3 \,\s}{1-\s} \, \ln \!\left(\frac{1 + \sqrt{\s}}{2}\right) \right. \nnb\\
&&\left. - \frac{\s \,\ln \s}{\left( 1 - \s \right)} \right] \;,\\
\dps \omega^{\rm (em)}_{910}(\s)& = & \dps
\ln \left(\frac{m_b^2}{m_\ell^2}\right)\!\left[\ln (1 - \sqrt{\s}) -\frac{5 - 16\,\sqrt{\s} + 11\,\s}{
 4\,\left( 1 - \s \right)} + \frac{1-5 \,\s}{1-\s} \, \ln \!\left(\frac{1 + \sqrt{\s}}{2}\right) \right. \nnb\\
&&\left. - \frac{(1-3 \,\s) \ln \s}{\left( 1 - \s \right)} \right] , 
\eea
and
\bea
\omega_{ 99 }^{\rm (em)} (\s)& = &
\ln \left(\frac{m_b^2}{m_\ell^2}\right)\,\left[ - \frac{1 + 4\,\s - 8\,\s^2}{
 6\,\left( 1 - \s \right)\,\left( 1 + 2\,\s \right) }
+ \ln (1 - \s)
- \frac{\left( 1 - 6\,\s^2 + 4\,\s^3 \right) \,\ln \s}{
   {2\,\left( 1 - \s \right) }^2\,\left( 1 + 2\,\s \right) } \right]
\nonumber \\ && \hskip -1cm
- \frac{1}{9} \,\text{Li}_2(\s) + \frac{4}{27} \pi^2 -
\frac{37 - 3\,\s - 6\,\s^2}{72\,\left( 1 - \s \right)\,\left( 1 + 2\,\s \right)}  -
  \frac{\left( 41 + 76\,\s \right) \,\ln (1 - \s)}{36(1 + 2\,\s)} \nonumber\\
&& \hskip -1cm
+ \left( \frac{6 - 10\,\s - 17\,\s^2 + 14\,\s^3}{
18\left( 1 -\s \right)^2\,\left( 1 + 2\,\s \right) } +
\frac{17\,\ln (1 - \s)}{18} \right) \,\ln \s -
  \frac{\left( 1 - 6\,\s^2 + 4\,\s^3 \right) \, \ln^2 \s}{
2\left( 1 -\s \right)^2\,\left( 1 + 2\,\s \right) } \; .
\label{omegaem}
\eea

The operators $P_{1,2}^c$ and $P_{1,2}^u$ both contribute to the electromagnetic corrections. 
We therefore split their contributions. Functions for the $c$-part were already know from previous studies in 
$b\to s \ell^+\ell^-$~\cite{Huber:2005ig,Huber:2007vv,Huber:2015sra}. They are not known analytically, but are given in terms of fit functions 
for fixed default values of $m_b$ and $m_c$ by

\bea
\omega_{2c9}^{\rm (em)}(\s) & = &
\ln \left(\frac{m_b^2}{m_\ell^2}\right)\,\left[\frac{\Sigma_1(\s)+ i \,\Sigma_1^I(\s)}{8 (1-\s)^2 (1+2\s)}\right] + \frac{16}{9} \,
\omega_{1010}^{\rm (em)}(\s)\,\ln\!\left(\frac{\mu_b}{5\gev}\right)\;,\\
\omega_{2c2c}^{\rm (em)}(\s) & = &
\ln \left(\frac{m_b^2}{m_\ell^2}\right)\,\left[\frac{\Sigma_2(\s)}{8 (1-\s)^2 (1+2\s)} +
\, \frac{\Sigma_1(\s)}{9 (1-\s)^2 (1+2\s)}\ln\!\left(\frac{\mu_b}{5\gev}\right)\right] \nnb\\ & & \nnb \\
    &&+ \, \frac{64}{81} \; \omega_{1010}^{\rm (em)}(\s)\, \ln^2\!\left(\frac{\mu_b}{5\gev}\right)
\;,\\ & & \nnb\\
\omega_{2c7}^{\rm (em)}(\s) & = &
\ln \left(\frac{m_b^2}{m_\ell^2}\right)\,\left[\frac{\Sigma_3(\s)+ i \, \Sigma_3^I(\s)}{96 (1-\s)^2}\right] + \frac{8}{9} \,
\omega_{79}^{\rm (em)}(\s) \, \ln\!\left(\frac{\mu_b}{5\gev}\right)\;,
\label{om79em} \\
\dps\omega^{\rm (em)}_{2c10}(\s) &=& \dps
\ln \left(\frac{m_b^2}{m_\ell^2}\right)\,\left( -\frac{\Sigma_7(\s) +i \Sigma^I_7(\s) }{24\s(1-\s)^2} \right)
+ \frac{8}{9} \, \omega^{\rm (em)}_{910,A}(\s)\,\ln\!\left(\frac{\mu_b}{5\gev}\right)\, .
\eea
\bea
\dps\omega^{\rm (em)}_{2c10}(\s) &=&\ \dps
\ln \left(\frac{m_b^2}{m_\ell^2}\right)\,\left( -\frac{\Sigma_7(\s) +i \Sigma^I_7(\s) }{24\s(1-\s)^2} \right)
+ \frac{8}{9} \, \omega^{\rm (em)}_{910,A}(\s)\,\ln\!\left(\frac{\mu_b}{5\gev}\right)\, ,
\eea
The $\Sigma$ functions read 
\bea
\Sigma_1(\s) &=&\left\{ \begin{array}{ll} 23.787 - 120.948\, \s + 365.373\, \s^2 - 584.206\, \s^3,
&   \text{low-$\s$} ; \\
-153.673 \, \delta^2 + 498.823 \, \delta^3 - 1146.74 \, \delta^4 + 1138.81 \, \delta^5,\;  &   \text{high-$\s$}, \\
\end{array}\right.  \\
\Sigma_1^I(\s) &=&\left\{ \begin{array}{ll} 1.653 + 6.009\, \s - 17.080\, \s^2 + 115.880\, \s^3, & \text{low-$\s$}; \\
- 255.712 \, \delta^2 + 1139.10 \, \delta^3 - 2414.21 \, \delta^4 + 2379.91 \, \delta^5, \; &    \text{high-$\s$}, \\
\end{array}\right. \\
\Sigma_2(\s) &=&\left\{ \begin{array}{ll} 11.488 - 36.987\, \s + 255.330\, \s^2 - 812.388 \, \s^3  + 1011.791 \, \s^4,
&  \text{low-$\s$}; \\
- 220.101 \, \delta^2 + 875.703 \, \delta^3 - 1920.56 \, \delta^4  + 1822.07 \, \delta^5 ,\; &    \text{high-$\s$}, \\
\end{array}\right.  \\
\Sigma_3(\s) &=&\left\{ \begin{array}{ll} 109.311 - 846.039\, \s + 2890.115\, \s^2 - 4179.072\, \s^3,
&  \text{low-$\s$} ; \\
- 310.113 \, \delta^2 + 834.253 \, \delta^3 - 2181.94 \, \delta^4 + 2133.78 \, \delta^5, \; &    \text{high-$\s$}, \\
\end{array}\right. \\
\Sigma_3^I(\s) &=&\left\{ \begin{array}{ll} 4.606 + 17.650\, \s - 53.244\, \s^2 + 348.069\, \s^3 \;,
&    \text{low-$\s$}; \\
- 518.180 \, \delta^2 + 2047.18 \, \delta^3 - 4470.04 \, \delta^4 + 4827.74 \, \delta^5, \; &    \text{high-$\s$}, \\
\end{array}\right.  \\
\Sigma_7(\s) &=&\left\{ \begin{array}{ll} 351.322 \s^4-378.173 \s^3+160.158 \s^2 -24.2096 \s  -0.305176,
&   \text{low-$\s$}; \\
77.0256 \, \delta^2 - 264.705 \, \delta^3 + 595.814 \, \delta^4 - 610.1637 \, \delta^5,\;  &   \text{high-$\s$},  \\
\end{array}\right.  \\
\Sigma_7^I(\s) &=&\left\{ \begin{array}{ll} [{-7.98625 -238.507 \, b +766.869\, b^2}]\, b^2\, \theta(b), &   \text{low-$\s$}; \\
135.858\, \delta^2 - 618.990\, \delta^3 +1325.040\, \delta^4 - 1277.170\, \delta^5, \; &   \text{high-$\s$}, \\
\end{array}\right.
\eea
with $\delta \equiv 1- \s$ and $b=\s- (4 m_c^2/m_b^2)^2$. 
The polynomials in the high-$\s$-region were obtained such as to have a double zero at $\s=1$.

The new contributions induced by $P_{1,2}^u$ can be obtained analytically for all but the $2u2c$ interference term. We find

\bea
\omega_{2u7}^{\rm (em)}(\s) & = &  \
\omega_{79}^{\rm (em)}(\s)  \left(\frac{8}{9} \ln \left(\frac{\mu_b}{m_b}\right)+\frac{8}{27}+\frac{4 i \pi }{9}\right) + \ln \left(\frac{m_b^2}{m_\ell^2}\right)\,\\
&&  \times  \left(\frac{4 \text{Li}_2(\s)}{9}+\frac{\left(2 \s^2-2 \s+1\right) \ln ^2(\s)}{9 (\s-1)^2}-\frac{5 \s-3}{18 (\s-1)}  +\frac{(2 \s-1) \ln (\s)}{9 (\s-1)^2}-\frac{2 \pi ^2}{27}\right), \nonumber \\
\omega_{2u9}^{\rm (em)}(\s) & = & 2 \omega_{1010}^{\rm (em)}(\s)  \left(\frac{8}{9} \ln \left(\frac{\mu_b}{m_b}\right)+\frac{8}{27}+\frac{4 i \pi }{9}\right) +  \ln \left(\frac{m_b^2}{m_\ell^2}\right)\, \Sigma_4(\s)\; ,
\eea

\bea
\omega_{2u2u}^{\rm (em)}(\s) & = &  \frac{8}{9} \ln \left(\frac{\mu_b}{m_b}\right) \Sigma_4(\s) \ln \left(\frac{m_b^2}{m_\ell^2}\right)
 +\frac{64}{81}  \omega_{1010}^{\rm (em)}(\s)  \left(\ln ^2\left(\frac{\mu_b}{m_b}\right)+\frac{2}{3} \ln \left(\frac{\mu_b}{m_b}\right)\right)  \\
&& + \ln \left(\frac{m_b^2}{m_\ell^2}\right) \left[ \frac{32 \zeta (3)}{81} + \frac{64 \text{Li}_2(\s)}{243}-\frac{32 \text{Li}_3(\s)}{81}-\frac{8 \pi ^2 \left(32 \s^2-16 \s-7\right)}{729 (\s-1) (2 \s+1)}  \right. \nonumber \\
&& -\frac{8 (2 \s-1) \left(2 \s^2-2 \s-1\right) \ln ^3(\s)}{243 (\s-1)^2 (2 \s+1)}
+\frac{4 \left(202 \s^3-87 \s^2-56\right) \ln (\s)}{729 (\s-1)^2 (2 \s+1)}
 \nonumber \\
&&  -\frac{8 \pi ^2 \left(8 \s^3-12 \s^2+1\right) \ln (\s)}{243 (\s-1)^2 (2 \s+1)}
-\frac{4 \left(24 \s^2-7\right) \ln ^2(\s)}{243 (\s-1)^2 (2 \s+1)} 
+\frac{16}{81} \pi ^2 \ln (1-\s) \qquad \nnb\\
&&\left. +\frac{64}{729} \ln (1-\s)
-\frac{2 \left(1742 \s^2-1159 \s-229\right)}{2187 (\s-1) (2 \s+1)}  \right], \nnb
\eea

\bea
\omega_{2u10}^{\rm (em)}(\s) & = & \ln \left(\frac{m_b^2}{m_\ell^2}\right)
\left(-\frac{4\text{Li}_2\left(1-\sqrt{{\s}}\right)}{9} + \frac{4 (5 {\s}-1)\text{Li}_2\left(-\sqrt{{\s}}\right)}{9 ({\s}-1)}
+\frac{2 \left(3 \sqrt{{\s}}-1\right)}{9\left(\sqrt{{\s}}+1\right)} \right. \nonumber\\
&&+\frac{\pi ^2 (5 {\s}-1)}{27 ({\s}-1)}+\frac{(3 {\s}-1)\ln ^2({\s})}{3 ({\s}-1)}
- \frac{\left(7 {\s}-8 \sqrt{{\s}}+3\right) \ln({\s})}{9 \left({\s}-1\right) } -\frac{4}{9} \ln\left(1-\sqrt{{\s}}\right) \ln ({\s}) 
 \nonumber\\
&&\left. 
-\frac{2 (5 {\s}-1) \ln \left(\frac{1}{4} \left(\sqrt{{\s}}+1\right)\right) \ln ({\s})}{9 ({\s}-1)}
-\frac{8}{9} \ln \left(\frac{1}{2} \left(\sqrt{{\s}}+1\right)\right)\right)
\nonumber\\
&&+\left(\frac{8}{9} \ln\left(\frac{\mu_b}{m_b}\right)+\frac{4 i \pi }{9}+\frac{8}{27}\right) \ln \left(\frac{m_b^2}{m_\ell^2}\right)
\left(-\frac{11 {\s}-16 \sqrt{{\s}}+5}{4(1-{\s})} \right. \nnb\\ 
&& \left.
   +\ln \left(1-\sqrt{{\s}}\right)+\frac{(1-5 {\s}) \ln \left(\frac{1}{2}
   \left(\sqrt{{\s}}+1\right)\right)}{1-{\s}}-\frac{(1-3 {\s}) \ln
   ({\s})}{1-{\s}}\right),  
\eea
\bea
\omega_{2u2c}^{\rm (em)}(\s) & = &  \omega_{2c9}^{\rm (em)*}(\s)
\left(\frac{8}{9} \ln \left(\frac{\mu_b}{m_b}\right)+\frac{8}{27}+\frac{4 i \pi }{9}\right)
+ \ln \left(\frac{m_b^2}{m_\ell^2}\right) \frac{3 \left(\Sigma_5({\s}) +i\Sigma_5^I(\s) \right)}{16(1-{\s})^2 (2 {\s}+1)} \nonumber\\
&&+\ln \left(\frac{m_b^2}{m_\ell^2}\right)\left(\frac{8}{9} \ln \left(\frac{\mu_b}{m_c}\right)+\frac{8}{27}\right)\Sigma_4(\s), 
\eea
where
\bea
\Sigma_4(\s) &=& \frac{8 \text{Li}_2(\s)}{9} -\frac{4 \pi ^2}{27} +\frac{2 \left(16 \s^3-3\right) \ln (\s)}{27 (\s-1)^2 (2 \s+1)}
-\frac{170 \s^2-109 \s+17}{81 (\s-1) (2 \s+1)}  \nonumber\\
&&+\frac{2 (2 \s-1) \left(2 \s^2-2 \s-1\right) \ln ^2(\s)}{9 (\s-1)^2 (2 \s+1)}\; , \\
\Sigma_5(\s) &=&\left\{ \begin{array}{ll} -3.057\, -1139.24 {\s}^4+1123.24 {\s}^3 & \\
-466.284 {\s}^2+90.282 {\s}-{0.181}/{{\s}},\; &    \text{low-$\s$}; \\
-11.1584 \delta ^5-0.00857151 \delta ^2 \ln (\delta ) & \\
-25.0743 \delta ^4-10.5518 \delta ^3 \ln (\delta), &    \text{high-$\s$}, \\
\end{array}\right.  \\
\Sigma_5^I(\s) &=&\left\{ \begin{array}{ll} -0.226\, -54.066 {\s}^4+22.676 {\s}^3 & \\
-6.460 {\s}^2+0.0955{\s}+{0.000263}/{{\s}},\; &    \text{low-$\s$}; \\
-106.547 \delta ^5-0.130667 \delta ^2 \ln (\delta ) & \\
+37.6882 \delta ^4-22.5916 \delta ^3 \ln (\delta ), &    \text{high-$\s$}. \\
\end{array}\right. 
\eea

\end{appendix}


\bibliographystyle{JHEP} 

\providecommand{\href}[2]{#2}\begingroup\raggedright\endgroup

\end{document}